%% file: uc_review.tex
\documentclass[fleqn,10pt]{wlscirep}
\usepackage[utf8]{inputenc}
\usepackage[T1]{fontenc}
\usepackage{bm}

\usepackage{makecell}
\usepackage[most]{tcolorbox}
\usepackage[normalem]{ulem}

\newcommand{\bew}{\begin{widetext}}
\newcommand{\ew}{\end{widetext}}

\newcommand{\ii}{{\rm i}}
\newcommand{\bx}{\mathbf{x}}
\newcommand{\by}{\mathbf{y}}
\newcommand{\bz}{\mathbf{z}}
\newcommand{\bp}{\mathbf{p}}
\newcommand{\bq}{\mathbf{q}}
\newcommand{\bv}{\mathbf{v}}

\newcommand{\br}{\mathbf{r}} 

\newcommand{\bG}{\mathbf{G}}
\newcommand{\bH}{\mathbf{H}}
\newcommand{\bg}{\mathbf{g}}
\newcommand{\bff}{\mathbf{f}}
\newcommand{\bu}{\mathbf{u}}

\newcommand{\bw}{\mathbf{w}}

\newcommand{\bk}{\mathbf{k}}

\newcommand{\bP}{\mathbf{P}}

\newcommand{\sep}{ \ \ \ , \ \ \ }

\newcommand{\beq}{\begin{equation}}
\newcommand{\eeq}{\end{equation}}
\newcommand{\beqn}{\begin{eqnarray}}
\newcommand{\eeqn}{\end{eqnarray}}
\newcommand{\pp}{\partial}
\newcommand{\dd}{{\rm d}}
\newcommand{\ee}{{\rm e}}

\newcommand{\cO}{{\cal O}}

\newcommand{\la}{\langle}

\newcommand{\ra}{\rangle}

\newcommand{\vnab}{{\bf \nabla}}

\newcommand{\diagram}[1]{\begin{array}{l}\includegraphics[page=#1,trim={0 3.2pt 0 0}]{diagrams.pdf}\end{array}}

\title{Hydrodynamics, Renormalization Group, and Universality Classes Far from Equilibrium}

\author[1,*]{Patrick Jentsch}
\author[2,$\dagger$]{Chiu Fan Lee}
\affil[1]{1Cell Biology and Biophysics Unit, European Molecular Biology Laboratory, Meyerhofstraße 1, 69117 Heidelberg, Germany}
\affil[2]{Department of Bioengineering, Imperial College London, South Kensington Campus, London SW7 2AZ, United Kingdom}
\affil[*]{e-mail: patrick.jentsch@embl.de}
\affil[$\dagger$]{e-mail: c.lee@imperial.ac.uk}

\begin{abstract}
Universality is one of the central organising principles of modern physics, explaining why systems with vastly different microscopic constituents can exhibit identical large-scale behaviour. While the classification of equilibrium critical phenomena through hydrodynamics and the renormalization group (RG) is now well established, our understanding of universality far from equilibrium remains far less developed. In recent years, however, rapid progress -- driven in large part by developments in active and living matter -- has uncovered a growing range of genuinely nonequilibrium universality classes (UCs) with no equilibrium counterparts.
In this review, we present a pedagogical and unified introduction to hydrodynamic and RG approaches to nonequilibrium many-body systems. We first show how hydrodynamic theories can be systematically constructed from symmetry and conservation laws alone. We then introduce perturbative dynamic RG methods and demonstrate how hydrodynamic theories are organised into distinct UCs according to their scaling behaviour. Building on these foundations, we review the diverse nonequilibrium UCs uncovered since 2015, while emphasizing the conceptual connections and unifying physical principles underlying their emergence. We conclude by discussing open theoretical and experimental challenges for the field.
\end{abstract}
\begin{document}

\flushbottom
\maketitle

\thispagestyle{empty}

\section{Introduction}

How is physics different from other scientific disciplines? To many, the defining feature is the generality of its findings. Physical theories often reveal principles that apply across systems whose microscopic constituents and mechanisms differ widely, allowing seemingly disparate phenomena to exhibit remarkably similar large-scale behaviour. This capacity to uncover common laws governing qualitatively different systems distinguishes physics among the natural sciences.

This perspective is reflected in the strategy often adopted by theoretical physicists when attempting to understand a phenomenon. Rather than beginning with a detailed description of every component of a system, the aim is typically to identify a {\bf minimal model} containing only the essential ingredients required to reproduce the behaviour of interest. Once such a model is constructed, the next step is to derive predictions from it and confront those predictions with empirical observations.

The search for such minimal descriptions is commonly justified through the principle of {\bf Occam’s razor}, according to which the simplest explanation consistent with the observations should be preferred. In this sense, constructing a model capable of reproducing a phenomenon is not itself surprising; after all, the model has been designed precisely to generate the behaviour under consideration.
Far more remarkable, however, is when such minimal models yield predictive and testable consequences extending beyond the observations used to motivate them, and when these predictions are subsequently borne out experimentally.

This recurring success raises a deeper conceptual question: {\it why do relatively simple theoretical descriptions capture the behaviour of complex natural systems so effectively?}

In this review, through the lens of nonequilibrium many-body systems, we argue that the remarkable success of this programme can be understood through the mathematical frameworks of {\bf hydrodynamic theory} and the {\bf renormalization group (RG)}.  Hydrodynamic descriptions capture the emergent dynamics of systems at large spatial and temporal scales by focusing on a small set of slow collective variables dictated by symmetries and conservation laws. Complementarily, RG theory explains why many microscopic details become irrelevant at long wavelengths, such that macroscopic behaviour depends only on a limited number of effective parameters. Together, these frameworks explain why simple models can often faithfully reproduce, organise, and predict the behaviour of systems whose microscopic constituents and interactions may be extraordinarily complex. From this perspective, the progression from empirical observation to minimal modelling and ultimately to quantitative prediction reflects a deeper structural feature of many-body physics: emergent collective behaviour is often governed primarily by a small set of qualitative ingredients --- hydrodynamic variables, symmetries, and conservation laws --- while microscopic details play only a secondary role.

These ideas have enabled physicists to classify equilibrium critical phenomena into distinct {\bf universality classes (UC)} through the RG.\cite{hohenberg_rmp77} For much of the period following the development of RG theory in the 1970s, research on universality focused predominantly on equilibrium and near-equilibrium passive systems, with notable exceptions such as the Kardar-Parisi-Zhang equation.\cite{kardar_prl86} Over the past decade, however, the universality landscape has expanded rapidly with the discovery of a growing number of genuinely nonequilibrium UCs. These arise in systems ranging from active and living matter to driven diffusive systems, reaction-diffusion processes, growing interfaces, and absorbing-state transitions. Among these, active matter has emerged as a particularly fertile arena for uncovering new forms of universal behaviour, a development that can arguably be traced back to the seminal discovery of long-range order in two-dimensional flocks by Vicsek {\it et al.}\cite{vicsek_prl95} and by Toner and Tu.\cite{toner_prl95}

The primary aim of this review is to provide a comprehensive account of these recent developments while systematically elucidating the scientific chain linking empirical observation, hydrodynamic theory, RG analysis, and ultimately the identification of UCs. To this end, we begin in {\bf Section~\ref{sec:universal_models}} with a pedagogical exposition of how a hydrodynamic model can be constructed systematically from symmetry principles and conservation laws alone. In {\bf Section~\ref{sec:rg}}, we introduce the dynamic RG and show how hydrodynamic theories can be organised into distinct UCs according to their large-scale behaviour. Building on these foundations, {\bf Section~\ref{sec:new_ucs}} surveys the diverse nonequilibrium UCs uncovered since 2015 and highlights the connections between them. Finally, in {\bf Section~\ref{sec:conclusion}}, we discuss open theoretical questions, experimental challenges, and promising future directions.

Ultimately, we hope this review will serve three complementary purposes:

\begin{enumerate}
\item
to provide readers with the tools needed to construct hydrodynamic theories for a wide range of nonequilibrium systems;
\item
to equip readers with the practical knowledge required to perform perturbative dynamic RG calculations; and
\item
to offer a comprehensive overview of recently discovered nonequilibrium UCs and the relationships between them.
\end{enumerate}

\include{SectionTwo_v2.tex}

\include{SectionThree_v3.tex}

\include{SectionFour_v2.tex}

\section{Conclusion and outlook}
\label{sec:conclusion}
As outlined in the Introduction, this review has pursued three primary objectives: to provide readers with the tools needed to construct hydrodynamic theories, to introduce the practical application of perturbative dynamic renormalization group (RG) methods, and to present a coherent overview of recently identified nonequilibrium universality classes (UCs) and the connections between them. Together, these elements establish a unified framework that links empirical observations to the universal large-scale behaviour they exhibit, bridging the gap between phenomenology, minimal modelling, and UC-based classification. 
This framework can be summarised schematically as
\[ 
\begin{array}{c} \text{Phenomenon}\\ \text{(observation)} 
\end{array} \rightarrow 
\begin{array}{c} 
\text{Hydrodynamic variables}\\ \text{+ SCC} 
\end{array} 
\rightarrow 
\begin{array}{c} \text{Hydrodynamic}\\ \text{EOM} 
\end{array} \rightarrow 
\begin{array}{c} \text{Phases / }\\ \text{critical points} 
\end{array} \rightarrow \begin{array}{c} \text{RG} 
\end{array} \rightarrow 
\begin{array}{c} \text{Fixed points / }\\ \text{universality classes} 
\end{array}. 
\]

Looking ahead, we will focus on what we view as the two major frontiers of the field: the experimental exploration of emergent scaling behaviour in active and living systems, and the theoretical discovery and classification of new nonequilibrium UCs. To make these challenges concrete, we conclude each of the following sections by highlighting one specific {\bf grand challenge} that, in our view, captures a central open problem at the current experimental or theoretical frontier of nonequilibrium universality.

\subsection{Experimental frontiers}

As an empirical science, physics ultimately requires that theoretical predictions be confronted with experiments. Although many of the nonequilibrium universality classes reviewed here are now theoretically well established, systematic experimental tests remain comparatively limited. Nevertheless, a growing body of evidence suggests that collective scale-invariant behaviour can indeed emerge in active systems. Notable examples include natural starling flocks, in which velocity correlations remain scale-free across the entire flock size \cite{cavagna_pnas10}, and midge swarms, where finite-size scaling and dynamic scaling analyses revealed behaviour consistent with proximity to a scale-invariant critical state \cite{attanasi_prl14,cavagna_natphys17}. These observations demonstrate that scale-invariant signatures are experimentally accessible in living active systems. 
However, a direct connection to a specific UC, including the measurement of all predicted scaling exponents, has yet to be established.

Among controlled laboratory systems, Quincke roller suspensions provide perhaps the closest experimental realization of a polar active matter system currently available \cite{bricard_nature13}. Consisting of dielectric colloids driven into self-propelled motion by an external electric field, these systems exhibit robust flocking behaviour and support propagating collective modes whose properties can be measured quantitatively \cite{geyer_prx19}. Their high degree of experimental control makes them particularly attractive for quantitative tests of Toner--Tu-type predictions. Although immersed in a fluid, strong substrate friction and confinement strongly screen hydrodynamic interactions, rendering the fluid velocity a heavily damped mode over experimentally accessible scales. Nevertheless, the fluid velocity  remains an additional conserved field, so these systems are more properly viewed as screened wet active matter rather than genuinely dry active fluids.

A complementary experimental platform is provided by active nematics assembled from microtubules and kinesin motors \cite{sanchez_nature12}. These systems have enabled quantitative studies of active turbulence, self-propelled topological defects, and defect ordering \cite{decamp_natmat15,keber_science14}, establishing one of the most successful experimental programmes in active matter. Although no new UCs have yet been found in active nematic systems, these experiments have repeatedly uncovered qualitatively new forms of collective behaviour and remain a promising arena for future discoveries. More broadly, their success demonstrates how carefully engineered active materials can reveal emergent phenomena inaccessible in equilibrium systems.

Some other promising opportunities are living active matter embedded within structured environments. As discussed in {\bf Section~\ref{sec:new_ucs}}, suspensions of motile cells, amoebae, or bacteria in porous media provide a natural platform for probing incompressible, Malthusian, and chemotactic active-fluid regimes. The porous matrix acts as a momentum sink, suppressing long-range hydrodynamic interactions and promoting effectively dry behaviour, while confinement and substrate structure can often be tuned independently. More generally, recent studies of collective migration in structured extracellular matrices demonstrate the feasibility of studying active systems under controlled confinement and mechanical feedback \cite{alert_annrev20}. 

Particularly intriguing in this context is the recent observation that a strain of {\it Dictyostelium discoideum} deficient in chemotaxis exhibits persistent motion, weak chiral symmetry breaking, and alignment interactions consistent with a two-dimensional chiral Vicsek model \cite{hayakawa_a25}. Although two-dimensional chiral flocks were recently argued not to support a true flocking phase \cite{chen_pre26}, suitable chirally symmetric active species may nevertheless be identified or engineered in the near future. Such systems would provide an attractive route toward direct experimental tests of incompressible, Malthusian, and chiral active fluid UCs.

Ultimately, one of the central challenges for the field is to develop experimental platforms in which hydrodynamic interactions, confinement, dissipation, alignment, and conservation laws can all be tuned independently. Such systems would enable not only stringent quantitative tests of existing universality classes, but also the discovery of entirely new nonequilibrium scaling regimes. Achieving this level of experimental control will be essential if the study of nonequilibrium universality is to evolve from a predominantly theoretical framework into a quantitatively predictive experimental science.

\bigskip
\noindent
{\bf Grand challenge:} Experimentally identify and fully characterise scaling behaviour of a new or theoretically predicted universality class in living or man-made active matter systems.

\subsection{Theoretical frontiers}

While experimental tests are still catching up, theoretical developments also remain far from complete. Indeed, the rapid emergence of new nonequilibrium UCs over the past decade suggests that current classifications remain far from exhaustive. Many additional UCs likely await discovery, particularly in systems with larger numbers of hydrodynamic variables, competing symmetries, and more complex nonlinear couplings (see, e.g., the missing entries in Fig.~\ref{fig:periodictab}). However, meaningful progress within this expanding landscape will likely depend on physically motivated models, rather than purely formal generalisations.

In this regard, one particularly promising direction is the extension of polar active-fluid theories to multi-species systems. Examples include mixtures of fast and slow self-propelled particles, predator--prey active agents, chemically interacting species, and active colloids with distinct alignment rules or asymmetric and nonreciprocal couplings. Recent studies have already shown that such ingredients can qualitatively alter collective dynamics, producing phenomena absent in single-component flocks, including chasing states, oscillatory patterns, chiral motion, species segregation, and unconventional propagating modes. From the perspective of universality, however, the most important feature of these systems is the emergence of additional hydrodynamic fields, conservation laws, broken symmetries, and nonlinear mode couplings. These ingredients can fundamentally modify the RG flow structure and therefore provide natural routes toward entirely new forms of scale-invariant behaviour and new nonequilibrium universality classes.

At the same time, several fundamental problems associated with established active-fluid phases remain unresolved. These include determining the UCs governing the ordered phases of generic compressible polar flocks (the Toner--Tu phase) and two-dimensional Malthusian flocks \cite{chate_prl24,chen_a25}. Among these, the generic Toner--Tu problem remains one of the foremost open challenges in active matter and nonequilibrium statistical physics. Unlike most universality classes discussed in this Review, it involves multiple coupled hydrodynamic fields, 
making even the linear theory considerably more complex while introducing a large number of relevant nonlinearities
 that continue to resist existing analytical approaches. In particular, the physically most relevant two-dimensional case is expected to be perturbatively inaccessible. Its resolution would not only complete the classification of a central active-fluid phase, but also provide a stringent test of current RG methods and advance our understanding of scale invariance in strongly interacting nonequilibrium systems.

The analytical grand challenge is therefore clear:

\bigskip
\noindent
{\bf Grand challenge:} Identify the universality class governing the ordered phase of the generic compressible Toner--Tu model and determine its asymptotic scaling behaviour  in two and three dimensions.

\subsection{Final perspective}
The growing recognition of biological and active systems as sources of fundamentally new physical phenomena echoes the mid-twentieth-century realization that everyday materials harbour deep and unexpected collective behaviour -- a development that ultimately gave rise to condensed matter physics, now the largest subfield of physics. However, the emergence of condensed matter physics was driven not only by the discovery of a remarkably fertile landscape of new physics, but also by rapid experimental advances and the immense technological relevance of materials science. In many ways, these same ingredients are now in place in the study of active and living matter: a wealth of novel nonequilibrium phenomena, increasingly quantitative experimental tools spanning molecular to organismal scales, and direct relevance to life, health, and biotechnology.

Just as condensed matter physics reshaped both technology and our conceptual understanding of nature, we anticipate that the study of living and active matter will play an analogous role in the decades ahead. In this sense, active and living matter may become one of the next central frontiers of physics, reshaping  our understanding of matter, life, and the principles that govern both.

\section*{Acknowledgements}

C.F.L. thanks John Toner and Leiming Chen for more than a decade of inspiring collaborations and for generously sharing their insights into renormalization group theory, universality, and active matter.

\bibliography{references}

\newpage

\appendix

\section{Graphical Corrections}
\label{appendix}

We present here the 1-loop calculations of the remaining couplings for the Navier-Stokes equation not shown in {\bf Section~\ref{sec:rg}}.

\subsection*{Renormalization of field expectation value $\bv_0$}

We start with the easiest diagram, which is the renormalization of the expectation value of the field $\bv_0$ whose bare value we set to zero without loss of generality. In the general case, the field expectation value can be shifted due to nonlinear fluctuations. Here its loop correction is given by,
\begin{equation}
    \diagram{49}  =\int_{\tilde \bp} (-\ii \lambda) \Big(p_j \delta_{ik}-p_k\delta_{ij}\Big) \frac{2D P_{jk}(\bp)}{|-\ii\omega_p+\mu |\bq/2-\bp|^2|^2} = 0 \ ,
\end{equation}
where the last  equality follows directly from contracting the $\bp$'s with the transverse projector. Therefore the flow equation of the expectation value is trivially,
\begin{equation}    
    \frac{\dd \bv_0}{\dd \ell} = 0 \ ,
\end{equation}
which is simply reflecting conservation of momentum.

\subsection*{Renormalization of propagator}

The propagator has only one parameter $\mu$ which can be obtained by projecting the inverse propagator onto its transverse component,
\begin{equation}
     \frac{P_{ij}(\bq)}{d-1} \diagram{1}^{-1} = -\ii \omega + \mu \bq^2
\end{equation}
and taking the term of second order in $\bq$.

Applying the same projection to the loop correction and expanding to second order in $\bq$ yields,
\begin{align}
    \label{eq:calc_loop1_detail}
    \nonumber
    &-4\diagram{25} \frac{P_{in}(\bq)}{d-1} \\
    \nonumber
    &=- \frac{P_{in}(\bq)}{d-1}\int_{\tilde \bp} (-\ii \lambda) \Big(\Big(p_j+\frac{q_j}{2}\Big) \delta_{ik}+\Big(\frac{q_k}{2}-p_k\Big)\delta_{ij}\Big) \frac{P_{jl}(\bq/2-\bp)}{-\ii \omega_q/2+\ii\omega_p+\mu |\bq/2-\bp|^2} \\
    \nonumber
    &\hspace{1.1cm} \times (-\ii \lambda) \Big(-\Big(p_n+\frac{q_n}{2}\Big)\delta_{ml}+q_m\delta_{nl}\Big) \frac{2D \Big|\frac{\bq}{2}+\bp\Big|^2 P_{mk}(\bq/2+\bp)}{|-\ii\omega_q/2-\ii \omega_p+\mu |\bq/2+\bp|^2|^2} \\
    \nonumber
    &=- \frac{P_{in}(\bq)}{d-1}\int_{\tilde \bp} (-\ii \lambda) \Big(q_j \delta_{ik}+q_k\delta_{ij}\Big) \frac{P_{jl}(\bq/2-\bp)}{-\ii \omega_q/2+\ii\omega_p+\mu |\bq/2-\bp|^2} \\
    \nonumber
    &\hspace{1.1cm} \times (-\ii \lambda) \Big(-p_n\delta_{ml}+q_m\delta_{nl}\Big) \frac{2D \Big|\frac{\bq}{2}+\bp\Big|^2 P_{mk}(\bq/2+\bp)}{|-\ii\omega_q/2-\ii \omega_p+\mu |\bq/2+\bp|^2|^2} \\
    \nonumber
    &= \frac{P_{in}(\bq)}{d-1} \lambda^2 \Big(q_j \delta_{ik}+q_k\delta_{ij}\Big)  \int_{\tilde \bp} \Bigg\{  \frac{P_{jn}(\bp)}{\ii\omega_p+\mu |\bp|^2}   \frac{2D p^2 q_mP_{mk}(\bp)}{|-\ii \omega_p+\mu |\bp|^2|^2} - \frac{P_{jk}(\bp)}{\ii\omega_p+\mu |\bp|^2}   \frac{2D \bp\cdot\bq p_n }{|-\ii \omega_p+\mu |\bp|^2|^2} \\
    \nonumber
    &\hspace{1.1cm} +\frac{1}{\ii\omega_p+\mu |\bp|^2} \frac{D p^2}{|-\ii \omega_p+\mu |\bp|^2|^2} \frac{ \Big(q_j p_n p_k-q_k p_n p_j\Big)}{p^2} +\frac{2D p^2\mu  }{|\ii\omega_p+\mu |\bp|^2|^4} p_n \bq\cdot \bp P_{jk}(\bp)
    \Bigg\} + \mathcal{O}(q^4,q^2\omega_q)\\
    \nonumber
    &=  \frac{1}{4}\frac{D \lambda^2}{ (d-1)\mu^2} \int_{\bp} \frac{1}{p^2} \Big(q_j \delta_{ik}+q_k\delta_{ij}\Big)\Bigg\{ 2 q_m P_{mk}(\bp)P_{jn}(\bp)P_{in}(\bq)+P_{in}(\bq) \frac{1}{p^2}(q_j p_n p_k-q_kp_np_j)
    \Bigg\} + \mathcal{O}(q^4,q^2\omega_q)\\
    \nonumber
    &=  \frac{1}{2}\frac{D \lambda^2}{ (d-1)\mu^2} \int_{\bp} \frac{1}{p^2} \Bigg\{ \Big(q_j \delta_{ik}+q_k\delta_{ij}\Big) q_m P_{mk}(\bp) P_{jn}(\bp)  P_{in}(\bq) 
    \Bigg\} + \mathcal{O}(q^4,q^2\omega_q)\\
    \nonumber
    &=  \frac{1}{2}\frac{D \lambda^2}{ \mu^2} \int_{\bp} \frac{1}{p^2} \Bigg\{ \bq\cdot \bP(\bp)\cdot\bP(\bq)\cdot\bP(\bp)\cdot\bq + \bq\cdot \bP(\bp)\cdot\bp {\rm Tr}(\bP(\bp)\bP(\bq))
    \Bigg\} + \mathcal{O}(q^4,q^2\omega_q)\\
    \nonumber
    &=\frac{1}{2}\frac{D \lambda^2}{(d-1) \mu^2} \int_{\bp} \frac{1}{p^2} \Bigg\{ \Bigg(p^2-\frac{(\bq\cdot\bp)^2}{q^2}\Bigg)\frac{(\bq\cdot\bp)^2}{p^4} + \Bigg(d-2+\frac{(\bq\cdot\bp)^2}{p^2q^2}\Bigg)\Bigg(q^2-\frac{(\bq\cdot\bp)^2}{p^2}\Bigg)
    \Bigg\} + \mathcal{O}(q^4,q^2\omega_q)\\
    \nonumber
    &=  \frac{1}{2}\frac{D \lambda^2}{ (d-1)\mu^2} \int_{\bp} \frac{1}{p^2} \Bigg\{ (d-2)q^2+ (4-d) \frac{(\bq\cdot \bp)^2}{p^2}-2\frac{(\bq\cdot\bp)^4}{q^2p^4}
    \Bigg\} + \mathcal{O}(q^4,q^2\omega_q)\\
    \nonumber
    &= q^2 \frac{1}{2}\frac{D \lambda^2 }{ (d-1)\mu^2} \int_{\bp} \frac{1}{p^2} \Bigg\{ (d-2)+ \frac{(4-d)}{d} - \frac{6}{d(d+2)}
    \Bigg\} + \mathcal{O}(q^4,q^2\omega_q)\\
    &= q^2 \frac{S_d}{(2\pi)^d} \frac{1}{2} \frac{d^2-2}{d^2+2d} \frac{D \lambda^2}{\mu^2} \int_{\Lambda(\ell+\dd\ell)}^{\Lambda(\ell)} \dd p \frac{p^{d-1}}{p^2} + \mathcal{O}(q^4,q^2\omega_q)
\end{align}
where in the third line, some simplifications 
using the transverse projectors were made, in the fourth line, the expansion to leading order in momentum was made, and in the fifth line
Cauchy’s residue theorem was applied.

Its flow equation is therefore given as,
\begin{equation}
    \frac{\dd\mu}{\dd \ell} = \frac{S_d}{(2\pi)^d} \frac{1}{2} \frac{d^2-2}{d^2+2d} \frac{D \lambda^2}{\mu^2}  \Lambda(\ell)^{d-2} \ .
\end{equation}

\subsection*{Noise renormalization}

The renormalization of the noise can be obtained in a similar fashion, by projecting to the transverse component and expanding to second order in $\bq$,
\begin{align}
    \label{eq:calc_loop2}
    \nonumber
    &\frac{P_{in}(\bq)}{d-1} \diagram{26} \\
    \nonumber
    &= \frac{P_{in}(\bq)}{d-1}\int_{\tilde \bp} (-\ii \frac{\lambda}{2}) \Big[\Big(p_j+\frac{q_j}{2}\Big) \delta_{ik}+\Big(\frac{q_k}{2}-p_k\Big)\delta_{ij}\Big] \frac{2 D |\bq/2-\bp|^2 P_{jl}(\bq/2-\bp)}{|-\ii \omega_q/2+\ii\omega_p+\mu |\bq/2-\bp|^2|^2} \\
    \nonumber
    &\hspace{1.1cm} \times (-\ii \frac{\lambda}{2}) \Big[\Big(-p_l-\frac{q_l}{2}\Big) \delta_{nm}+\Big(p_m-\frac{q_m}{2}\Big)\delta_{nl}\Big] \frac{2 D |\bq/2+\bp|^2 P_{mk}(\bq/2+\bp)}{|-\ii \omega_q/2-\ii\omega_p+\mu |\bq/2+\bp|^2|^2} \\
    \nonumber
    &=\frac{P_{in}(\bq)}{d-1}\int_{\tilde \bp} (-\ii \frac{\lambda}{2}) \Big[q_j \delta_{ik}+q_k\delta_{ij}\Big] \frac{2 D |\bq/2-\bp|^2 P_{jl}(\bq/2-\bp)}{|-\ii \omega_q/2+\ii\omega_p+\mu |\bq/2-\bp|^2|^2} \\
    \nonumber
    &\hspace{1.1cm} \times (-\ii \frac{\lambda}{2}) \Big[-q_l \delta_{nm}-q_m\delta_{nl}\Big] \frac{2 D |\bq/2+\bp|^2 P_{mk}(\bq/2+\bp)}{|-\ii \omega_q/2-\ii\omega_p+\mu |\bq/2+\bp|^2|^2}+ \mathcal{O}(q^4,q^2\omega_q) \\
    \nonumber
    &=\frac{P_{in}(\bq)}{d-1}\frac{\lambda^2}{4} \int_{\tilde \bp} \Big[q_j \delta_{ik}+q_k\delta_{ij}\Big] \frac{4 D^2 p^4 P_{jl}(\bp) P_{mk}(\bp)}{|\ii\omega_p+\mu p^2|^4}  \Big[q_l \delta_{nm}+q_m\delta_{nl}\Big]  + \mathcal{O}(q^4,q^2\omega_q) \\
    \nonumber
    &=\frac{1}{d-1}\frac{\lambda^2D^2}{4\mu^3 p^2} \int_{\tilde \bp} P_{in}(\bq)\Big[q_j \delta_{ik}+q_k\delta_{ij}\Big] \frac{1}{ p^2}  P_{jl}(\bp) P_{mk}(\bp)  \Big[q_l \delta_{nm}+q_m\delta_{nl}\Big]  + \mathcal{O}(q^4,q^2\omega_q) \\
    \nonumber
    &=\frac{1}{d-1}\frac{\lambda^2D^2}{2\mu^3 p^2} \int_{\tilde \bp} \frac{1}{p^2} \Bigg\{ \bq\cdot \bP(\bp)\cdot\bP(\bq)\cdot\bP(\bp)\cdot\bq + \bq\cdot \bP(\bp)\cdot\bp {\rm Tr}(\bP(\bp)\bP(\bq)) \Bigg\} + \mathcal{O}(q^4,q^2\omega_q)\\
    &= q^2 \frac{S_d}{(2\pi)^d} \frac{1}{2} \frac{d^2-2}{d^2+2d} \frac{D^2 \lambda^2}{\mu^3} \int_{\Lambda(\ell+\dd \ell)}^{\Lambda(\ell)} \dd p \frac{p^{d-1}}{p^2} + \mathcal{O}(q^4,q^2\omega_q) \ ,
\end{align}
which follows analogously to Eq.~\eqref{eq:calc_loop1_detail}. The flow equation of the noise is thus,
\begin{equation}
    \frac{\dd D}{\dd \ell} = \frac{S_d}{(2\pi)^d} \frac{1}{2} \frac{d^2-2}{d^2+2d} \frac{D^2 \lambda^2}{\mu^3}  \Lambda(\ell)^{d-2} \ .
\end{equation}
Only with symmetric momentum routing do we find that the ratio $\mu/D$ remains constant, which encodes the fluctuation  dissipation theorem of (effective) thermodynamic equilibrium.

\subsection*{Renormalization of $\lambda$}

To project out the renormalization of $\lambda$ from Eq.~\eqref{eq:triangle_diags}, note that we can rewrite the bare vertex as,
\begin{equation}
     \diagram{50}  = -\frac{ \ii \lambda}{2} [q_a\delta_{in} +q_n \delta_{ia} ] \ ,
\end{equation}
since it can only couple to the transverse modes. 
Keeping this in mind, we can calculate the loop corrections,
\begin{align}
    &8\diagram{47}   \\
    \nonumber
    &=\int_{\tilde \bp} (-\ii \lambda) \Big(\Big(\frac{q_j}{2}+p_j\Big) \delta_{ik}+\Big(\frac{q_k}{2}-p_k\Big)\delta_{ij}\Big) \frac{P_{jl}(\bq/2-\bp)}{-\ii \omega_q/2+\ii\omega_p+\mu |\bq/2-\bp|^2} \\
    &\hspace{1.1cm} \times (-\ii \lambda) \Big(-\Big(k_n+p_n\Big)\delta_{lb}+\Big(\frac{q_b}{2}+k_b\Big)\delta_{ln}\Big)  \frac{P_{bc}(\bk+\bp)}{\ii \omega_k+\ii\omega_p+\mu |\bk+\bp|^2}\\
    &\hspace{1.1cm}\times (-\ii \lambda) \Big(\Big(\frac{q_m}{2}-k_m\Big)\delta_{ca}-\Big(\frac{q_a}{2}+p_a\Big)\delta_{cm}\Big)  \frac{2D {|\bq/2+\bp|^2} P_{mk}(\bq/2+\bp)}{|-\ii\omega_q/2-\ii \omega_p+\mu |\bq/2+\bp|^2|^2} \\
    &=\int_{\tilde \bp} (-\ii \lambda) \Big( q_j \delta_{ik}+q_k\delta_{ij}\Big) \frac{P_{jl}(\bp)}{\ii\omega_p+\mu p^2}  (-\ii \lambda) \Big(-p_n\delta_{lb}\Big)  \frac{P_{bc}(\bp)}{\ii\omega_p+\mu p^2}\\
    &\hspace{1.1cm}\times (-\ii \lambda) \Big(-p_a\delta_{cm}\Big)  \frac{2D p^2 P_{mk}(\bp)}{|-\ii \omega_p+\mu p^2|^2} +\mathcal O(q^2,qk,q\omega_q,q\omega_k) \\
    &= \ii \lambda^3 \Big( q_j \delta_{ik}+q_k\delta_{ij}\Big) \int_{\tilde \bp} \frac{P_{jk}(\bp) p_n p_a}{(\ii\omega_p+\mu p^2)^2}  \frac{2D  p^2}{|-\ii \omega_p+\mu p^2|^2} +\mathcal O(q^2,qk,q\omega_q,q\omega_k) \\
    &= \ii \lambda^3 \Big( q_j \delta_{ik}+q_k\delta_{ij}\Big) \frac{2D}{8\mu^3} \int_{\bp}   \frac{ (p^2\delta_{jk}-p_j p_k) p_n p_a }{ p^{6}} +\mathcal O(q^2,qk,q\omega_q,q\omega_k) \\
    &= \ii \lambda^3 \Big( q_j \delta_{ik}+q_k\delta_{ij}\Big) \frac{2D}{8\mu^3} \Bigg(\frac{\delta_{jk}\delta_{na}}{d}-\frac{\delta_{jk}\delta_{na}+\delta_{jn}\delta_{ka}+\delta_{ja}\delta_{nk}}{d^2+2d}\Bigg) \frac{S_d}{(2\pi)^d} \int_{\Lambda(\ell+\dd\ell)}^{\Lambda(\ell)} \dd p \frac{ p^{d-1}}{ p^{2}} +\mathcal O(q^2,qk,q\omega_q,q\omega_k) \\
    &= \ii \lambda^3  \frac{2D}{4\mu^3} \frac{(d+1)q_i\delta_{na}-q_a\delta_{in}-q_n\delta_{ia}}{d^2+2d} \frac{S_d}{(2\pi)^d} \int_{\Lambda(\ell+\dd\ell)}^{\Lambda(\ell)} \dd p \frac{ p^{d-1}}{ p^{ 2}} +\mathcal O(q^2,qk,q\omega_q,q\omega_k) \ , 
\end{align}
where in the third line, we simplified the first vertex via the projection operators, which shows that the entire diagram is proportional to $\bq$. Terms of higher order in the external momenta and frequencies are thus neglected since they do not contribute to the nonlinearity characterized by $\lambda$.  The contribution to the renormalization of $\lambda$ can be identified by realizing that the term proportional to $q_i$, does not contribute to the transverse mode.

Similarly, for the second diagram we get,
\begin{align}
    & 4\diagram{48}  \\
    \nonumber
    &=\frac{1}{2}\int_{\tilde \bp} (-\ii \lambda) \Big(\Big(\frac{q_j}{2}+p_j\Big) \delta_{ik}+\Big(\frac{q_k}{2}-p_k\Big)\delta_{ij}\Big) \frac{P_{jl}(\bq/2-\bp)}{-\ii \omega_q/2+\ii\omega_p+\mu |\bq/2-\bp|^2} \\
    &\hspace{1.1cm} \times (-\ii \lambda) \Big(-\Big(k_n+p_n\Big)\delta_{lb}+\Big(\frac{q_b}{2}+k_b\Big)\delta_{ln}\Big)  \frac{2D { |\bk+\bp|^2} P_{bc}(\bk+\bp)}{|\ii \omega_k+\ii\omega_p+\mu |\bk+\bp|^2|^2}\\
    &\hspace{1.1cm}\times (-\ii \lambda) \Big(\Big(\frac{q_c}{2}-k_c\Big)\delta_{ma}+\Big(k_a+p_a\Big)\delta_{mc}\Big)  \frac{P_{mk}(\bq/2+\bp)}{-\ii\omega_q/2-\ii \omega_p+\mu |\bq/2+\bp|^2} \\
    &=\frac{1}{2}\int_{\tilde \bp} (-\ii \lambda) \Big( q_j \delta_{ik}+q_k\delta_{ij}\Big) \frac{P_{jl}(\bp)}{\ii\omega_p+\mu p^2}  (-\ii \lambda) \Big(-p_n\delta_{lb}\Big)  \frac{2D p^2 P_{bc}(\bp)}{|\ii\omega_p+\mu p^2|^2}\\
    &\hspace{1.1cm}\times (-\ii \lambda) \Big(p_a\delta_{cm}\Big)  \frac{P_{mk}(\bp)}{-\ii \omega_p+\mu p^2} +\mathcal O(q^2,qk,q\omega_q,q\omega_k) \\
    &=-\frac{1}{2} \ii \lambda^3 \Big( q_j \delta_{ik}+q_k\delta_{ij}\Big) \int_{\tilde \bp} \frac{2D { p^2} P_{jk}(\bp) p_n p_a}{|-\ii \omega_p+\mu p^2|^4}   +\mathcal O(q^2,qk,q\omega_q,q\omega_k) \\
    &= -\ii \lambda^3 \Big( q_j \delta_{ik}+q_k\delta_{ij}\Big) \frac{2D}{8\mu^3} \int_{\bp}   \frac{ (p^2\delta_{jk}-p_j p_k) p_n p_a }{ p^{ 6}} +\mathcal O(q^2,qk,q\omega_q,q\omega_k) \\
    &= -\ii \lambda^3 \Big( q_j \delta_{ik}+q_k\delta_{ij}\Big) \frac{2D}{8\mu^3} \Bigg(\frac{\delta_{jk}\delta_{na}}{d}-\frac{\delta_{jk}\delta_{na}+\delta_{jn}\delta_{ka}+\delta_{ja}\delta_{nk}}{d^2+2d}\Bigg) \frac{S_d}{(2\pi)^d} \int_{\Lambda(\ell+\dd\ell)}^{\Lambda(\ell)} \dd p \frac{ p^{d-1}}{ p^{ 2}} +\mathcal O(q^2,qk,q\omega_q,q\omega_k) \\
    &= - \ii \lambda^3  \frac{2D}{4\mu^3} \frac{(d+1)q_i\delta_{na}-q_a\delta_{in}-q_n\delta_{ia}}{d^2+2d} \frac{S_d}{(2\pi)^d} \int_{\Lambda(\ell+\dd\ell)}^{\Lambda(\ell)} \dd p \frac{ p^{d-1}}{ p^{2}} +\mathcal O(q^2,qk,q\omega_q,q\omega_k) \ ,
\end{align}
which exactly cancels out the contribution from the first diagram. We therefore find the trivial flow equation,
\begin{equation}    
    \frac{\dd \lambda}{\dd \ell} = 0 \ , 
\end{equation}
for $\lambda$. This is a consequence of Galilean symmetry. If $\lambda$ were to change, Galilean symmetry would no longer hold. Therefore it cannot renormalize.

\end{document}

%% file: SectionTwo_v2.tex
\section{Universal hydrodynamic models}
\label{sec:universal_models}

At the heart of universality in physical systems lies the central limit theorem: the sum of a large number of independent random variables approaches a Gaussian distribution characterised by only two parameters: the mean and the variance, regardless of the detailed form of the underlying distributions.

In the hydrodynamic limit of a many-body system, \emph{i.e.}, at sufficiently large length and time scales, one likewise observes collective variables, such as the velocity field in fluids, that emerge from averaging over many microscopic degrees of freedom. Unlike in the central limit theorem, however, the microscopic constituents of a physical system generally interact with one another, so their probability distributions are not statistically independent. These interactions can therefore generate nontrivial collective behaviour and lead to different, yet still universal, large-scale distributions. Nevertheless, even in interacting systems, most microscopic details become irrelevant at long wavelengths, and the resulting large-scale behaviour can often be characterised by only a small set of relevant parameters.

Since our interest lies precisely in these large-scale, universal  properties, the most efficient approach is to discard as much irrelevant microscopic information as possible. In practice, this means integrating out the ``fast'' and ``microscopic" degrees of freedom (DOF), while retaining only the ``slow'' and ``large-scale'' hydrodynamic DOF together with the fundamental qualitative properties of the system: {\bf S}ymmetries, {\bf C}onservation laws, and exact {\bf C}onstraints ({\bf SCC}). One then constructs equations of motion for the hydrodynamic fields by including \emph{all} terms consistent with the SCC. 
Finally, the influence of the eliminated microscopic DOF is represented by stochastic forcing terms, typically assumed to have zero mean and short-range ($\delta$-correlated) statistics in both space and time, again reflecting the central-limit-theorem idea that the cumulative effect of many microscopic fluctuations approaches Gaussian white noise.

We now make these, perhaps abstract, ideas concrete through the example of a passive incompressible fluid, leading to the incompressible Navier--Stokes equation. 
The general procedure for constructing a universal hydrodynamic model is summarised in {\bf Recipe 1} below.

\begin{tcolorbox}[colback=yellow!5!white,colframe=yellow!50!black,
  colbacktitle=yellow!75!black,title=Recipe 1: Building a universal hydrodynamic model]
  1.~Choose a set of dynamical variables, $\{ \Psi_\alpha \}$, relevant to the phenomenon under consideration.
\\
2.~Identify all i) {\bf S}ymmetries, ii) {\bf C}onservation laws, and iii) exact {\bf C}onstraints ({\bf SCC}) associated with these dynamical variables.
\\
3.~Construct the most general equations of motion (EOM) for the dynamical variables,
\begin{equation}
\partial_t \Psi_\alpha = F_\alpha ,
\end{equation}
such that the EOM preserve the tensorial character (scalar, vector, rank-2 tensor, \emph{etc.}) specified in {\bf Step 1}, and satisfy all SCC identified in {\bf Step 2}.
\\
4.~Perform a mean-field analysis to identify potentially distinct phases, \emph{i.e.}, stable spatially homogeneous configurations characterised by uniform hydrodynamic fields $\{ \bar{\Psi}_\alpha \}$. Choose a particular phase and rewrite the EOM in terms of fluctuations about this mean-field state:
\begin{equation}
\Phi_\alpha \equiv \Psi_\alpha - \bar{\Psi}_\alpha .
\end{equation}
\\
5.~Expand $F_\alpha$ as a series in the fluctuating fields $\Phi_\alpha$ and their spatial derivatives. In practice, this expansion is truncated at some finite order according to a systematic procedure, which will be described in {\bf Section~\ref{sec:rg}}.
\\
6.~Add stochastic forcing terms to $F_\alpha$, typically taken to be Gaussian and $\delta$-correlated in space and time, such that they also respect all SCC identified in {\bf Step 2}.
\end{tcolorbox}

\subsection{Constructing a universal hydrodynamic model: Incompressible Navier-Stokes equations}
\label{ssec:constructing_eom}

{\bf Step 1} is to identify the relevant hydrodynamic variables. A natural choice is usually given by the key observables of the system. For an incompressible fluid, this is the velocity field,
$\bv = v_x \hat{\bx} + v_y \hat{\by} + v_z \hat{\bz}$.
Although this choice may appear obvious, it is only so because we are already familiar with the Navier--Stokes equations governing simple fluids. In general, identifying the appropriate hydrodynamic variables is a highly nontrivial task that requires genuine physical insight into the phenomenon under study. In many cases, the relevant variables correspond to conserved quantities and/or order parameters associated with spontaneously broken symmetries. Determining these variables therefore constitutes the first essential step in constructing a universal hydrodynamic theory.

Moreover, while $\bv$ is a natural choice for an incompressible fluid, it is not unique. For example, one could equivalently formulate the theory in terms of the momentum density $\bg = \rho \bv$, where $\rho$ is the mass density field.

{\bf Step 2} is to identify the relevant {\bf S}ymmetries, {\bf C}onservation laws, and exact {\bf C}onstraints ({\bf SCC}). We begin with the most fundamental symmetries of space and time: translational invariance, rotational invariance, temporal invariance, and chiral symmetry. To these we add Galilean invariance, which applies to all Newtonian systems.

Turning to conservation laws, we impose momentum conservation, which holds for any isolated Newtonian fluid. Although total energy is also strictly conserved in a closed system, we do not explicitly enforce energy conservation here. The justification is twofold: (i) microscopic dissipation converts kinetic energy into local heating through viscosity, and (ii) these small-scale temperature fluctuations are assumed not to feed back onto the macroscopic velocity dynamics. While standard, this assumption is nontrivial: if such feedback becomes important, the total energy density must itself be promoted to an additional hydrodynamic variable (see {\bf Pitfall 1}).

Finally, we consider exact constraints not already implied by symmetries or conservation laws. For an incompressible fluid, the defining constraint is the incompressibility condition: $\nabla \cdot \bv = 0$,
which holds identically.

There is also particle-number conservation. In a generic fluid, this implies that the density field $\rho$ is itself a slow hydrodynamic variable obeying the continuity equation,
\begin{equation}
\partial_t \rho + \nabla \cdot (\rho \bv) = 0 .
\end{equation}
Here, however, we restrict attention to an incompressible fluid, for which density fluctuations are suppressed exactly, \emph{i.e.}, $\delta \rho = 0$, so that the density remains fixed, $\rho(\br,t)=\rho_0$.
Under this assumption, the continuity equation reduces directly to the incompressibility constraint $\nabla \cdot \bv = 0$.

At this stage, the essential physical input is complete: the hydrodynamic variables, the set of SCC has now been identified.

{\bf Step 3} is to construct the most general equation of motion (EOM) for an incompressible fluid consistent with the chosen SCC. Since the only hydrodynamic variable is the velocity field $\bv(\br,t)$, we begin with
\begin{equation}
\partial_t \bv = {\bf A}\ ,
\end{equation}
where ${\bf A}$ may be interpreted as the local acceleration field.

Before specifying the explicit form of ${\bf A}$, it is important to note that the hydrodynamic limit is already implicit in this equation. In particular, we do not include higher-order time derivatives, such as $\partial_t^2 \bv$, because hydrodynamics describes the long-wavelength, low-frequency behaviour of the system.

To see this explicitly, consider the Fourier transform $\bv(\bq,\omega)
=
\int_{\br,t}
\bv(\br,t)\,
e^{\,i(\omega t-\bq\cdot\br)}$.
In Fourier space, the magnitudes of $\partial_t \bv$ and $\partial_t^2 \bv$ scale as
$
|\omega \, \bv(\bq,\omega)|$ and $
|\omega^2 \, \bv(\bq,\omega)|$,
respectively. The hydrodynamic limit corresponds to taking both $\omega \to 0$ and $|\bq| \to 0$. In this regime, terms involving $\partial_t^2 \bv$ (and higher-order time derivatives) are negligible compared to those involving $\partial_t \bv$. It is therefore sufficient to retain only the leading first-order time derivative in the EOM.

Returning to ${\bf A}$, we first impose the conservation law, namely momentum conservation. Since the momentum density $\rho \bv$ is conserved, and the density $\rho$ is constant under the incompressibility condition, the velocity field $\bv$ itself behaves as a conserved quantity. Consequently, the right-hand side of the EOM must be expressible as the divergence of a rank-2 tensor: $\partial_t \bv = \nabla \cdot \bH$.
To make the tensorial structure explicit, we write the EOM in index notation,
\begin{equation}
\partial_t v_i = \partial_j H_{ji} .
\end{equation}

Having incorporated momentum conservation, we next enforce the exact incompressibility constraint, $\nabla \cdot \bv = 0$,
using a Lagrange multiplier. The EOM therefore becomes $\partial_t v_i = \partial_j H_{ji} - C_i$,
where $C_i$ denotes the term constraining the system to be incompressible. 

To determine the form of $C_i$, we invoke the Helmholtz decomposition, which states that any sufficiently smooth vector field can be decomposed into an irrotational part and a divergence-free part:
$ \nabla \cdot \bH
=
\nabla p
+
\nabla \times \mathbf{B}$,
where $p$ is a scalar field and $\mathbf{B}$ is a vector field~\cite{stone_b09}.
Since incompressibility requires the dynamics to remain divergence-free, 
the irrotational component must be projected out. We therefore identify the constraining term as
$C_i = \partial_i p$,
where the scalar field $p$ plays the role of the pressure field. The EOM thus becomes
\begin{equation}
\partial_t v_i
=
\partial_j H_{ji}
-
\partial_i p .
\end{equation}
   
   We now impose the stated symmetries to further constrain the form of $\bH$.
\begin{enumerate}
	\item
	{\bf Temporal invariance}: $H_{ij}$ cannot depend explicitly on time $t$, thereby forbidding terms such as $t\,\partial_i v_j$. Physically, this symmetry implies that the same instantaneous velocity configuration $\bv(\br)$ cannot experience different forces merely because it is realised at a later time.
	
	\item
	{\bf Translational invariance}: $H_{ij}$ cannot depend explicitly on the spatial coordinate $\br$, thereby forbidding terms such as $r_i v_j$. This reflects the absence of preferred spatial locations within the bulk of the system. Strictly speaking, translational invariance is always broken in a finite system by the presence of boundaries; here we consider the hydrodynamic limit in which boundary effects are negligible deep within the fluid.
	
	\item
	{\bf Rotational invariance}: the EOM must remain invariant under rotations of the reference frame, thereby forbidding terms such as $w_i v_j$ involving a fixed external vector $\bw$, which would single out a preferred direction in space. This symmetry therefore expresses the absence of intrinsic directional bias in the system.
	
	\item
	{\bf Chiral invariance}: the EOM must remain invariant under parity transformations, \emph{i.e.}, spatial reflections. This forbids terms involving the antisymmetric tensor $\epsilon_{ijk}$, such as $\epsilon_{ijk}v_k$. Physically, this means that the fluid possesses no intrinsic handedness, thereby excluding terms such as $\vnab \times \bv$ from the EOM for $\bv$.
\end{enumerate}

Having imposed these symmetries, we now expand $\bH$ systematically in powers of spatial gradients. The generic EOM is then obtained by writing down all symmetry-allowed terms up to a given order in $\bv$ and its spatial derivatives. Retaining terms up to $\mathcal{O}(\partial^2)$, we obtain
\beq
\label{eq:EOM0}
\partial_t v_i
=
\partial_j
\Big[
\lambda(|\bv|) v_i v_j
+
\mu(|\bv|) \partial_j v_i
+
\nu(|\bv|) \partial_i v_j
\Big]
-
\partial_i p
\ ,
\eeq
where, for complete generality, the coefficients are allowed to depend on the rotationally invariant scalar quantity $|\bv|$.

Note, however, that the symmetry analysis is not yet complete, as one important symmetry remains: Galilean invariance.
\begin{enumerate}
	\setcounter{enumi}{4}
	\item
	{\bf Galilean invariance}: in the absence of external forces, the EOM must remain invariant under transformations to reference frames moving at constant velocity relative to one another.
\end{enumerate}

Under this symmetry, the EOM remains invariant under the simultaneous transformations $\br \mapsto \br-\bw t$ and $\bv(\br,t) \mapsto \bv(\br-\bw t,t) +\bw$, where $\bw$ is an arbitrary constant vector.

A crucial consequence of Galilean invariance is that the magnitude $|\bv|$ is itself not invariant under such transformations. Instead, only spatial derivatives, {\it e.g.}, $\partial_i v_j$, and the combination $\partial_t \bv + (\bv\cdot\nabla)\bv$ remain invariant. Consequently, the coefficients $\lambda$, $\mu$, and $\nu$ cannot depend explicitly on $|\bv|$, but only on scalar combinations constructed from Galilean-invariant quantities. 

At the order retained in our gradient expansion, this restriction implies that $\lambda$, $\mu$, and $\nu$ must all be constants. Galilean invariance further fixes $\lambda=-1$, such that the time derivative and nonlinear advective term combine into the material (or convective) derivative.

Using the incompressibility condition, enforced through the Lagrange multiplier $p$, the EOM can thus be brought into the standard form,
\begin{equation}
\label{EOM}
\partial_t \bv + (\bv \cdot \nabla)\bv
=
-\nabla p
+
\mu \nabla^2 \bv
\ ,
\end{equation}
i.e., precisely the incompressible Navier--Stokes equation.

Let us pause to appreciate how much has emerged already from this remarkably simple construction. Starting from a single hydrodynamic variable, $\bv$, and imposing only a set of seemingly innocuous and physically unavoidable symmetries, conservation laws, and constraints, we arrive at an extremely constrained EOM involving just a single phenomenological parameter: the viscosity $\mu$. The resulting equation applies universally to all incompressible momentum-conserving fluids precisely because it depends only on the hydrodynamic variables and SCC, and not on the microscopic details of the underlying system.

Underlying this derivation is the crucial hydrodynamic limit, which we have implicitly invoked throughout. For example, why is the EOM local in time, \emph{i.e.}, ``Markovian''? In principle, one could imagine memory terms such as $\bv(\br,t-\tau)$, which would make the dynamics depend explicitly on past times. However, upon Fourier transformation, such a term becomes $\exp(-i\omega\tau)\bv(\bq,\omega)$, which approaches $\bv(\bq,\omega)$ as $\omega \to 0$ for any finite memory time $\tau$. Thus, in the hydrodynamic limit, finite-time memory effects become irrelevant, and the dynamics reduce to a local-in-time form. This assumption can fail in systems with intrinsically long memory, such as glassy systems.

A similar argument applies to spatial gradients. Under Fourier transformation, spatial derivatives become factors of $i\bq$, which vanish in the long-wavelength limit $|\bq|\to0$. Consequently, terms involving higher-order spatial derivatives are progressively suppressed at large scales, thereby justifying the truncation of the gradient expansion.

In {\bf Step 4}, we perform a mean-field analysis of the EOM. Specifically, we assume that the velocity field $\bv$ is spatially homogeneous and determine which homogeneous configurations satisfy the equation. In the present case, it is immediately clear that \emph{any} constant vector $\bv_0$ constitutes a solution.

Invoking Galilean invariance, however, we can always transform to the comoving reference frame in which this constant velocity vanishes. Consequently, all spatially homogeneous states are physically equivalent, and without loss of generality we may take the mean-field solution to be $\bv_0=\mathbf{0}$.

In {\bf Step 5}, the EOM is expanded in powers of the fluctuations $\delta \bv \equiv \bv - \bv_0$.
In the comoving frame, where $\bv_0=\mathbf{0}$, the velocity field itself therefore directly represents the fluctuations, \emph{i.e.}, $\bv=\delta \bv$.
This  expansion is again a consequence of the hydrodynamic limit. As one probes progressively larger length scales, the velocity field is effectively averaged over increasingly large coarse-graining volumes. By analogy with the central limit theorem, one expects that averaging over many microscopic degrees of freedom suppresses fluctuations, even in interacting systems. Consequently, higher powers of the fluctuating fields become progressively less important at large scales.

In the present case, however, Galilean invariance has already strongly constrained the EOM~\eqref{EOM}, such that it contains terms only up to quadratic order in $\bv$. Thus, {\bf Step 5} is effectively already complete.

This simplification would no longer hold if Galilean invariance were broken. For example, in a fluid moving relative to a fixed substrate or through a porous medium, there exists a preferred reference frame, and the more general EOM~\eqref{eq:EOM0} would apply. In that case, the coefficients $\lambda$, $\mu$, and $\nu$ would themselves need to be expanded in powers of $\delta \bv$. The question of where such expansions may consistently be truncated will be discussed in {\bf Section~\ref{sec:rg}}.

At this stage, we have completed {\bf Steps 1}--{\bf 5} of {\bf Recipe 1}. The only remaining ingredient is to incorporate fluctuations into the EOM. These fluctuations represent the fast microscopic degrees of freedom that evolve on time and length scales much shorter than those associated with the hydrodynamic field $\bv$, which by construction is assumed to be the slowest degrees of freedom in the system.
To account for these unresolved microscopic fluctuations, we add a stochastic forcing term to the EOM. In accordance with the SCC, this noise term must preserve all symmetries, conservation laws, and constraints of the system. It is typically taken to be Gaussian and $\delta$-correlated in both space and time.

Why Gaussian white noise? Once again, the answer lies in the hydrodynamic limit. Hydrodynamics focuses on long times and large distances, which necessarily involves averaging over many microscopic degrees of freedom. By the central limit theorem, the cumulative effect of these many microscopic fluctuations approaches a Gaussian random process. The mean of these fluctuations contributes to the deterministic part of the EOM and hence to the value of the parameter $\mu$, whereas fluctuations about the mean contribute to the stochastic noise. For the same reason, any finite temporal or spatial correlations in the microscopic noise become effectively local when viewed on sufficiently large scales. Thus, in the hydrodynamic limit, the noise becomes universally short-ranged, \emph{i.e.}, $\delta$-correlated in both space and time. Conversely, if the noise cannot be approximated as $\delta$-correlated, this signals the presence of an additional slow degree of freedom, implying that not all hydrodynamic modes have been correctly identified.

Here, momentum conservation implies that the fluctuating EOM takes the form
\begin{equation}
\partial_t \bv + (\bv \cdot \nabla)\bv
=
-\nabla p
+
\mu \nabla^2 \bv
+
\bff
\ ,
\label{eq:EOM}
\end{equation}
where
\begin{equation}
\label{eq:noise}
\la \bff (\br,t) \ra = {\bf 0} \sep
\langle f_i(\br,t) f_j(\br^\prime,t^\prime) \rangle
=
-2D \nabla^2 \delta^3(\br-\br^\prime)\delta(t-t^\prime)
\ ,
\end{equation}
and $D$ denotes the noise strength, which constitutes an additional phenomenological parameter to be determined experimentally.
The Laplacian appearing in the noise correlator ensures that the stochastic forcing itself respects momentum conservation, {\it i.e.},  $\bff$ is the divergence of a random field. One may further verify that the full stochastic EOM~\eqref{eq:EOM} satisfies all remaining elements of the SCC.

At this point, one may wonder why the noise strength $D$ is taken to be constant, rather than depending on $\bv$, its spatial derivatives, or other local quantities. In principle, $D$ could indeed admit an expansion in the hydrodynamic fields, much like the tensor $\bH$ introduced earlier. However, in the hydrodynamic limit, the leading constant contribution generically dominates unless symmetry forces it to vanish. This justifies treating $D$ as a constant to leading order.

The same reasoning also clarifies why the coefficients appearing in the EOM are themselves not taken to fluctuate. Effectively, such fluctuations are already encoded within the stochastic forcing: fluctuations of the coarse-grained parameters can be absorbed into higher-order contributions to the noise term. The parameters appearing explicitly in the EOM therefore represent coarse-grained averages, while their residual fluctuations are subsumed into the stochastic noise.

Finally, since our focus throughout this Review is on intrinsically nonequilibrium systems, we will not impose the fluctuation--dissipation relation, which in equilibrium systems would relate the noise strength $D$ to the viscosity $\mu$.

\bigskip

We have now completed our construction recipe, and the EOM~\eqref{eq:EOM} constitute the universal hydrodynamic equations for generic incompressible fluids. Their universality stems from the fact that any system governed by the same set of SCC will necessarily be described by the same hydrodynamic EOM at large scales. Since the SCC were chosen precisely to characterise incompressible momentum-conserving fluids in general, all such systems fall into the same universal hydrodynamic class, irrespective of their microscopic details. 

Although the construction is conceptually straightforward, several subtle pitfalls can arise in practice. Some of the most common are summarized in the box {\bf Pitfall 1}.

\begin{tcolorbox}[colback=red!5!white,colframe=red!50!black,
  colbacktitle=red!75!black,title=Pitfall 1]
  1.~Incorrect identification of the hydrodynamic variables. Determining the appropriate hydrodynamic degrees of freedom is arguably the most important step in constructing a universal theory. Unfortunately, a general recipe is difficult to codify  
 as this identification is highly problem-dependent and requires genuine physical insight into the phenomenon under consideration.
\\
2.~Incorrect identification of the relevant set of symmetries, conservation laws, and constraints (SCC). As with the choice of hydrodynamic variables, this is where the essential scientific content of the theory enters.
\\
3.~The hydrodynamic limit may not coincide with the experimentally accessible regime. In conventional fluids, which typically contain Avogadro numbers of molecules, the hydrodynamic limit is often readily realised experimentally. In contrast, for active-matter systems composed of comparatively small numbers of organisms or agents (cells, animals, \emph{etc.}), it is sometimes less clear whether the asymptotic hydrodynamic regime is experimentally attainable.
\end{tcolorbox}

\subsection{Caveats: microscopic versus macroscopic theories}

The route from a minimal set of SCC to universal hydrodynamic equations may appear almost too powerful to be true. What, then, is the catch? In fact, one of the profound insights of twentieth-century physics was precisely that symmetry principles alone can strongly constrain the behaviour of physical systems --- a realization that fundamentally reshaped modern theoretical physics. In this sense, the success of hydrodynamic reasoning is not ``too good to be true'', but rather a reflection of the deep organising power of symmetry and scale separation.

The real limitation lies elsewhere. By focusing exclusively on the universal hydrodynamic limit, we necessarily discard information about microscopic and mesoscopic length scales, as well as short-time dynamics.  Hydrodynamic theories are therefore not intended to provide a faithful microscopic description of a system, but rather an effective description of its large-scale behaviour. Consequently, their parameters, as well as their relationship to the underlying microscopic parameters, must generally be determined phenomenologically.

%% file: SectionThree_v3.tex
\section{Perturbative renormalization group analysis}
\label{sec:rg}

In {\bf Section~\ref{sec:universal_models}}, we discussed how to systematically construct a universal hydrodynamic theory. Such a theory, however, is not itself a universality class (UC). A single hydrodynamic equation of motion (EOM) can exhibit multiple phases and phase transitions, each with distinct large-scale behaviour. When these phenomena display long-range correlations and scale invariant behaviour, they can be assigned to UCs characterised by universal scaling properties. The paradigmatic example is the Ising UC, which describes both the ferromagnetic ordering transition in uniaxial magnets and the critical point of the liquid--gas transition.

To determine these scaling properties, and thereby identify UCs, we now introduce the renormalization group (RG). The RG provides a systematic framework for determining which nonlinear terms generated in Step 5 of {\bf Recipe 1} remain relevant at large scales and which become irrelevant. More importantly, it quantifies how these nonlinearities modify the scaling behaviour predicted by the linear theory. In this sense, the RG serves as a classification scheme that organises hydrodynamic theories into UCs. The essential steps of this procedure are summarised in {\bf Recipe 2}.

\begin{tcolorbox}[colback=yellow!5!white,colframe=yellow!50!black,
  colbacktitle=yellow!75!black,title={Recipe 2: Identifying the UC using perturbative RG at the {\bf one-loop} level}]
1.~Starting from the EOM, compute the correlation functions within the linear theory in the hydrodynamic limit, and attempt to cast them into the scaling form of the type
\begin{equation}
\label{eq:correlationform}
\la \Phi_\alpha (\br_\parallel, \br_\perp, t)\Phi_\alpha (\mathbf 0,\mathbf 0,0) \ra=
|\br_\perp|^{2\chi_\alpha} c_\alpha
S_{\alpha,{\rm lin}}\left(
\frac{ c_\parallel|\br_\parallel'|}{|\br_\perp|^{\zeta}},
\frac{  c_t t}{ |\br_\perp|^z}
\right)\ ,
\end{equation}
where the scaling function $S_{\alpha,\mathrm{lin}}$ is dimensionless, and the constant $c$'s ensure that both the correlation function and its arguments have the correct physical dimensions.
In the simplest cases, $S_{\alpha,{\rm lin}}$ approaches a nonzero constant as its arguments tend to zero, yielding power-law scaling in $|\br_\perp|$.
More generally, $S_{\alpha,{\rm lin}}$ may contain additional angular, propagating, or oscillatory structure, e.g., through the variables $t/|\br_\perp|$ (see, e.g., Eq.~\eqref{eq:compNVcorr}). Such dependencies modify the detailed form of the correlation functions but do not alter the scaling exponents.
Further, we have explicitly separated a distinguished direction $\br_\parallel$ to accommodate spatial anisotropy; for example, $\br_\parallel'$ may denote a comoving (boosted) coordinate, $\br_\parallel'=\br_\parallel \pm vt$, introduced to account for collective motion.

\vspace{.08in}

If such a scaling form exists, the linearised theory is scale invariant, and the associated hydrodynamic variables are termed soft (or ``massless''), meaning that they lack a finite correlation length. We may then proceed to Step~2.

2.~Using the linear exponents ${\chi_\alpha,\zeta,z}$, perform zeroth-order RG power counting to determine the upper critical dimension $d_c$, below which the linear theory can become unstable to nonlinearities.

3.~Augment the linearised EOM by retaining all nonlinear terms that are relevant just below $d_c$. Construct the corresponding one-loop diagrams and derive the RG flow equations for all  parameters.

4.~Rewrite the flow equations in terms of suitably defined dimensionless couplings (the $g$'s), and derive the associated RG flow equations.

5.~If the RG flow generically approaches a fixed point (FP), then that FP governs the asymptotic scaling behaviour, with the scaling exponents determined from the flow equations. The set of systems controlled by the same RG FP defines a universality class (UC).
\end{tcolorbox}

\subsection{Step 1: Linear theory}
\label{ssec:linear_theory}
Solving  the hydrodynamic EOM at the linear level can typically be  accomplished by transforming the equations into Fourier space in both space and time.
We thus start from the linearised incompressible equation
\begin{equation}
\partial_t \mathbf v = \mu \nabla^2 \mathbf v -\vnab p+ \mathbf f,
\qquad
\nabla \cdot \mathbf v = 0 .
\end{equation}

In Fourier space, the linear solution is
\begin{equation}
v_i(\mathbf q,\omega)=G^0_{ij}(\mathbf q,\omega)\,f_j(\mathbf q,\omega) \quad , \quad \langle f_i(\mathbf q,\omega) f_j(\mathbf q',\omega')\rangle
=
2D q^2 (2\pi)^{d+1}\delta^d(\mathbf q+\mathbf q')\delta(\omega+\omega')\delta_{ij} \ ,
\end{equation}
where the `bare' propagator is
\begin{equation}
G^0_{ij}(\mathbf q,\omega)=\frac{P_{ij}(\mathbf q)}{-i\omega+\mu q^2}\ , 
\end{equation}
in which $P_{ij}(\mathbf q)=\delta_{ij}-\frac{q_i q_j}{q^2}$ is  the transverse projector that enforces
the incompressibility. 

We thus readily obtain the velocity correlator in Fourier space:
\begin{align}
\langle v_i(\mathbf q,\omega)v_j(\mathbf q',\omega')\rangle
&=
G^0_{i\ell}(\mathbf q,\omega)\,
G^0_{jm}(\mathbf q',\omega')\,
\langle f_\ell(\mathbf q,\omega)f_m(\mathbf q',\omega')\rangle \\
&=
2D q^2 (2\pi)^{d+1}\delta^d(\mathbf q+\mathbf q')\delta(\omega+\omega')
\frac{P_{ij}(\mathbf q)}{\omega^2+\mu^2 q^4}.
\end{align}
Therefore the  spatiotemporal correlation function is
\begin{equation}
C_{ij}(\mathbf r,t)
\equiv
\langle v_i(\mathbf r,t)v_j(\mathbf 0,0)\rangle
=
\int \frac{d^d q}{(2\pi)^d}\frac{d\omega}{2\pi}\,
e^{i\mathbf q\cdot \mathbf r-i\omega t}
\frac{2D q^2 P_{ij}(\mathbf q)}{\omega^2+\mu^2 q^4}.
\end{equation}
To extract the scaling form, let $\mathbf q=\frac{\tilde{\mathbf q}}{r},
\omega=\frac{\mu \tilde\omega}{r^2},
r\equiv |\mathbf r|$, and thus
\begin{equation}
    \label{eq:corr_fourier}
C_{ij}(\mathbf r,t)
=
\frac{D}{\mu}\,r^{-d}
\int \frac{d^d \tilde q}{(2\pi)^d}\frac{d\tilde\omega}{2\pi}\,
e^{i\tilde{\mathbf q}\cdot \hat{\mathbf r} -i\tilde\omega\, \mu t/ r^2}
\frac{2\tilde q^2 P_{ij}(\tilde{\mathbf q})}{\tilde\omega^2+\tilde q^4}.
\end{equation}
Equivalently, after tracing over indices or projecting onto transverse components in a rotationally invariant manner, one may write
\begin{equation}
\label{eq:NVcorr}
\langle \mathbf v(\mathbf r,t)\cdot \mathbf v(\mathbf 0,0)\rangle
=
r^{2\chi}\, \frac{D}{\mu}
S\!\left(\frac{\mu t}{ r^z}\right)
\end{equation}
with the linear dynamic and roughness exponents are, respectively,
\begin{equation}
z=2,\qquad \chi=-\frac d2\ ,
\end{equation}
thus establishing the scale-invariant nature of the correlation function.

\paragraph{Compressibility $\rightarrow$ sound modes $\rightarrow$ more intricate scaling functions.}

As already hinted at in {\bf Recipe 2}, the simple scaling form shown in Eq.~\eqref{eq:NVcorr} is the exception rather than the rule. Indeed, if the incompressibility constraint is relaxed, the linearised EOM become
\begin{equation}
\label{eq:compNV}
\pp_t \rho + \rho_0 \nabla \cdot \bv = 0 \ , 
\qquad
\pp_t \bv =- \frac{ v_s^2}{\rho_0} \nabla \rho + \mu_T \nabla^2 \bv + \mu_L \nabla (\vnab \cdot \bv)+ \bff \ ,
\end{equation}
where $v_s$ is the sound speed and $\rho_0$ is the mean mass density.
The coupling of the density and velocity modes leads naturally to the emergence of sound modes, and the velocity correlation function is cast into a more general scaling form \cite{kadanoff_annphys00}
\begin{equation}
\label{eq:compNVcorr}
\langle \mathbf v(\mathbf r,t)\cdot \mathbf v(\mathbf 0,0)\rangle
=
r^{2\chi}
\left[
\frac{D}{\mu_T+\mu_L}
S_L\!\left(
\frac{(\mu_T+\mu_L)t}{ r^z},
\frac{v_s t}{r}
\right)
+
\frac{D}{\mu_T}
S_T\!\left(
\frac{\mu_{ T} t}{r^z}
\right)
\right],
\end{equation}
where $S_L$ and $S_T$ denote the longitudinal and transverse contributions to the velocity correlation function, respectively. The additional dependence on the propagation scaling variable $v_s t/r$ modifies only the detailed form of the scaling functions and does not alter the asymptotic scaling exponents, which remain identical to the incompressible case.

\subsection{Step 2: Identification of the upper critical dimension and relevant nonlinearities}
If the system is scale-invariant, we expect that the theory would remain  invariant under rescaling of space, time and fields with an arbitrary scaling factor $\ee^{\dd \ell}>1$,
\begin{equation}
    \br \rightarrow \ee^{\dd \ell} \br \ , \ \ \ \ t \rightarrow \ee^{z\dd \ell} t \ , \ \ \ \ v \rightarrow \ee^{\chi\dd \ell} v \ .
\end{equation}
Performing such a rescaling to our EOM with the nonlinearity restored, we have
\begin{equation}
   \ee^{(-z+\chi)\dd \ell}   \pp_t \bv +\lambda \ee^{(2\chi-1)\dd \ell}  (\bv\cdot \nabla)\bv=\mu \ee^{(-2+\chi)\dd \ell}    \nabla^2 \bv + \ee^{(\frac{-z-d-2}{2})\dd \ell}  \bff \ ,
\end{equation}
where we have added the coefficient $\lambda$ as  book-keeping parameter even though we have established in {\bf Section~\ref{sec:universal_models}} that $\lambda=1$ due to Galilean invariance.  This added parameter will be useful for our  perturbative expansion of the EOM later. 
Note that we have also ignored the Lagrange multiplier term ($\vnab p$) for simplicity since, as we saw in the previous section, the incompressibility will be enforced by using the tranverse projector $P_{ij}$ in Fourier space, which is manifestly scale invariant. We further note that the noise scaling follows from the fact under the scaling, 
the delta functions in the noise statistics scale as
$\delta^d(\br)\delta(t)
\to
e^{-(d+z)\dd\ell}
\delta^d(\br)\delta(t)$. 
Therefore the noise field must scale as the square root of this factor:
$\bff \to e^{-(d+z)\dd\ell/2}\bff$, 
which gives the factor
in the EOM.

We now divide both sides by $\ee^{(-z+\chi)\dd\ell}$ to get
\begin{equation}
 \pp_t \bv +\lambda \ee^{(\chi -1+z)\dd\ell}(\bv\cdot \nabla)\bv= \mu \ee^{(-2+z)\dd\ell}  \nabla^2 \bv + \ee^{(\frac{z-d-2-2\chi}{2})\dd\ell} \bff \ .
\end{equation}
Using the exponents we obtained from our linear theory, in the absence of the nonlinearity, the EOM is unchanged, thus confirming again that  the system is scale-invariant, as indicated by our linear theory (however, see {\bf Pitfall 2}).

\begin{tcolorbox}[colback=red!5!white,colframe=red!50!black,
  colbacktitle=red!75!black,title=Pitfall 2: Inferring scaling exponents directly from the rescaled EOM.]

A common shortcut in the literature is to rescale the EOM directly and infer the scaling exponents by demanding that some of the  coefficients remain scale invariant. While this procedure can sometimes yield the correct result, it may become misleading in systems containing both propagating and diffusive sectors.

\vspace{.08in}

To illustrate this, consider again the linearised {\it compressible} Navier-Stokes equation \eqref{eq:compNV}, for which one cannot simultaneously keep the coefficients of both the propagating  ($v_s^2/\rho_0$) and diffusive ($\mu$'s) terms invariant under rescaling. This ambiguity may then lead one to the erroneous conclusion that the diffusive terms are asymptotically negligible.

\vspace{.08in}

The flaw in this reasoning is that the asymptotic equal-time scaling structure of the fluctuations is generally not determined  by the propagating dynamics, but rather by the stationary equal-time correlation function. Indeed, linear compressible Navier-Stokes fluids 
exhibit the same equal-time scaling structure as the incompressible case analysed above, despite the presence of propagating sound modes.

\vspace{.08in}

The same subtlety also appears in the Toner--Tu equations discussed in 
{\bf Section~\ref{sec:rev_tt}}, where propagating sound-like modes coexist with fluctuation-controlling diffusive dynamics, yet the asymptotic scaling behaviour is not dictated by the propagating modes.

\end{tcolorbox}

Using the same exponents, the nonlinear term itself scales as follows:
\begin{equation}
    (\bv\cdot\nabla)\bv \rightarrow \ee^{(\chi -1+z)\dd\ell} (\bv\cdot\nabla)\bv = \ee^{(-d/2 +1)\dd\ell} (\bv\cdot\nabla)\bv\ .
\end{equation}
As a result, whether the nonlinear term grows or shrinks upon rescaling critically depends on the dimensionality of the problem. If $d>2$,  by rescaling to larger and larger length scales, $\ee^{ \dd  \ell}\gg1$, the effective coefficient characterising this nonlinearity becomes smaller and smaller. So whether we can consider the nonlinearity as small and, therefore, whether it is valid to perturbatively expand the theory around the linear theory, is not a question of its initial value $\lambda=1$, but rather the scale we are considering. In particular, in the hydrodynamic limit, where $\ell\rightarrow\infty$, $\lambda$ will always become a small parameter. In this case we speak of $\lambda$ as being an irrelevant parameter which can be neglected when considering the scaling behaviour of the system.

Conversely, if $d<2$, $\lambda$ is a relevant parameter. As one takes to the hydrodynamic limit, the nonlinearity will always become large eventually and {\it cannot} be neglected. This will modify the scaling exponents.

The behaviour switches over at the  upper critical dimension $d_c = 2$. Here, $\lambda$ seems to neither grow nor shrink. One speaks of a marginal variable. Nonlinear effects can still lead to a logarithmic divergence of this term at the critical dimension so it cannot be neglected in this case either, even if it is initially small.

This observation resolves the previously arbitrary truncation procedure introduced in {\bf Step 5} of {\bf Recipe 1}: the expansion in fields and derivatives may be terminated once all relevant nonlinear terms have been retained. Importantly, the set of relevant nonlinearities generally depends on spatial dimension.

Recapitulating, this power-counting analysis shows that: (i) the upper critical dimension is $d_c=2$, and (ii)  just below $d_c$, the only relevant nonlinearity is $\lambda(\bv\cdot\nabla)\bv$. As can be verified straightforwardly, all higher-order nonlinear terms, {\it e.g.,} $\nabla^2 [(\bv \cdot \vnab) \bv]$, are irrelevant in $d=2-\epsilon$ dimensions, where $\epsilon$ is a small positive  number.

\subsection{Step 3: Nonlinear analysis via the RG}
Through the analysis above, we find that for the fluctuating incompressible Navier–Stokes equations, the upper critical dimension is $d_c = 2$. Consequently, the linear theory is sufficient to describe the scaling behaviour  of the system in all physical dimensions $d \geq 2$. At first glance, this may seem surprising, since most studies in fluid mechanics do not operate within a regime where nonlinearities are irrelevant.

This apparent paradox arises because the system we consider is highly idealised: it has no boundaries, no external forcing, and describes a fluid at rest subject only to momentum-conserving fluctuations.  In this setting, nonlinear advective effects are indeed subdominant. However, this situation is far removed from the typical contexts of interest in fluid mechanics, where driving, boundary conditions, and large-scale flows play a central role.

Our focus on this simplified system is therefore primarily pedagogical. It provides a clean setting in which to illustrate the renormalization group analysis without additional complications. With this understanding, we now proceed to examine how fluctuations modify the behaviour of the system in dimensions below the upper critical dimension, $d < d_c = 2$.

The key method we will use is a graphical method that facilitates the perturbative incorporation of the nonlinear term into the analysis. 
But before we dive into the formalism, we will first briefly discuss the key RG logic in the perturbative scheme that we focus on here.

\subsubsection{RG concept}
\label{sec:rgconcept}
To limit the scope of this review, our discussion of the renormalization group (RG) will focus on  the perturbative dynamic renormalization group\cite{forster_pra77}, in particular the $\epsilon$-expansion, with the goal of analysing RG flows and identifying their fixed points. We emphasise that the RG framework is considerably more general, and both restrictions can be relaxed using, for example, nonperturbative functional RG methods\cite{dupuis_pr21}.

\begin{figure}

  \center
\includegraphics[scale=.6]{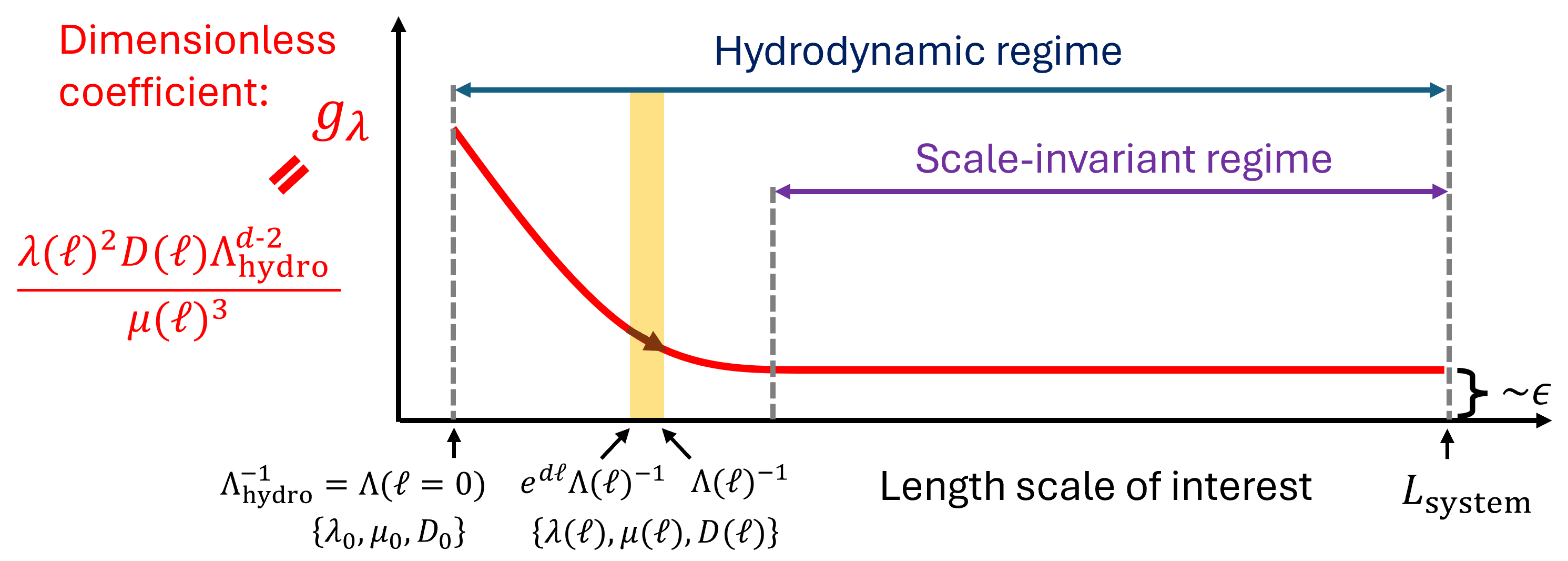}
\caption{
{\it
Schematic illustration of the concept underlying the Wilsonian renormalization group (RG).} Starting from a hydrodynamic theory defined at the microscopic hydrodynamic cutoff scale $\Lambda_{\rm hydro}^{-1}$, fluctuations within a thin momentum shell (yellow highlight) are integrated out,
 thereby increasing the effective minimum length scale  to $\Lambda(\ell)^{-1}$. This coarse-graining procedure changes the value of the effective dimensionless coupling $g_\lambda$ in a scale-dependent manner, a process referred to as RG flow.
The subsequent rescaling step (through changing the measurement units) restores the cutoff to its original value, enabling direct comparison between theories at different scales. 
If the RG flow approaches a scale-invariant regime, $g_\lambda$ converges to a fixed-point value $g_\lambda^\ast$, which defines the corresponding universality class. In perturbative RG schemes based on the $\epsilon$-expansion, the nontrivial fixed point satisfies $g_\lambda^\ast \sim \epsilon \equiv d_c-d$, thereby providing the small parameter justifying  the perturbative expansion and loop-based truncation.
}
\label{fig:sect3cartoon}
\end{figure}

With this preamble, we now elucidate the conceptual steps underlying a Wilsonian RG transformation\cite{wilson_physrep74}:
\begin{enumerate}
\item
We begin with the hydrodynamic model itself, so the shortest length scale of interest is $\Lambda_{\rm hydro}^{-1}$, namely the scale beyond which the hydrodynamic description becomes valid. Here, we use the conventional notation $\Lambda$ to denote an inverse length scale (see Fig.~\ref{fig:sect3cartoon}). Importantly, this scale is already far removed from the microscopic regime. The parameters appearing in the equations of motion are therefore defined at this cutoff scale and are referred to as the \textit{bare} parameters.

For example, in the incompressible Navier--Stokes equation with noise, the relevant parameters include the viscosity $\mu$, the noise strength $D$, and the bookkeeping parameter $\lambda$ associated with the nonlinear advective term.

\item
We then perform a coarse-graining step by integrating out fluctuations with wavevectors in a thin momentum shell
\beq
\Lambda_{ \rm hydro} e^{-\dd \ell} \leq |\bp| \leq \Lambda_{ \rm hydro} \ .
\eeq
This procedure corresponds to eliminating short-wavelength modes. The cumulative effect of these eliminated modes is to renormalize the parameters of the theory, thereby generating an effective equation of motion with modified coefficients. In this way, the procedure systematically captures how small-scale fluctuations of the hydrodynamic fields influence the emergent large-scale behaviour.

\item
Following coarse-graining, we restore the original cutoff by rescaling space, time, and fields according to
$ \br \mapsto e^{\dd \ell} \br,  
t \mapsto e^{z \dd \ell} t, 
\bv \mapsto e^{\chi \dd \ell} \bv,
$
where $z$ and $\chi$ are the dynamic and roughness exponents, respectively. 

Physically, after integrating out short-distance degrees of freedom, the system genuinely has a larger effective minimum length scale $\Lambda=e^{-\dd \ell}\Lambda_{\rm hydro}$ (i.e., a reduced UV cutoff). The rescaling step is therefore a change of units that restores the cutoff to its original value $\Lambda_{ \rm hydro}$, allowing the resulting theory to be compared directly with the original one.

A useful way to think about this procedure is the following. Suppose the initial cutoff corresponds to $\Lambda_{ \rm hydro}^{-1}=1\,\mathrm{mm}$. After coarse graining, we integrate out fluctuations on length scales between $1\,\mathrm{mm}$ and $10\,\mathrm{mm}$, so that the new minimal length scale becomes $10\,\mathrm{mm}$. We then redefine the unit of length from millimetres to centimetres; under this change of units, $10\,\mathrm{mm}$ is represented as $1\,\mathrm{cm}$, thereby restoring the numerical value of the cutoff. Physically, the system has become coarser, but when expressed in the rescaled units, the theory again takes the same form as before.

To make this process more transparent, we parametrise the RG flow by the logarithmic scale variable $\ell$, such that the running momentum cutoff is $\Lambda(\ell)=\Lambda_{\rm hydro} e^{-\ell}$.  When the unit of measurement is fixed, $\Lambda(\ell)^{-1}$ increases as shown in Fig.~\ref{fig:sect3cartoon}.

Throughout the remainder of this section, parameters appearing in the coarse-grained theory are understood to depend on the RG scale $\ell$. We therefore write $\mu(\ell), D(\ell), \lambda(\ell)$, while the corresponding bare parameters at the hydrodynamic cutoff scale are denoted $\mu_0, D_0, \lambda_0$.

\item
Iterating this combined coarse-graining and rescaling procedure generates a flow of the model parameters under successive RG transformations.

\end{enumerate}

In summary, starting from a hydrodynamic model, an RG transformation consists of two key steps:
(i) coarse-graining, implemented by integrating out short-wavelength modes within a thin momentum shell, and
(ii) rescaling, which restores the cutoff to its original value and allows for direct comparison between theories at different scales.
 
Regarding (i), the thin-shell construction is one of Wilson’s key conceptual advances, since it allows the RG transformation to be formulated as differential flow equations for the running couplings. As we shall see, integrating out fluctuations only within an infinitesimal shell $\Lambda e^{-\dd\ell}<|\bp|<\Lambda$ introduces the small parameter $\dd\ell$, enabling the flow of the couplings to be tracked continuously with scale. Moreover, because only modes near the cutoff are eliminated at each step, the resulting corrections can be systematically expanded in external momenta, ensuring that the effective hydrodynamic theory retains the same local structure as the original EOM, with only its coefficients being renormalized. By contrast, in real-space RG approaches, coarse graining is typically performed over finite spatial blocks, corresponding to integrating out fluctuations across a broad range of length scales in a single step. In that case, contributions from different scales become mixed, making systematic and controlled approximations considerably more difficult.

Regarding (ii), although the rescaling step is often viewed as essential, in practice one can frequently extract large-scale scaling behaviour directly by tracking how coarse-graining modifies the parameters as the cutoff is lowered, without explicitly performing the rescaling at each stage. In this perspective, universal scaling is encoded in the asymptotic dependence of the renormalized parameters on the running cutoff.

That said, the rescaling step plays an important conceptual role: it restores the cutoff, renders the couplings dimensionless, and makes scale invariance explicit through the appearance of fixed points in the RG flow equations. It therefore provides a systematic and broadly applicable framework for identifying scaling regimes and universality classes, particularly in more complex settings.

\subsubsection{Graphical Method}
Having hopefully prepared our readers for what lies ahead, we are ready to perform the perturbative RG analysis. We will
 build on the formulation introduced in  {\bf Section~\ref{ssec:linear_theory}}, where the incompressible Navier--Stokes equation was expressed in Fourier space and the bare propagator $G^0_{ij}(\tilde{\bq})$ was defined. Our goal here is to systematically incorporate the nonlinear term $\lambda(\bv\cdot\nabla)\bv$ using a perturbative expansion organised by graphical methods.

We therefore return to the full EOM already introduced in {\bf Section~\ref{ssec:constructing_eom}},
\begin{equation}
    \partial_t \bv + \lambda (\bv \cdot \nabla ) \bv 
    = - \nabla p + \mu \nabla^2 \bv + \bff \ ,
\end{equation}
together with the incompressibility constraint and noise correlations specified there. As before, we eliminate the pressure term by using the transverse projector $P_{ij}(\bq)$, so that all dynamics can be expressed solely in terms of $\bv$.

Using the same Fourier-space representation as in {\bf Section~\ref{ssec:linear_theory}}, the EOM can be written in the compact form
\begin{equation}
    \label{eq:eomfourier}
    v_i(\tilde \bq) 
    = G^0_{ij}(\tilde \bq)\, f_j(\tilde \bq)
    -  G^0_{ij}(\tilde \bq)  \frac{\ii\lambda}{2} \int_{\tilde \bp} \Big[\bv(\tilde \bq-\tilde \bp)\cdot \bq\,  v_j(\tilde \bp) +\bv(\tilde \bp)\cdot \bq\,  v_j(\tilde \bp-\tilde \bq)  \Big]\ ,
\end{equation}
where $\tilde \bq=(\bq,\omega)$, and $G^0_{ij}(\tilde \bq)$ is precisely the bare propagator obtained from the linear theory in  {\bf Section~\ref{ssec:linear_theory}}. 
This equation defines a recursive relation for the velocity field: the first term is the linear response to the noise, while the second term describes mode-coupling induced by the nonlinearity. Iterating this relation generates a perturbative expansion in powers of $\lambda$, in which each order involves integrals over internal momenta and frequencies.

The graphical method provides a convenient way to organise this expansion. Each occurrence of the propagator $\bG^0$, the noise correlation, and the nonlinear interaction can be represented as diagrammatic elements, allowing perturbative corrections to be constructed and evaluated systematically. We now introduce these diagrammatic rules and use them to compute corrections to the linear theory.

Specifically, we introduce the following graphical notation,
\begin{align}
\diagram{1} &= G^0_{ij}(\tilde \bq) \ , \\
\diagram{2} &= -\frac{ \ii \lambda}{2} [q_j\delta_{ik} +p_k \delta_{ij} ] \ , \\
\diagram{3} &= v_i(\tilde \bq)  \ , \\
\diagram{4} &= f_i(\tilde \bq) \ , \\
\label{eq:noise_diag}
\diagram{5} &= 2D q^2 \delta_{ij}   \ ,
\end{align}
 with which we can represent Eq.~\eqref{eq:eomfourier} as,
\begin{equation}
    \label{eq:eom_graph}
    \diagram{3} = \diagram{6} + \diagram{7} \ .
\end{equation}
 When connecting two elements to form a diagram, one needs to ensure conservation of ``momenta\footnote{The term momentum is often used by analogy with quantum field theory. In the context of classical statistical field theory, however, these quantities are simply wavevectors and should not be interpreted as physical momenta. Their conservation at each vertex follows directly from translational invariance.}'', i.e., wavenumbers at each vertex, and contract the indices of two connected lines. Further, one needs to integrate over each field momentum. As a consequence all indices and momenta other than $\tilde \bq$ and $i$ are dummy indices which is why we haven't specified them. Since there is only one external line, the label $(\tilde \bq,i)$ can also be omitted, since it is unambiguous and arbitrary.

\subsubsection{Perturbative Expansion}

Eq.~\eqref{eq:eom_graph} defines a recursion relation for $\bv$, which can in principle be solved by recursively inserting Eq.~\eqref{eq:eom_graph} into itself. This defines a perturbative series which only converges if the coupling $\lambda$ is small.  What it means for $\lambda$ to be small will be made more precise later. If  this is indeed the case, we can justify a truncation of the series. For the purposes of this review, we expand the series to  third order in $\lambda$,

\begin{align}
    \label{eq:eom_recursive}
    \diagram{8} = \diagram{9} +& \diagram{10} + 2\diagram{13}  + 4\diagram{31} +\diagram{32} \ .
\end{align}

\subsubsection{One-loop diagrams and graphical corrections}

We now construct the corresponding one-loop diagrams by projecting from Eq.~\eqref{eq:eom_recursive} the structures already present in the original EOM and averaging over the remaining noise terms. For example, the propagator can be obtained by taking a functional derivative of Eq.~\eqref{eq:eom_recursive} with respect to $\bff$. Setting $\bff=0$ then yields the noiseless limit and hence the bare propagator. To obtain the renormalized propagator, however, we instead average over the noise field. Since $\bff$ is a zero-mean Gaussian noise, this procedure amounts to discarding all odd powers of $\bff$, while for even powers one performs all possible pairwise contractions of the noise fields using the replacement,
\begin{equation}
    \diagram{9}\diagram{33} \to \diagram{28} \ ,
\end{equation}
i.e., for the renormalized propagator we find,
\begin{equation}
    \label{eq:red_prop_diags}
    G  = \left\langle\frac{\delta \bv}{\delta\bff} \right\rangle = \diagram{20} = \diagram{21} + 4 \diagram{22} + 2 \diagram{23} \ .
\end{equation}

Similarly, to obtain the renormalized vertex $\lambda$ we take two functional derivatives with respect to $\bff$, corresponding to the system's response to two fluctuations of the forcing field, i.e.,
\begin{align}
    \label{eq:red_triangle_diags}
    \nonumber
    \left\langle\frac{\delta^2 \bv}{\delta\bff^2} \right\rangle =& 2\diagram{34} +8 \diagram{35} +16 \diagram{36} +8 \diagram{37} \\
    & + 16\diagram{38} + 8\diagram{39} + 4\diagram{40} \ ,
\end{align} 
where it is now understood that the incoming lines are symmetrized.

At first glance, this may appear to involve a large number of diagrams. However, the expansion contains a considerable amount of redundant information. For example, four of the diagrams in Eq.~\eqref{eq:red_triangle_diags} merely represent the renormalization of the propagator, which has already been derived in Eq.~\eqref{eq:red_prop_diags}. Consequently, to order $\lambda^3$, Eq.~\eqref{eq:red_triangle_diags} can be rewritten as,
\begin{equation}
    \label{eq:triangle_diags}
    \left\langle\frac{\delta^2 \bv}{\delta\bff^2} \right\rangle = 2\diagram{41} +16 \diagram{42} +8 \diagram{43} \ .
\end{equation} 
To project this expression onto $\lambda$ itself, the remaining step is to remove, \emph{i.e.}, amputate, the external propagators from the diagrams. We therefore see that, to the chosen order in $\lambda$, the renormalization of the parameter $\lambda$ requires the evaluation of only two amputated diagrams. In the Appendix, we show that the two diagrams exactly cancel, so that $\lambda$ is not renormalized, as required by Galilean invariance.

Furthermore, the last diagram in Eq.~\eqref{eq:red_prop_diags} should also be excluded from the renormalization of the propagator, since it instead corresponds to the renormalization of the expectation value of $\bv$, i.e.,
\begin{equation}
    \left\langle\bv \right\rangle = \diagram{44} + \mathcal O(\lambda^2) \ .
\end{equation} 
Indeed, fluctuations can modify the observed expectation value of the hydrodynamic field relative to its mean-field prediction. Strictly speaking, this effect should already be accounted for in the expansion of the hydrodynamic EOM about the mean state. It can be incorporated through a simple shift of $\bv_{0}$ by the contribution represented by this diagram in the EOM, which precisely cancels the second diagram in Eq.~\eqref{eq:red_prop_diags}. In the Appendix, we show that, in the present case, this diagram also vanishes identically, as a consequence of momentum conservation.

Both of these observations exemplify a more general principle: graphical corrections that are already accounted for through the renormalization of another quantity should not be included separately. More generally, in the renormalization of a parameter, only {\it one-particle-irreducible (1PI) diagrams} need to be considered. These are diagrams that remain connected when any single internal line is cut.

Let us now apply this principle to the renormalization of the noise term. To do so, we expand the two-point correlation function of $\bv$ to order $\lambda^3$,
\begin{equation}
    \left\langle\bv \bv\right\rangle = \diagram{45} + 2\diagram{46} 
\end{equation}
such that, after discarding one-particle-reducible (1PR) diagrams ({\it i.e.}, diagrams that are {\it not} 1PI), only a single diagram remains. The omitted 1PR diagrams again merely account for the renormalization of the propagator and of the field expectation value.

\paragraph{Propagator graphical correction.} We are now in a position to calculate the RG corrections to the couplings. As shown above, only a single diagram contributes to the propagator at the chosen order in $\lambda$. To facilitate comparison with the parameter $\mu$ appearing in the propagator, we invert Eq.~\eqref{eq:red_prop_diags},
\begin{equation}
    G^{-1}_{ij}(\tilde \bq) = \diagram{20} {}^{-1} = \diagram{21}{}^{-1} - 4 \diagram{24} 
\end{equation}
which again holds to order $\mathcal O(\lambda^2)$. In the final diagram, the external lines without arrows indicate that the external propagators have been amputated.

We then project this diagram onto the transverse sector,
\begin{align}
    \label{eq:calc_loop1}
    \nonumber
    &-4\frac{P_{in}(\bq)}{d-1}\diagram{25}  \\
    \nonumber
    &=- \frac{P_{in}(\bq)}{d-1}\int_{\tilde \bp} (-\ii \lambda) \Big[\Big(p_j+\frac{q_j}{2}\Big) \delta_{ik}+\Big(\frac{q_k}{2}-p_k\Big)\delta_{ij}\Big] \frac{P_{jl}(\bq/2-\bp)}{-\ii \omega_q/2+\ii\omega_p+\mu |\bq/2-\bp|^2} \\
    &\hspace{1.1cm} \times (-\ii \lambda) \Big[-\Big(p_n+\frac{q_n}{2}\Big)\delta_{ml}+q_m\delta_{nl}\Big] \frac{2D \Big|\frac{\bq}{2}+\bp\Big|^2 P_{mk}(\bq/2+\bp)}{|-\ii\omega_q/2-\ii \omega_p+\mu |\bq/2+\bp|^2|^2} \\
    &= q^2 \frac{S_d}{(2\pi)^d} \frac{1}{2} \frac{d^2-2}{d^2+2d} \frac{D \lambda^2}{\mu^2} \int_{0}^{\Lambda_{\rm hydro}} \dd p \frac{p^{d-1}}{p^2} + \mathcal{O}(q^4,q^2\omega_q)
\end{align}
where, in the final line, we expanded the diagram to leading order in wavenumber and frequency so as to match the terms already present in the EOM, consistent with the hydrodynamic limit. We thus find that the bare coefficient $\mu$ appearing in the linear part of the EOM is renormalized by the  combined effects of nonlinearities and fluctuations,
\begin{equation}
    \label{eq:corrmu}
    \mu^\prime = \mu + \frac{S_d}{(2\pi)^d} \frac{1}{2} \frac{d^2-2}{d^2+2d} \frac{D \lambda^2}{\mu^2} \int_{0}^{\Lambda_{\rm hydro}} \dd p \frac{p^{d-1}}{p^2} \ .
\end{equation}

 A more detailed calculation of this diagram, as well as the calculation of the remaining diagrams is presented in the Appendix.

\subsubsection{DRG flow equations, fixed points \& universality classes}

Since we are considering the case $d<2$, the integral on the right-hand side is divergent. This divergence directly reflects the fact that $\lambda$ is a relevant nonlinearity. Here, the central idea of the RG becomes essential: instead of averaging over all fluctuations at once, as done above, we integrate out only the noise modes lying within a thin momentum shell, \emph{i.e.}, modes $f(\omega,\bp)$ with $\Lambda(\ell) \ee^{-\dd \ell}<|\bp|<\Lambda(\ell) $. Under this procedure, the integral in Eq.~\eqref{eq:corrmu} becomes
\begin{equation}
    \label{eq:thinshell}
    \int_{\Lambda(\ell)\ee^{-\dd \ell}}^{\Lambda(\ell)} \dd p \frac{p^{d-1}}{p^2}   = \Lambda(\ell)^{d-2} \dd \ell\ ,
\end{equation}
and so Eq.~\eqref{eq:corrmu} is transformed into an ordinary differential equation governing the flow of $\mu$,
\begin{equation}
    \frac{\dd\mu}{\dd \ell} = \frac{S_d}{(2\pi)^d} \frac{1}{2} \frac{d^2-2}{d^2+2d} \frac{D \lambda^2}{\mu^2}  \Lambda(\ell)^{d-2} \ ,
\end{equation}
which describes how $\mu$ is progressively modified by fluctuations originating from each scale individually. Applying the same procedure to the other couplings, $D$ and $\lambda$ (see the Appendix for the explicit calculations), yields,
\begin{equation}
    \frac{\dd D}{\dd \ell} = \frac{S_d}{(2\pi)^d} \frac{1}{2} \frac{d^2-2}{d^2+2d} \frac{D^2 \lambda^2}{\mu^3}  \Lambda(\ell)^{d-2} \ , \ \ \ \ \text{and} \ \ \ \ \frac{\dd \lambda}{\dd \ell} = 0 \ .
\end{equation}

Integrating these equations from $\ell=0$ to some larger coarse-graining scale $\ell$ systematically incorporates fluctuations between the microscopic cutoff $\Lambda_{\rm hydro}$ and the running cutoff $\Lambda(\ell)$. Although the couplings still diverge in the limit $ \ell\to\infty$, we are now in a much better position to interpret this divergence by comparing effective theories defined at different scales.

But how can fluctuations at, say, the millimetre scale be meaningfully compared with those at the centimetre scale? The answer is to render the EOM dimensionless with respect to the relevant scale. Within the RG framework, this is naturally achieved by expressing all quantities relative to the running cutoff scale $\Lambda(\ell)=e^{-\ell}\Lambda_{\rm hydro}$.

Rescaling length scales alone, however, is not sufficient. At larger length scales, the relevant time scales also change, as does the typical magnitude of fluctuations, characterised respectively by the dynamic exponent $z$ and the roughness exponent $\chi$. 
This is the rationale underlying the rescalings below,
\begin{equation}
    \br \rightarrow  \br e^{\ell} \ , \ \ \ t \rightarrow  t e^{z \ell } \ , \ \ \ \ \bv \rightarrow  \bv e^{\chi \ell} \ .
\end{equation}
thereby mapping  effective theories back to the original {\it numerical}  value of the cutoff $\Lambda_{\rm hydro}$, where they can be directly compared with one another. 
Expressed in terms of the resulting dimensionless nonlinear coupling, 
\begin{equation}
    g_\lambda \equiv \frac{S_d}{(2\pi)^d} \frac{\lambda(\ell)^2 D(\ell){ \Lambda_{\rm hydro}{}^{d-2}}}{\mu(\ell)^3} \ ,
\end{equation}
i.e., with the numerical value of the cutoff  fixed at $\Lambda_{\rm hydro}$ due to  the rescaling step,
the RG flow equations can be brought into the form,
\begin{align}
    \label{eq:flow_1}
\frac{1}{\mu}\frac{d\mu}{d\ell}
&= z - 2 + \frac{1}{2}\frac{d^2-2}{d^2+2d}\,g_\lambda \ ,
\\
    \label{eq:flow_2}
\frac{1}{D}\frac{dD}{d\ell}
&= z - d - 2\chi + \frac{1}{2}\frac{d^2-2}{d^2+2d}\,g_\lambda \ ,
\\
    \label{eq:flow_3}
\frac{1}{\lambda}\frac{d\lambda}{d\ell}
&= z + \chi - 1 \ .
\end{align}
Again, taking the flow equation for $\mu$ as an example, the term $(z-2)$ originates from the rescaling step, whereas the term proportional to $g_\lambda$ arises from the one-loop graphical correction.
At first sight, the fixed-point problem appears overconstrained: there are three RG flow equations, but only two scaling exponents, $z$ and $\chi$, available to adjust. This apparent mismatch is resolved by recognising that the flow equations for the dimensionful parameters contain redundancies associated with arbitrary choices of units. The physically meaningful RG flow must therefore be formulated in terms of dimensionless combinations of parameters.

\begin{figure}

  \center
\includegraphics[scale=.6]{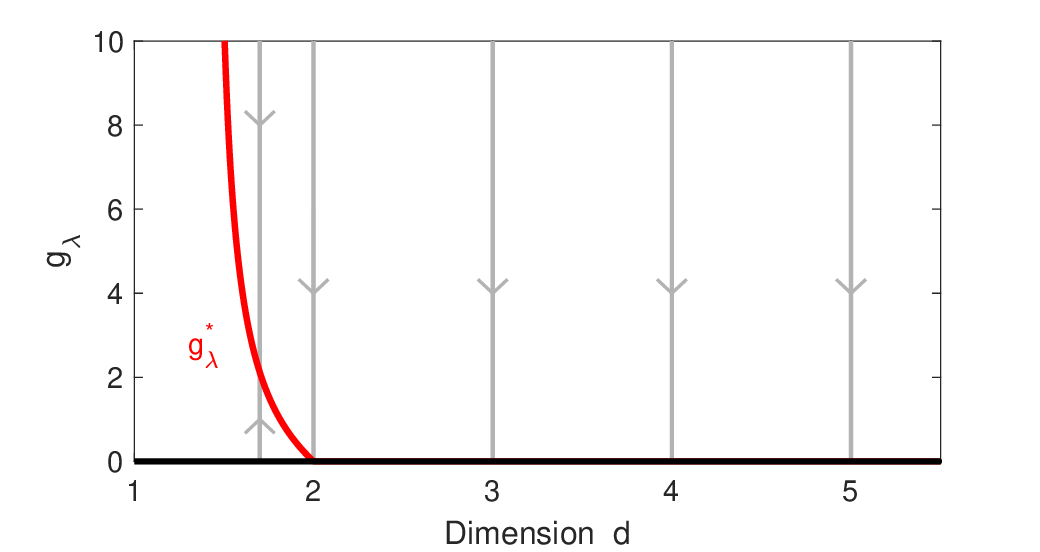}
\caption{
{\it RG flow of the fluctuating incompressible Navier--Stokes equation.}
Renormalization group (RG) flow of the dimensionless coupling $g_\lambda$ as a function of spatial dimension $d$, deduced from our perturbative calculation to the one-loop level. The red curve denotes the nontrivial fixed-point value $g_\lambda^\ast$  [Eq.~\eqref{sect3eq:g*}], which exists only below (and { strictly speaking,} close to) the upper critical dimension $d_c=2$. For $d>2$, the only stable fixed point is the Gaussian fixed point $g_\lambda^\ast=0$ (black horizontal line). Grey vertical lines illustrate the RG flow at representative integer dimensions, with arrows indicating the direction of the  RG flow. For $d<2$, the coupling flows toward a nonzero fixed point, signalling the breakdown of the linear theory and the emergence of a non-Gaussian universality class. For $d>2$, the coupling flows to zero, implying that the asymptotic scaling behaviour is governed by the linear theory. The merger of the nontrivial and Gaussian fixed points at $d_c=2$ gives rise to the controlled $\epsilon=d_c-d$ expansion.
}
\label{fig:ns}
\end{figure}

In the present problem, the quantities entering the RG equations are
$\mu$, $D$, $\lambda$, and $\Lambda$. Their dimensions are constructed from three independent units:
$[t]$, $[\br]$, and $[\bv]$. Although $\bv$ physically has the dimension of a velocity, this was an arbitrary choice of normalization for the hydrodynamic field. Within the RG framework it carries an independent field-normalisation exponent $\chi$. Since the equation of motion is insensitive to the choice of units used to measure the field amplitude, $[\bv]$ must therefore be treated as an independent unit for dimensional analysis.
Consequently, by the Buckingham $\Pi$ theorem, four dimensional quantities built from three independent units can generate only a single independent dimensionless combination. Up to arbitrary numerical prefactors and powers, this combination is precisely $g_\lambda$. 
Indeed, the fixed-point structure is most naturally analysed in terms of this dimensionless coupling $g_\lambda$, rather than the individual dimensionful parameters themselves. From the definition of $g_\lambda$,
\begin{align}
\frac{1}{g_\lambda}\frac{d g_\lambda}{d\ell}
&=
\frac{2}{\lambda}\frac{d\lambda}{d\ell}
+
\frac{1}{D}\frac{dD}{d\ell}
-
\frac{3}{\mu}\frac{d\mu}{d\ell}
=
(2-d)
-
\frac{d^2-2}{d^2+2d}\,g_\lambda ,
\end{align}
where, in the second equality, the dependence on $z$ and $\chi$ cancels, as expected for a dimensionless coupling ---  the RG flow of a dimensionless coupling cannot depend on arbitrary choices of field and time units. Therefore,
\begin{align}
\frac{d g_\lambda}{d\ell}
=
\left[
(2-d)
-
\frac{d^2-2}{d^2+2d}\,g_\lambda
\right]g_\lambda .
\end{align}
For $d>2$, $g_\lambda$ flows to zero, corresponding to the linear theory. This confirms that above the upper critical dimension $d_c$, the scaling behaviour of the system is governed by the scaling exponents obtained from the linearised EOM. By contrast, below $d=2$, the RG flow approaches a stable nontrivial fixed point,
\begin{align}
\label{sect3eq:g*}
g_\lambda^*
=
\frac{(2-d)(d^2+2d)}{d^2-2}
=
\frac{d(d+2)(2-d)}{d^2-2}= 4 \epsilon + \cO(\epsilon^2) \ ,
\end{align}
where the final expression is in terms of $\epsilon \equiv d_c-d =2-d$. We will discuss  the origin of this controlled expansion in $\epsilon$ in the next section.  This behaviour is illustrated in Fig.~\ref{fig:ns}.

The stable fixed point $g_\lambda^*$
 defines a scale-invariant theory. All microscopic systems whose coarse-grained dynamics flow toward this fixed point therefore belong to the same {\bf universality class (UC)} ({\bf Step 5} of {\bf Recipe 2}).
The scaling exponents characterising this UC can be obtained by substituting the fixed-point value into the RG flow equations of the dimensionful parameters \eqref{eq:flow_1}-\eqref{eq:flow_3}. This yields
\begin{align}
z &= 1+\frac{d}{2} \quad , \quad \chi = -\frac{d}{2}
\quad , \quad (d<2,~{\rm uncontrolled~approximation})\ .
\end{align}
The qualifier ``uncontrolled approximation'' reflects the fact that the one-loop fixed point is strictly controlled only near $d_c$, where $g_\lambda^\ast \sim \epsilon \ll 1$. For non-small $\epsilon$, contributions from higher-loop corrections are no longer small, so the extrapolation of the one-loop result to arbitrary $d<d_c$ is not controlled. For completeness, we also quote the controlled result obtained from a systematic $\epsilon$-expansion about $d_c=2$:
\begin{align}
z &= 2 - \frac{\epsilon}{2} \quad , \quad 
\chi = -1 + \frac{\epsilon}{2}
\quad , \quad (d<2,~{\rm controlled~approximation~to}~\cO(\epsilon))\ ,
\end{align}
where $\epsilon = d_c - d$.

\subsection{Controlled nature of the $\epsilon$-expansion}
We now explain why a one-loop perturbative RG calculation yields the leading correction in an $\epsilon$-expansion, where $\epsilon \equiv d_c-d$. The argument is heuristic, but captures the essential reasoning.

At the level of dimensional analysis, a typical one-loop correction involves a single shell integral of the form \eqref{eq:thinshell}, and therefore scales as $\Lambda^{d-2}\dd\ell$. In the Navier--Stokes problem, this combines with the bare parameters into the dimensionless coupling $g_\lambda$. Thus, one-loop corrections to the equations of motion are proportional to $g_\lambda$. At the perturbative fixed point, $g_\lambda^\ast \sim \epsilon$, implying that the resulting corrections to scaling exponents are of order $\mathcal{O}(\epsilon)$.

To understand higher-loop corrections, consider a representative two-loop contribution of the form
\begin{equation}
\label{eq:sect3_2loop}
    \int_{\Lambda e^{-\dd \ell} < |\bp| < \Lambda} 
    \int_{ |\bk| < \Lambda} 
    \frac{d^d p}{p^2}
    \frac{d^d k}{k^2}
    \, B(\bp, \bk, \bq),
\end{equation}
where the factors $p^{-2}$ and $k^{-2}$ originate from propagators, and $B$ denotes the remaining structure of the integrand.

A crucial technical point is that only one of the loop momenta (say $\bp$) can be restricted to the shell, while the other must be integrated over the full domain. If both were restricted to the shell, the contribution would be of order $\dd\ell^2$ and would therefore vanish in the limit $\dd\ell \to 0$. Thus, genuine two-loop contributions that survive in the differential RG involve one shell integration and one bulk integration.

Neglecting the detailed structure of $B$ for the purpose of dimensional analysis, Eq.~\eqref{eq:sect3_2loop} scales schematically as
\begin{equation}
\Lambda^{2(d-2)} \dd\ell \sim g_\lambda^2 \ .
\end{equation}

While the precise numerical coefficient depends on the detailed structure of $B$, the overall scaling structure is generic: each additional loop introduces an additional power of the dimensionless coupling. In the present problem, this coupling is $g_\lambda$. More generally, in perturbative RG schemes controlled by an $\epsilon$-expansion, the nontrivial fixed point satisfies $g^\ast \sim \epsilon$, so higher-loop contributions are parametrically suppressed by correspondingly higher powers of $\epsilon$. This hierarchy underpins the controlled nature of the $\epsilon$-expansion near $d_c$, and justifies truncation at one-loop order to leading accuracy.

\subsection{Beyond $\epsilon$-expansion}

We conclude this section by briefly discussing approaches that extend beyond the perturbative RG framework emphasised here.

\subsubsection{`Exact' scaling exponents from non-renormalization conditions}
\label{sec:exactexps}
Suppose that, for a given hydrodynamic model, we have completed {\bf Steps 1 \& 2} of {\bf Recipe 2} and identified all nonlinearities that are relevant in the vicinity of the upper critical dimension $d_c$. It may then happen that the structure of these nonlinearities implies that certain coefficients cannot be renormalized.
For example, consider a nonlinearity that always carries a prefactor of an external momentum $q_x$. Any loop correction constructed from such a vertex must also retain this prefactor and therefore cannot generate graphical corrections to terms that do not contain $q_x$, such as the transverse diffusive term $\mu_y \partial_y^2$. In this case, $\mu_y$ is protected from renormalization by such nonlinearity in  the theory.

If the number of independent non-renormalization conditions equals the number of scaling exponents to be determined, then the scaling exponents can be obtained without explicitly evaluating loop integrals. Such exponents are often referred to as {\it exact} because they follow from symmetry or structural constraints rather than from a perturbative calculation.
The word ``exact'' should however be interpreted with some care. These results are rigorously controlled only in the vicinity of $d_c$, where the set of nonlinearities retained in the analysis is complete. Far below $d_c$, operators that are irrelevant near $d_c$ may become important and invalidate the original non-renormalization arguments. 

\subsubsection{Nonperturbative functional RG}
\label{sec:npfrg}
A variety of exact RG formalisms have been developed that, in principle, yield exact RG flow equations. In practice,
these formulations involve an infinite hierarchy of equations, which are generically intractable and thus require approximations.
Consequently, despite being based on formally exact RG equations, these methods do not usually produce exact results.

Nevertheless, substantial progress over the past two decades has demonstrated that suitably chosen approximation schemes can yield remarkably accurate predictions for scaling exponents \cite{dupuis_pr21}, even far away from the upper critical dimension in some cases. Among the various formulations, the nonperturbative functional RG based on the Wetterich equation has emerged as a particularly powerful and widely used framework. As we shall see in the next section, this approach has recently been employed to uncover novel UCs.

%% file: SectionFour_v2.tex
\section{New universality classes uncovered since 2015}
\label{sec:new_ucs}
Having established how universal hydrodynamic descriptions can be constructed from minimal physical ingredients ({\bf Section~\ref{sec:universal_models}}), and how their large-scale behaviour can be analysed using renormalization group (RG) methods ({\bf Section~\ref{sec:rg}}), we are now in a position to survey recently discovered universality classes (UCs) through this unified framework.

To keep the discussion focused, we restrict attention to UCs that (i) were identified from 2015 onwards, (ii) describe intrinsically nonequilibrium, classical ({\it i.e.}, non-quantum) systems, and (iii) are governed by non-Gaussian RG fixed points --- their scaling behaviour lies beyond linearised hydrodynamics.

Under these criteria, the majority of currently known examples arise from active-matter systems. This predominance reflects the central role that active matter has played in the recent expansion of nonequilibrium UCs, providing one of the most fertile settings for the discovery of genuinely new large-scale collective behaviour.

Even within our restricted scope, the number of new UCs uncovered over the past decade is substantial. Not all of them, however, are equally accessible in practice. A UC that governs a stable phase and emerges without parameter fine-tuning is generally far more likely to be realised experimentally than one associated with a finely tuned critical or multicritical system. To make this distinction precise, we begin by clarifying the concepts of \emph{phases}, \emph{critical points}, and \emph{multicritical points}.

\subsection{Phases, critical points, and multicritical points}
\label{sec:phases_crit}

We have seen in {\bf Section~\ref{sec:universal_models}} how one begins by writing down the most general equations of motion (EOM) consistent with symmetries in order to identify potential phases of a given class of systems. The qualifier ``potential'' is essential: the phases identified via the mean-field analysis in {\bf Recipe 1} may be destabilised by linear instabilities, fluctuations, and/or nonlinear effects.

Here, we focus exclusively on dynamical systems that possess stable RG fixed points and therefore exhibit well-defined scale-invariant behaviour. In this setting, {\bf Recipe 3} provides a systematic framework for identifying scale-invariant phases, as well as any associated critical or multicritical points.

\begin{tcolorbox}[colback=yellow!5!white,colframe=yellow!50!black,
  colbacktitle=yellow!75!black,
  title={Recipe 3: Identifying scale-invariant phases, critical, and multicritical points}]
1.~Starting from the EOM obtained in {\bf Recipe 1} for a chosen mean-field phase, determine the RG fixed points (FPs) using {\bf Recipe 2}.
\\
2.~Classify the RG FPs by their stability with respect to small perturbations in {\bf all} model parameters. RG FPs that are stable against all such perturbations correspond to scale-invariant phases of the hydrodynamic theory. RG FPs that are unstable with respect to perturbations in {\bf any} parameters describe critical or multicritical behaviour.
\\
3.~For each unstable RG FP, determine the number of unstable directions in the RG flow. Fixed points with the smallest number of unstable direction(s) correspond to critical points, while those with more unstable directions are identified as multicritical points.
\end{tcolorbox}

\subsubsection{Example: Surface growth}

We illustrate the above classification principles using the Kardar-Parisi-Zhang (KPZ) equation for surface growth\cite{kardar_prl86}. Here, the hydrodynamic field is the interfacial height $h$ between two phases. This can represent, for example, the growing front of a bacterial colony invading empty space, or the interface between two immiscible liquids. The position of the interface is \emph{a priori} not special, leading to translational invariance on the interfacial positions. Therefore, the dynamics must obey the shift symmetry $h \to h + w$ for an arbitrary constant $w$. Following {\bf Recipe 1}, we therefore arrive at the hydrodynamic equation of motion
\begin{equation}
  \label{eq:KPZ}
  \pp_t h = v + \lambda |\nabla h|^2 + \mu \nabla^2 h + f \ ,
\end{equation}
with nonconserving noise satisfying $\langle f(0,0)f(t,\br)\rangle
=
2D\delta^{d}(\br)\delta(t)$.
Upon shifting to the comoving frame, $h\to h+vt$, the constant growth term can be removed. The resulting deterministic mean-field equation admits only a spatially uniform stationary solution corresponding to a flat interface.

Restricting first to $d<2$, application of {\bf Recipe 2} shows that the mean-field scaling behaviour, corresponding to the linear Edwards-Wilkinson (EW) UC\cite{edwards_rspa82}, is unstable against the nonlinear coupling $\lambda$. Instead, the RG flow approaches a nontrivial fixed point with nonzero dimensionless coupling $ g_{\rm KPZ}
    \sim
 \lambda^2 D \Lambda_{\rm hydro}^{d-2}/\mu^3$,
which is stable against generic symmetry-allowed perturbations. According to {\bf Recipe 3}, this fixed point -- the KPZ UC -- generically governs the large-scale behaviour of the model, {\it i.e.}, it is classified as a phase, whereas the EW UC corresponds to a fine-tuned unstable limit obtained when $\lambda=0$ and is thus classified as a critical point.

Had we instead started from a different hydrodynamic model in which the two phases are physically identical, e.g., two identical but immiscible fluids, the dynamics would additionally possess an up-down symmetry $h\to -h$. This symmetry forbids the nonlinear term in Eq.~\eqref{eq:KPZ}, reducing the dynamics to the Edwards-Wilkinson model\cite{edwards_rspa82}. In this case, the EW UC generically describes the phase of the system, rather than a fine-tuned critical point.

\begin{figure}
  \center
  \includegraphics[scale=.6]{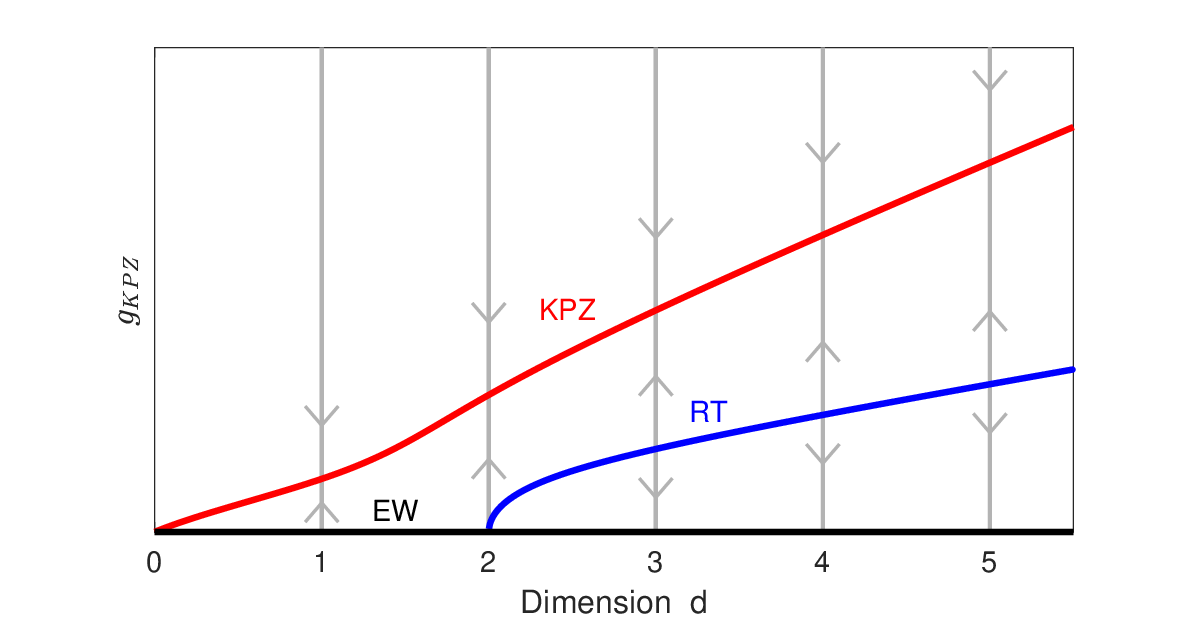}
  \caption{\textit{Schematic RG flow structure of the Kardar-Parisi-Zhang (KPZ) equation.} 
RG flow as a function of spatial dimension $d$, shown in terms of the dimensionless coupling $g_{\rm KPZ}$. Grey lines with arrows indicate the corresponding RG flow directions. For $d<2$, the perturbative KPZ fixed point (red) is the only stable fixed point and therefore generically governs the large-scale behaviour of the system. In contrast, the Gaussian Edwards--Wilkinson (EW) fixed point (black) is unstable and can only be realised by fine-tuning the nonlinear coupling to zero; it therefore corresponds to a critical point. For $d>2$, the EW fixed point becomes stable in the weak-coupling regime and thus governs a phase of the KPZ equation. The transition between the weak-coupling EW phase and the strong-coupling KPZ phase is controlled by an additional critical fixed point associated with the roughening transition (RT, blue). Figure adapted from Ref.~\cite{canet_jstatmech25}.}
  \label{fig:kpz}
\end{figure}

For $d\geq 2$, matters become even subtler. Perturbative RG predicts that, for sufficiently small bare $\lambda$, the nonlinear coupling flows to zero, such that the EW UC governs a weak-coupling smooth phase even for the KPZ equation with nonzero $\lambda$. It is, however, now well established that sufficiently large bare $\lambda$ instead drives the system towards a strong-coupling rough phase governed by a distinct strong-coupling RG fixed point\cite{canet_jstatmech25}. Accessing this regime requires nonperturbative methods extending beyond the perturbative RG framework discussed in {\bf Section~\ref{sec:rg}}. Further, the roughening transition separating the weak- and strong-coupling phases belongs to yet another UC. The different fixed points, their stabilities, and their dependence on spatial dimension are illustrated schematically in Fig.~\ref{fig:kpz}.

This example highlights the intrinsic fluidity in determining whether a given RG fixed point -- and its associated UC -- describes a phase, a critical point, or a multicritical point. Ultimately, this distinction depends sensitively on the {\bf underlying physics} encoded in the original hydrodynamic model.

\bigskip
\noindent
We now turn to a selection of key new universality classes (UCs) uncovered since 2015. To keep the discussion streamlined, 
  we restrict attention primarily to the more generic UCs governing genuinely new phases and critical points, rather than multicritical phenomena. Broadly speaking, these UCs can be organised into three archetypal classes of nonequilibrium systems: {\it interfacial dynamics}, {\it polar active fluids}, and {\it chemotactic systems}. The corresponding UCs, together with their relevant nonlinearities, upper critical dimensions, and scaling exponents, are summarised in three tables: Table~\ref{tab1} lists phase-governing UCs, Table~\ref{tab2} summarises critical-point UCs, and Table~\ref{tab:multi} compiles multicritical UCs, which will not be discussed in detail in the main text.

 \begin{table}
 \renewcommand{\arraystretch}{1.5}
\begin{tabular}{c|c|c|c|c|c} 
Sect. & System & Relevant terms & $d_c$ & Exponents & Method
\\
\hline
\hline
\ref{ssec:q-kpz} &
\makecell{Interfacial dynamics with  \\ quasi-static bulk phases\cite{besse_prl23}}
&$
|\nabla|\left(\nabla h\right)^2$   & 2 & $z=3-\frac{\epsilon}{3} \ , \
 \chi=\frac{\epsilon}{3}$ & 1-loop
 \\
\hline
\ref{ssec:tt_quenched_phase}&
PAF with quenched disorder\cite{toner_prl18,toner_pre18} &$(\bv_T \cdot \nabla)\,\bv_T$  & 5 & $z=\zeta=\frac{4}{3}\ ,\ \chi=-\frac{1}{3}$ & Exact (3D) 
 \\
\hline
\ref{ssec:malth_phase} &
Malthusian PAF\cite{chen_prl20,chen_pre20} &$(\bv_\perp \cdot \nabla_\perp)\,\bv_\perp$  & 4 & $z=2-\frac{6 \epsilon}{11} \ , \
\zeta =1-\frac{3\epsilon}{11}\ , \ 
 \chi=-1 +\frac{6\epsilon}{11}$ & 1-loop 
 \\
\hline
\ref{ssec:malth_ezpl} &
\makecell{Malthusian PAF with\\ an easy plane ($xy$-plane)\cite{toner_pre24}} &$\pp_yv_y^2$  & 4 & $z=\frac{3}{2} \ , \
\zeta =\frac{3}{4}\ , \ 
 \chi=-\frac{1}{2}$ & Exact (3D) 
 \\
\hline
\ref{ssec:incompr_phase} &
Incomp.~PAF in 3D\cite{chen_njp18}&$(\bv_T \cdot \nabla)\,\bv_T$   & 4 & $z=\frac{8}{5} \ , \
\zeta =\frac{4}{5}\ , \ 
 \chi=-\frac{ 3}{5}$ & Exact (3D) 
 \\
\hline
\ref{ssec:incompr_phase2d} &
Incomp.~PAF in 2D\cite{chen_natcomm16,chen_pre24}  &$v_y^3$   & $\frac{5}{2}$ &  $z=1+\frac{\epsilon}{11}  \ , \
\zeta =\frac{2}{3}\ , \ 
 \chi=-\frac{1}{3}$
& 1-loop 
 \\
 \hline
\ref{ssec:incompr_phase2d_quenched} &
 \makecell{Incomp.~PAF in 2D\\ with quenched disorder\cite{chen_prl22,chen_pre22}} &$v_y^3$   &  $\frac{7}{3}$ & $z=\frac{2}{3}-\frac{5\epsilon}{9} \ , \
\zeta =\frac{2+\epsilon}{3}\ , \ 
 \chi=-\frac{1-\epsilon}{3}$ & 1-loop
 \\
\hline
\ref{ssec:active_suspension} &
\makecell{Active suspension with\\ an easy axis\cite{dadhichi_prl26}} &
$
\bu \cdot \vnab c $   & 4 & $z=\frac{3}{2}  \ , \ \zeta = 1 
  \ , \ \chi=-\frac{3}{2}$  & Exact (3D) 
 \\
\hline
\ref{ssec:phase_chemo_growth} &
\makecell{Conserved chemotactic \\systems with \\
quasi-static chemotactic field\cite{mahdisoltani_prr21}} &
$\nabla \cdot \left[
\delta \rho \nabla c\right]
\ , \
\nabla^2(\nabla c)^2 $
& 4 & $z=\frac{d+2}{3} \ , \
 \chi=-\frac{d+2}{3}$ & Exact 
 \\
\hline
\end{tabular}
\caption{
{\it New universality classes (UC) governing stable nonequilibrium phases uncovered since 2015.}
The columns list: (i) the section of the review where the UC is discussed; (ii) the  hydrodynamic model in which the UC arises, with PAF referring to polar active fluids; (iii) the leading relevant nonlinear term(s) controlling the RG fixed point; (iv) the upper critical dimension $d_c$; (v) the scaling exponents characterising the asymptotic scale-invariant behaviour, where $z$ is the dynamic exponent, $\zeta$ the anisotropy exponent (when applicable), and $\chi$ the roughness (or field scaling) exponent, using the convention of Eq.~(\ref{eq:correlationform}); and (vi) the analytical method used to obtain the exponents, distinguishing perturbative one-loop calculations from exact (see {\bf Section~\ref{sec:exactexps}}).  Throughout the table, $\epsilon=d_c-d$. 
For entries \ref{ssec:incompr_phase2d} (Incomp.~PAF in 2D) \&  \ref{ssec:incompr_phase2d_quenched} (Incomp.~PAF in 2D with quenched disorder), several RG schemes with different upper critical dimensions were employed. The exponents quoted here are taken from the schemes whose $d_c$ 
is closest to $d=2$, except for $\zeta$ and $\chi$ in entry \ref{ssec:incompr_phase2d}, which are known exactly through a mapping onto the (1+1)-dimensional KPZ UC\cite{chen_natcomm16}.
}
\label{tab1}
\end{table}

\subsection{New UCs in interfacial dynamics}

\subsubsection{Tensionless interface growth (2023)}
\label{ssec:inviscid_KPZ}
Already within the well-established KPZ equation discussed in the previous section [Eq.~\eqref{eq:KPZ}], a new UC has surprisingly been uncovered in recent years. Motivated by numerical simulations\cite{cartes_prsa22,rodriguez_fernandez_pre22} reporting an unexpected scaling regime with dynamic exponent $z=1$ in the fine-tuned tensionless limit $\mu=0$, Fontaine, Vercesi, Brachet and Canet\cite{fontaine_prl23} demonstrated that, in $d=1$, this critical point is governed by a previously unknown RG fixed point accessible only through nonperturbative RG methods. This fixed point is controlled by the same KPZ nonlinearity,
\begin{equation}
|\nabla h|^2 \ ,
\end{equation}
as the conventional KPZ fixed point, but now in an intrinsically strong-coupling regime. Shortly thereafter, Gosteva, Tarpin, Wschebor and Canet\cite{gosteva_pre24} showed that this fixed point persists in all spatial dimensions. However, apart from $d=1$, where an accidental time-reversal symmetry fixes the roughness exponent $\chi$ exactly, determining the remaining scaling exponents requires more sophisticated nonperturbative approaches that have yet to be fully developed for this problem.

\subsubsection{Interface with bulk-mediated long-range interactions (2023)}
\label{ssec:q-kpz}

Consider a system of self-propelled particles interacting purely through steric repulsion. At the hydrodynamic level, the only conserved field is the mass density $\phi$, and following {\bf Recipe 1} one arrives at a universal coarse-grained description of such systems. A particular realisation of this class is commonly referred to as Active Model~B$^+$~\cite{wittkowski_natcomm14,tjhung_prx18}. Despite the absence of attractive interactions, such systems can nevertheless undergo phase separation: persistence in directed motion induces an effective tendency for particles to cluster. This phenomenon is now known as \emph{motility-induced phase separation} (MIPS)~\cite{fily_prl12,cates_annurev15}.

In the phase-separated regime, one may focus on the dynamics of the interface separating the dense and dilute phases by introducing a height field $h(\mathbf{x},t)$, defined by the locus where the density lies midway between the two bulk values. In contrast to standard surface growth models, a key distinction here is that particle number is conserved. Under the non-moving interface condition,  interfacial fluctuations induce particle transport within the surrounding bulk phases, which in turn feeds back onto the interface dynamics. This coupling generates long-ranged, nonlocal dynamics of the interfacial profile.

An analytically tractable limit arises when bulk dynamics is fast compared to interfacial motion, such that the bulk fields may be treated as quasi-static. In this limit, number conservation implies that the bulk field obeys Laplace's equation away from the interface. To make this concrete, consider a nearly flat interface embedded in a $d$-dimensional bulk, parametrised as $y = h(\mathbf{x},t)$ with $\mathbf{x} \in \mathbb{R}^{d-1}$. Let $\psi(\mathbf{x},y)$ denote the conserved bulk density field. In the quasi-static regime, $\nabla^2 \psi = 0$.
Fourier transforming along the interface directions gives
$
(\partial_y^2 - q^2)\psi_{\mathbf{q}}(y) = 0$, whose solution that decays away from the interface is
$
\psi_{\mathbf{q}}(y) = \psi_{\mathbf{q}}(0)\, e^{-|\bq||y|} \, .
$
Now, the diffusive flux in or out of the interface corresponds to the normal derivative at the interface:
$
\partial_n \psi_{\mathbf{q}} \big|_{y=0} \sim -|\bq|\, \psi_{\mathbf{q}}(0)$.

The remaining ingredient is to determine $\psi_{\mathbf q}(0)$, namely the value of the bulk field at the interface. In equilibrium phase separation~\cite{weber_rpp19}, curvature modifies the interfacial boundary condition through the Gibbs--Thomson relation, implying that for weak interface deformations,
$
\psi_{\mathbf q}(0)
\propto
-q^2 h_{\mathbf q}$.
For MIPS with active Brownian particles, an analogous curvature-dependent interfacial condition has been argued to emerge effectively~\cite{lee_soft17}.

Combining this interfacial condition with the quasi-static bulk solution yields the linear interfacial dynamics
\begin{equation}
\partial_t h_{\mathbf q}
\sim
\partial_n \psi_{\mathbf q}\big|_{y=0}
\sim
-|\bq|q^2 h_{\mathbf q}
=
-|\bq|^3 h_{\mathbf q} \ .
\end{equation}

Going beyond the linear theory, Besse, Fausti, Cates, Delamotte, and Nardini~\cite{besse_prl23} showed that symmetry allows a KPZ-type nonlinearity, leading to the effective EOM
\begin{equation}
\partial_t h(\mathbf q,t)
=
-\nu |\mathbf q|^3 h(\mathbf q,t)
-
\frac{\lambda}{2}
|\mathbf q|
\int
\frac{d^{d-1}k}{(2\pi)^{d-1}}
\,
\mathbf k\cdot(\mathbf q-\mathbf k)\,
h(\mathbf k,t)\,
h(\mathbf q-\mathbf k,t)
+
f(\mathbf q,t) \ ,
\end{equation}
with noise correlations
\begin{equation}
\langle f(\mathbf q,t)f(\mathbf q',t') \rangle
=
2D|\mathbf q|
(2\pi)^{d-1}
\delta^{d-1}(\mathbf q+\mathbf q')
\delta(t-t') \ ,
\end{equation}
where the nonlocal $|\bq|$ dependence reflects the projection of conserved bulk fluctuations onto the interface.

Using perturbative RG, the authors showed that the upper critical interfacial dimension is $d_c=2$, corresponding to a bulk dimension of 3. Below this dimension, the leading relevant nonlinearity takes the schematic form
\begin{equation}
|\nabla|(\nabla h)^2 \ ,
\end{equation}
whose associated nontrivial RG fixed point defines a new UC, referred to as the $|q|$-KPZ universality class. The corresponding scaling exponents were obtained perturbatively to one-loop order.

\subsection{Compressible polar active fluids}
\label{sec:rev_tt}

We now turn to the largest family of physical systems in which new UCs   have been uncovered over the past decade: ``dry'' polar active fluids {\bf (PAF)}. These are intrinsically nonequilibrium systems with long-range polar order in which momentum is not conserved due to frictional interactions with an underlying substrate --- hence the term ``dry'', in contrast to ``wet'' active systems where motile agents instead swim within momentum-conserving fluids.

Of particular importance is the Toner--Tu model\cite{toner_prl95,toner_pre98}, originally introduced to describe the collective motion of bird flocks. To streamline the discussion that follows, however, we consider instead a different physical realisation: a collection of motile agents, such as cells, bacteria, or synthetic robots, crawling within a rigid polymeric matrix. This formulation naturally accommodates both two- and three-dimensional systems. Unless stated otherwise, we assume that the agents are much larger than the microscopic mesh size of the substrate, such that the surrounding medium may be treated as spatially homogeneous. In this case, the dominant fluctuations arise from the stochastic motion of the agents themselves, corresponding to a time-dependent annealed noise.

Despite the stark microscopic differences from flying flocks, the symmetry principles outlined in {\bf Section~\ref{sec:universal_models}} imply that the same universal hydrodynamic framework --- namely the Toner--Tu theory --- applies.

In this model, particle number is conserved, such that the continuity equation for the density $\rho$ forms the first hydrodynamic equation,
\begin{equation}
  \label{eq:tt1}
  \partial_t \rho + \nabla\cdot(\rho \bv) = 0 \ .
\end{equation}
Following {\bf Recipe 1}, and retaining the leading symmetry-allowed terms consistent with (time-) translational, rotational, and chiral symmetry, the equation of motion for the velocity field takes the form\footnote{Technically, several nonlinear terms of second order in spatial gradients have been omitted here. However, these terms neither contribute to the linear correlations of the soft mode nor constitute relevant nonlinearities, which is why they are typically neglected in the literature.
}
\begin{equation}
  \label{eq:tt2}
  \partial_t \bv + \lambda_1 (\bv\cdot \nabla)\bv+\lambda_2(\nabla\cdot\bv)\bv+\lambda_3 \nabla(|\bv|^2)
  =
  U \bv - \nabla P_1  -\bv(\bv\cdot\nabla P_2)
  + \mu_1  \nabla(\nabla\cdot \bv)
  + \mu_T\nabla^2\bv
  +\mu_2(\bv\cdot\nabla)^2 \bv
  + \bff \ ,
\end{equation}
with noise statistics
\begin{equation}
 \langle \bff(\br,t) \rangle = {\bf 0}, \qquad
 \langle f_i(\br,t) f_j(\br',t') \rangle
 =
 2D\,\delta_{ij}\,\delta^d(\br-\br') \delta(t-t') ,
\end{equation}
where all parameters, including $\lambda_i$, $U$, $P_i$, $\mu_i$, and $\mu_T$, may generically depend on both $|\bv|$ and $\rho$.

Because particle number is conserved, the mean-field theory admits arbitrary uniform densities $\rho_0$. Depending on the functional form of $U$, the velocity dynamics typically possesses two homogeneous steady states: a disordered state with $\bv=0$, and an ordered state with $\bv=\bv_0$. While the magnitude of $\bv_0$ is fixed by the condition $U(\rho_0,|\bv_0|)=0$, its direction is selected spontaneously. The spontaneous breaking of continuous rotational symmetry generically implies that fluctuations perpendicular to the velocity field -- the Goldstone modes -- become  hydrodynamically soft and therefore constitute a hydrodynamic mode.

Despite having played a foundational role in the emergence of active matter physics as a field, the UC governing the ordered phase of the Toner--Tu model remains controversial. Indeed, while Toner and Tu initially believed that the problem had already been solved in 1995\cite{toner_prl95}, Toner later realised that the original analysis was incomplete and revisited the problem in 2012\cite{toner_pre12}.

The key subtlety overlooked in the 1995 treatment is that fluctuations in the velocity magnitude $|\bv|$, although massive and therefore formally non-hydrodynamic, are linearly coupled to density and gradient fluctuations. As a result, integrating out the massive mode generates additional nonlinearities that are just as relevant under the RG as those retained in the original analysis. A more careful reanalysis in 2012\cite{toner_pre12} showed that the correct hydrodynamic equations governing the ordered phase are
\begin{align}
  \nonumber
  \label{eq:tt_ordered_1}
  \partial_t \bv_\bot + \gamma \partial_\parallel \bv_\bot + \lambda_1^0(\bv_\bot\cdot\nabla_\bot)\bv_\bot
  ={}&
  -g_1\delta\rho\,\partial_\parallel \bv_\bot
  -g_2\bv_\bot\partial_\parallel \delta \rho
  - \kappa \nabla_\bot \delta \rho
  - g_3 \nabla_\bot(\delta \rho^2)
  \\
  &
  +\mu_B\nabla_\bot(\nabla_\bot\cdot \bv_\bot)
  +\mu_T\nabla_\bot^2\bv_\bot
  +\mu_\parallel\partial_\parallel^2\bv_\bot
  + \nu_\parallel \partial_\parallel \nabla_\bot \delta\rho
  + \bff_\bot ,
  \\
  \nonumber
  \label{eq:tt_ordered_2}
  \partial_t \delta \rho
  + \rho_0\nabla_\bot \cdot \bv_\bot
  + w_1 \nabla_\bot\cdot (\bv_\bot\delta\rho)
  + v_2\partial_\parallel \delta\rho
  ={}&
  \mu_{\rho\parallel}\partial_\parallel^2\delta\rho
  +\mu_{\rho\bot}\nabla_\bot^2\delta \rho
  + \mu_{\rho v}\partial_\parallel (\nabla_\bot\cdot \bv_\bot)
  \\
  &
  +w_2 \partial_\parallel (\delta\rho)^2
  + w_3 \partial_\parallel(|\bv_\bot|^2) \ ,
\end{align}
where $\delta \rho = \rho-\rho_0$ and $
\bv_\bot
=
\bv - \bv_0(\bv_0\cdot\bv)/|\bv_0|^2$
denotes fluctuations  perpendicular to the ordered direction $\bv_0$. The symbols $\parallel$ and $\bot$ indicate derivatives parallel and perpendicular to $\bv_0$, respectively, reflecting the anisotropy generated spontaneously by collective motion.

Compared to the original analysis\cite{toner_prl95}, six additional nonlinearities are now known to be relevant. As a result, no controlled perturbative RG solution of the generic ordered Toner--Tu phase is currently available. Nevertheless, recent work has shown that imposing additional constraints can render the problem tractable, leading to the discovery of a new UC within a restricted limit of the theory\cite{jentsch_prl24}. The generic case, however, remains unresolved and continues to be the subject of active research\cite{chate_prl24,chen_a25}.

\subsubsection{Flocking phase with quenched disorder (2018)}
\label{ssec:tt_quenched_phase}

We now consider a situation in which the local orientation of a polymeric matrix biases the motion of the agents. In particular, this is expected when the agent size becomes comparable to the characteristic filament scale of the substrate. Since the polymer matrix is assumed to remain static, these random effects are quenched in time. To model this, the noise statistics in the Toner--Tu (TT) model are modified to
\begin{equation}
 \langle \bff(\br) \rangle = {\bf 0}, \qquad
 \langle f_i(\br) f_j(\br') \rangle
 =
 2D\,\delta_{ij}\,\delta^d(\br-\br') ,
\end{equation}
where the angular brackets now denote an average over disorder realizations. The absence of any time dependence reflects the fact that the disorder is spatially random but frozen in time, a form of disorder known as {\it quenched disorder}. Since quenched disorder is typically more relevant than conventional annealed noise at long wavelengths, the latter may be neglected asymptotically. Expanding the ordered-phase TT equations [Eqns.~\eqref{eq:tt_ordered_1} and \eqref{eq:tt_ordered_2}] with this modified noise prescription completes the hydrodynamic model.

Performing an RG analysis of this system, Toner, Guttenberg, and Tu~\cite{toner_prl18,toner_pre18} showed that, in the regime where density fluctuations and  perpendicular velocity fluctuations propagate with opposite characteristic velocities along the mean flocking direction, a remarkable simplification occurs: the density field, the velocity component parallel to the direction of collective motion, and the component of the  velocity field  orthogonal to the direction of collective motion, but parallel to the wavevector $\bq$, i.e., the longitudinal component of $v_\bot$, all become asymptotically irrelevant under RG.

As a result, the long-wavelength behaviour is governed solely by the transverse velocity field
\begin{equation}
\bv_T
=
\bv_\bot
-
\frac{\bq(\bq\cdot\bv_\bot)}{q^2} \ .
\end{equation}
Remarkably, incompressibility is not imposed microscopically but instead emerges dynamically at long wavelengths through the RG irrelevance of compressible fluctuations. The resulting effective hydrodynamic theory is therefore asymptotically incompressible, with $\nabla\cdot\bv_T=0$ by construction.

With these simplifications, the only remaining symmetry-allowed relevant nonlinearity is
\begin{equation}
(\bv_T \cdot \nabla)\,\bv_T \ .
\end{equation}
An RG analysis of the resulting incompressible hydrodynamic theory reveals a new UC governing the scaling behaviour of flocking systems with quenched disorder for $2<d<5$.

\subsubsection{Critical point within the ordered phase (2024)}
\label{ssec:tt_crit_in_order}

Coming back to the standard situation with annealed noise,
and remaining within the ordered phase, Miller and Toner recently showed that a critical transition can be induced by fine-tuning two model parameters\cite{miller_pre24}, namely $\kappa$ and $g_3$ in Eq.~\eqref{eq:tt_ordered_1}. Since the system is already ordered, this transition is not of the order–disorder type. Instead, it is analogous to a critical phase separation transition in thermal systems, with the important distinction that the two coexisting phases are both moving, with distinct speeds, and propagate parallel to one another. 

 Performing a perturbative RG analysis using an $\epsilon$-expansion, the authors found that the most relevant fluctuations are given by the density field. 
The transverse and longitudinal velocity fluctuations cannot, however, be neglected, as they mediate long-range interactions in the density dynamics, profoundly altering the scaling behaviour already at the mean-field level, in contrast to, for example, critical phase separation  in thermal equilibrium.
 In particular,  the upper critical dimension $d_c$ is found to be $5$. Close to this dimension, all of the nonlinear terms in Eqns.~\eqref{eq:tt_ordered_1} and \eqref{eq:tt_ordered_2} are irrelevant and instead the   single relevant nonlinearity,
\begin{equation}
\nabla_\perp \, \delta\rho^{\,3} \, ,
\end{equation}
 governs the nonlinear scaling behaviour.

Crucially, the analysis reveals a new UC controlling this critical transition, and the associated scaling exponents were computed to one-loop order.

\subsubsection{Critical order--disorder transition with an easy axis (2025)}
\label{ssect:crit_aim}

We now consider a situation in which  active agents preferentially move along a predetermined direction.  Physically, this may be realised by stretching an initially isotropic polymeric matrix along a single direction, thereby biasing motion along that direction, which we refer to as the easy axis. Because motion along the remaining directions is suppressed, the only hydrodynamic fields are the density fluctuation $\delta\rho$ and the velocity component along the easy axis, $v_x$.  Instead of velocity, equivalently momentum density $g_x=\rho v_x $ was chosen as a hydrodynamic variable\cite{wong_a25}. This minimal description is commonly referred to as the active Ising model \cite{solon_prl13,solon_pre15},  and can be described by the following hydrodynamic EOM,
\begin{align}
  \partial_t \rho  &= -\gamma \partial_x g_x + K \nabla_\bot^2\rho \ , \\
  \partial_t g_{x} &=-\lambda g_x \pp_x g_x  -\kappa_1\pp_x \rho - \kappa_2 \pp_x \rho^2 + \mu_x \pp_x^2 g_x + \mu_\bot \nabla_\bot^2 g_x + \left(-\alpha_0-\alpha_1\rho-\alpha_2 \rho^2-\beta g_x^2\right)g_x + f_x \ .
\end{align}

A critical transition from order to disorder can be realized in this model by fine-tuning two parameters, corresponding to setting the bare coefficients $\alpha_0$ and $\alpha_1$ to zero\cite{nesbitt_njp21}. Using a perturbative RG analysis based on an $\epsilon$-expansion, Wong and Lee\cite{wong_a25} uncovered a new UC governing this critical point. The upper critical dimension was found to be $d_c=4$. Although four nonlinearities are relevant below $d_c$ in the EOM for $g_x$ --- namely $g_x^3$, $\delta\rho^2 g_x$, $\partial_x \delta\rho^2$, and $g_x \partial_x g_x$ --- two of these can be shown to be marginally irrelevant. As a result, the generic RG fixed point depends only on the two nonlinearities 
\begin{equation}
g_x^3  \qquad \text{ and} \qquad \delta\rho^2 g_x \, .
\end{equation}
Values of the critical exponents were computed to the 1-loop level. The authors also showed that by further fine-tuning, two additional, distinct multicritical points emerge in this system.

\subsection{Malthusian PAFs}
We now consider active systems in which agents can reproduce and die. This so-called \emph{Malthusian} case was first introduced by Toner in 2012~\cite{toner_prl12}. Owing to birth and death processes, the density field is no longer conserved and therefore ceases to be a hydrodynamic soft mode. Consequently, density fluctuations can be integrated out and neglected at long wavelengths.

Eqns.~\eqref{eq:tt1} and \eqref{eq:tt2} then reduce to a single EOM for the velocity field:
\begin{equation}
  \label{eq:mf1}
  \partial_t \bv
  + \lambda_1 (\bv\cdot \nabla)\bv
  + \lambda_2(\nabla\cdot\bv)\bv
  + \lambda_3 \nabla(|\bv|^2)
  =
  U \bv
  + \mu_1  \nabla(\nabla\cdot \bv)
  + \mu_T\nabla^2\bv
  + \mu_2(\bv\cdot\nabla)^2 \bv
  + \bff \ ,
\end{equation}
where the phenomenological coefficients now depend only on $|\bv|$.

Expanding around the ordered state and integrating out the massive longitudinal fluctuation $\delta v_\parallel$, one obtains an effective EOM for the  Goldstone mode:
\begin{equation}
  \label{eq:mf2}
  \partial_t \bv_\bot
  + \gamma \partial_\parallel \bv_\bot
  + \lambda_1(\bv_\bot \cdot \nabla_\bot) \bv_\bot
  =
  \mu_T \nabla_\bot^2\bv_\bot
  + \mu_B\nabla_\bot (\nabla_\bot \cdot \bv_\bot)
  + \mu_\parallel \partial_\parallel^2\bv_\bot
  + \bff_\bot \ .
\end{equation}

\subsubsection{Flocking phase (2020)}
\label{ssec:malth_phase}

In the ordered phase, Chen, Lee, and Toner carried out a perturbative RG analysis using an $\epsilon$-expansion about the upper critical dimension $d_c=4$~\cite{chen_prl20,chen_pre20}. They showed that, near $d=4$, the only relevant nonlinearity in the effective long-wavelength theory is
\begin{equation}
(\bv_\perp \cdot \nabla)\,\bv_\perp \, .
\end{equation}
This led to the identification of a new nonequilibrium UC governing the ordered phase of Malthusian polar active flocks, with scaling exponents computed to one-loop order.

We note that, in a closely related earlier study, Toner argued that in $d=2$, where $\bv_\perp$ reduces to a scalar field, the scaling exponents could be determined exactly~\cite{toner_prl12}. More recently, however, it was recognised that rotational symmetry allows an additional relevant nonlinearity in two dimensions:
\begin{equation}
  \nabla v_\perp^3 \ .
\end{equation}
The presence of this term calls into question the exactness of the original scaling predictions, and whether the scaling exponents remain exactly calculable is currently the subject of ongoing debate~\cite{chate_prl24,chen_a25}.

\subsubsection{Flocking phase in 3D with an easy plane (2024)}
\label{ssec:malth_ezpl}

In 2024, Toner considered Malthusian flocks whose motion is preferentially confined to an \emph{easy plane}~\cite{toner_pre24}. Physically, such anisotropy may arise, for example, by compressing an initially isotropic polymeric matrix along one direction, thereby favouring motion within the remaining uncompressed dimensions. Choosing the easy plane to coincide with the $xy$-plane, and the mean flocking direction to lie along $\hat{\bx}$, the only remaining hydrodynamic degree of freedom is the perpendicular  velocity fluctuation $v_y$. The other mode components become hard (i.e., not soft): fluctuations in $v_z$ are suppressed because they point out of the easy plane, while fluctuations in $v_x$ correspond to  parallel  distortions of the ordered state.

Using a perturbative RG analysis, Toner showed that the only relevant nonlinearity in the effective long-wavelength equation of motion is
\begin{equation}
\partial_y v_y^{\,2} \, .
\end{equation}
This leads to a new nonequilibrium UC governing the ordered phase, for which the scaling exponents can be obtained exactly, in the sense that is described in {\bf Section~\ref{sec:exactexps}}.

\begin{table}
 \renewcommand{\arraystretch}{2}
\begin{tabular}{c|c|c|c|c|c}
Sect. & System & Relevant terms & $d_c$ & Exponents & Method 
\\
\hline
\hline
\ref{ssec:inviscid_KPZ} &
Tensionless KPZ\cite{fontaine_prl23,gosteva_pre24} &$|\nabla h|^2 $  & $\infty$ & $z=1\ ,\ \chi=\frac{1}{2} \  ({\rm for}\ d=1)$ & NP-FRG
 \\
\hline
\ref{ssec:tt_crit_in_order} &
PAF in the ordered phase\cite{miller_pre24} &$\vnab_\perp \delta \rho^3 $  & 5 & 
\makecell{$z=2 \ , \ \zeta=2$\ ,\\  $\chi=-2+\frac{\epsilon}{2} \ , \ \chi_\rho=-1 +\frac{\epsilon}{2} $}
& 1-loop
 \\
\hline
\ref{ssect:crit_aim} &
PAF with an easy axis\cite{wong_a25} &$ g_x^3 \ , \ \delta\rho^2 g_x$  & 4 & $z=2 \ , \
\chi=-1+\frac{\epsilon}{2}$ & 1-loop
 \\
\hline
\ref{ssec:incompr_crit} &
Incomp.~PAF\cite{chen_njp15} &$
(\bv \cdot \nabla)\,\bv \ ,\ |\bv|^2 \,\bv$  & 4 & $z=2-\frac{31 \epsilon}{113} \ , \
 \chi=-1+\frac{41\epsilon}{113}$ & 1-loop 
 \\
\hline
\ref{ssec:incompr_crit_quenched} &
\makecell{Incomp.~PAF with\\ quenched disorder\cite{zinati_pre22}} &$
(\bv \cdot \nabla)\,\bv$  & 6 & $z=2-\frac{8 \epsilon}{31} \ , \
 \chi=-1+\frac{15\epsilon}{62}$ & 1-loop 
 \\
\hline
\ref{ssec:incompr_crit_spin} &
\makecell{PAF with an additional\\  rank-2 spin field ${\bf s}$\cite{cavagna_natphys23}} &
9 terms, see Eqs.~(\ref{eq:spinfield_non1}, 
\ref{eq:spinfield_non2}) & 4 & \makecell{$z=2-0.65(2)\epsilon$, \\
 $\chi$ not calculated} & 1-loop 
 \\
\hline
\ref{ssec:chemo_growth} &
\makecell{Chemotactic bacterial growth \\ with quasi-static chemotactic field\cite{mahdisoltani_prr21}} &
$\nabla \cdot \left[
\delta \rho \nabla c\right]
\sep
\nabla^2(\nabla c)^2 $
& 6 & $z=\frac{d}{3} \ , \
 \chi=-\frac{d}{3}$ & Exact 
 \\
\hline
\ref{ssec:chemo_percolation} &
\makecell{Extinction of bacterial growth\\ with quasi-static chemotactic field\cite{vanderkolk_prl23} } &$\nabla \cdot (\rho \nabla c)
\ , \
\rho \nabla^2 c
\ , \
\rho^2$& 4 & 
\makecell{CA: $z=2+\frac{ \epsilon}{ 18}$\ ,  \\
CR: $z=2-\frac{ \epsilon}{ 2}$ } & 1-loop
 \\
\hline
\ref{ssec:driven_ON} &
Driven $O(N)$ with inertia\cite{daviet_prl24} &$|\vec{\phi}|^2 \vec{\phi}
\ , \
|\vec{\phi}|^2\,\partial_t \vec{\phi}
\ , \
\partial_t|\vec{\phi}|^2 \vec{\phi}$& 4 & 
\makecell{
$z=2+0.0072 \epsilon^2  ,$ \\
 $\chi=-1+\frac{\epsilon}{2}+0.1765 \epsilon^2$,\\
 (for $N=2)$
 } & 2-loop 
 \\
\hline
\end{tabular}
\caption{
{\it
New universality classes governing nonequilibrium critical phenomena uncovered since 2015.} The columns are defined as in Table~1. In the Method column, NP-FRG denotes the nonperturbative functional renormalization group approach, as, e.g., discussed in {\bf Section~\ref{sec:npfrg}}.
}
\label{tab2}
\end{table}

\subsection{Incompressible PAFs}

We now consider systems in which the number of active agents is conserved (i.e., there is no birth or death), but the density is sufficiently high --- or the agents avoid one another strongly enough --- that the system behaves effectively as an incompressible fluid, with the exact identity: $\nabla \cdot \bv = 0$.
We refer to this regime as \emph{incompressible} PAFs. As in incompressible Navier--Stokes fluids, enforcing this constraint introduces a nonlocal projection operator, rendering the EOM for the velocity field intrinsically long-ranged.

In this limit, both density fluctuations and longitudinal velocity modes cease to be hydrodynamic variables. Consequently, the same hydrodynamic equations as in the Malthusian case, Eqns.~\eqref{eq:mf1} and \eqref{eq:mf2}, continue to apply, supplemented by a Lagrange multiplier term analogous to that in Eq.~\eqref{EOM} to enforce the incompressibility constraint.

\subsubsection{Flocking phase in $2<d<4$ (2018)}
\label{ssec:incompr_phase}

Focusing first on the ordered phase, Chen, Lee, and Toner~\cite{chen_njp18} performed a perturbative RG analysis and showed that the only relevant nonlinearity is
\begin{equation}
(\bv_T \cdot \nabla)\,\bv_T \, .
\end{equation}
Remarkably, their analysis demonstrated that the original Toner--Tu (1995) UC~\cite{toner_prl95}, together with its associated {\it exact} scaling exponents governs the large-scale behaviour of incompressible flocks for $2<d<4$. 

We note that although the relevant nonlinearity has the same form as that appearing in flocks with quenched disorder and in Malthusian flocks, the resulting UC are distinct. In the former case, this is because the two systems are governed by different noise statistics, while in the latter, the exact scaling identities enforced by incompressibility are absent in Malthusian flocks. 

\subsubsection{Flocking phase in 2D (2024)}
\label{ssec:incompr_phase2d}

In two dimensions, the decomposition of the velocity field into longitudinal and transverse components used in higher dimensions is no longer available. Instead, the two velocity components are constrained by incompressibility: $\partial_x v_x + \partial_y v_y = 0$.
As a result, the hydrodynamic description reduces to an effectively long-ranged theory with a single scalar field.

Chen, Lee, and Toner~\cite{chen_natcomm16} performed an RG analysis and showed that the dominant nonlinearity in this reduced description originates from the cubic term in the Toner--Tu equations,
\begin{equation}
 v_y^3  \ .
\end{equation}
Introducing a stream function for the incompressible velocity field, they mapped the equal-time fluctuations of the flock onto a $(1+1)$-dimensional surface growth problem in the KPZ UC, with one spatial coordinate of the flock effectively playing the role of time in the KPZ model.

However, this construction only captures equal-time correlations and therefore discards the genuine dynamics of the flock. This limitation was recently overcome by Chen, Lee, Maitra, and Toner~\cite{chen_pre24}, who carried out a perturbative RG analysis directly on the hydrodynamic theory and computed the dynamic exponent. Since the dynamical scaling behaviour had not previously been determined for this system, their work identifies a new \emph{dynamic} UC.

\subsubsection{Flocking phase in 2D with quenched disorder (2022)}
\label{ssec:incompr_phase2d_quenched}

As in the compressible case, one may also consider incompressible flocks subject to quenched disorder, motivated, for example, by static defects in an underlying polymeric matrix. In three dimensions, Chen, Lee, Maitra, and Toner\cite{chen_prl22b} showed that the resulting ordered state belongs to the same UC as compressible flocks with quenched disorder.

In two dimensions, however, the situation becomes qualitatively different. As discussed above, incompressibility reduces the hydrodynamic description to an effectively long-ranged theory with a single scalar degree of freedom. Using a perturbative RG analysis, Chen, Lee, Maitra, and Toner~\cite{chen_prl22,chen_pre22} showed that --- much as in the case with annealed noise --- the cubic nonlinearity in the original Toner--Tu equations,
\begin{equation}
 v_y^3 \ ,
\end{equation}
is sufficient to generate a distinct UC governing two-dimensional incompressible flocks with quenched disorder. 

\subsubsection{Critical order--disorder transition (2015)}
\label{ssec:incompr_crit}

In addition to novel ordered phases, incompressible flocks also exhibit new critical behaviour. For incompressible PAFs subject to annealed noise, Chen, Toner, and Lee employed a perturbative RG analysis and showed that the order--disorder transition is governed, below the upper critical dimension $d_c=4$, by a new UC~\cite{chen_njp15}. The corresponding RG fixed point is controlled by two nonlinearities in the hydrodynamic equation of motion for the velocity field $\bv$:
\begin{equation}
(\bv \cdot \nabla)\,\bv , \qquad |\bv|^2 \,\bv \ .
\end{equation}
The associated critical exponents were computed to one-loop order.

\subsubsection{Critical order--disorder transition with quenched disorder (2022)}
\label{ssec:incompr_crit_quenched}

Reintroducing quenched disorder and focusing on the order--disorder transition, Ben Ali Zinati, Besse, Tarjus, and Tissier employed a perturbative RG analysis and showed that, below the upper critical dimension $d_c=6$, the transition is governed by a new UC~\cite{zinati_pre22}. The corresponding RG fixed point is controlled by a single relevant nonlinearity:
\begin{equation}
(\bv \cdot \nabla)\,\bv \ .
\end{equation}
The associated critical exponents were again computed to one-loop order.

\subsubsection{Critical transition with an additional spin field (2023)}
\label{ssec:incompr_crit_spin}

An experimental system studied in this context is the swarming behaviour of midges. Motivated by these experiments, Cavagna \emph{et al.} proposed that, for such systems, the Toner--Tu hydrodynamic description should be supplemented by an additional antisymmetric rank-2 spin field ${\bf s}$, which acts as the generator of rotations of the velocity field~\cite{cavagna_natphys23}. In the non-dissipative limit, this spin field is treated as conserved, thereby introducing an inertial component into the dynamics.

The construction of the EOM differs substantially from the universal symmetry-based hydrodynamic framework adopted in this Review. Rather than systematically enumerating all symmetry-allowed couplings, the authors employ a classical-mechanics-inspired Poisson-bracket formalism to determine the reversible couplings between the spin and velocity fields. The role of this Poisson structure is to impose a Hamiltonian structure that guarantees non-dissipative dynamics, i.e., dynamics in which entropy is not produced --- a criterion that is not central to our present focus on generic nonequilibrium systems.

Focusing on the incompressibility regime where the density field can be ignored, 
the authors performed a one-loop perturbative RG calculation to identify a new dynamic UC with upper critical dimension $d_c=4$, and computed  dynamic exponent $z$ to the same order.  The relevant nonlinearities retained in the theory are
\beq
\label{eq:spinfield_non1}
v_j \partial_j v_i \ , \ (v_j v_j)v_i \ , \ v_j s_{ij} \ ,
\eeq
for the EOM of $\bv$, and 
\beq
\label{eq:spinfield_non2}
v_k\partial_k s_{ij} \
, \
 \pp_k(s_{ik}v_j)
\ , \
 \partial_i(v_k s_{kj})
\ , \
v_i\pp_k \pp_k v_j
\ , \
\partial_i
(v_k\partial_k v_j)
\ , \
\partial_k
\!\left(
v_k
\partial_i v_j
\right)
\ , \ \ \ \text{ (antisymmetrised with respective to $i$ and $j$)}
\eeq
for the EOM of the rank-2 spin field ${\bf s}$.

\subsection{New UC in active suspensions with an easy axis (2025)}
\label{ssec:active_suspension}

We now consider the case in which the motile agents move not in a polymeric matrix (a ``dry'' system), but in an incompressible liquid (a ``wet'' system), {\it i.e.}, an \emph{active suspension}. From the perspective of symmetry, conservation laws and constraints (SCC), this change has two important consequences. First, momentum is now locally conserved: unlike a polymeric matrix, which acts as a momentum sink through friction, a fluid transports momentum over long distances. As a result, the velocity of the active particles is replaced by a combined velocity field $\mathbf u$ describing both the active particles and the suspending fluid. This hydrodynamic velocity field is therefore conserved and incompressible. The minimal description thus consists of a conserved concentration field $c$ coupled to an incompressible velocity field $\mathbf u$.

Momentum-conserving active systems are generically unstable at long wavelengths because active stresses tend to destabilise homogeneous states\cite{aditisimha_prl02}. However, this instability can be suppressed if the swimmers are aligned by an externally imposed easy axis, {\it e.g.}, by immersing them in the nematic phase of a liquid crystal that fixes their mean orientation. Dadhichi, Sahoo, Kumar and Ramaswamy~\cite{dadhichi_prl26} showed that the resulting homogeneous disordered phase is governed by a new nonequilibrium UC.

Specifically, in the stable regime and below the upper critical dimension $d_c = 4$, the only relevant nonlinearity is the advective coupling
\begin{equation}
\mathbf u \cdot \nabla c,
\end{equation}
where $\mathbf u$ is the incompressible fluid velocity and $c$ is the swimmer concentration. Owing to Galilean invariance, the coefficient of this term is not renormalized, which imposes strong constraints on the RG flow and scaling structure. 
A self-consistent one-loop analysis then yields exact scaling exponents for both the dynamics and the roughness of the concentration field.

\subsection{New UCs in chemotactic systems}

Beyond polar active fluids and interfacial dynamics, a third class of systems in which new UCs have been uncovered over the past decade consists of spatiotemporal population dynamics coupled to rapidly diffusing signalling fields. The canonical example is a colony of motile bacteria whose motion is biased by a chemical signal that they themselves produce and which diffuses much faster than the bacteria. In the absence of cell proliferation, death, and noise, such a system is described by the classic Keller--Segel equations\cite{keller_jtb70} ,
\begin{align}
\partial_t \rho
&= D_\rho \nabla^2 \rho
- \nabla \cdot \bigl(\beta \rho \nabla c \bigr) \ ,
\\
\partial_t c
&= D_c \nabla^2 c
- \lambda c
+ \alpha \rho \ ,
\end{align}
where $\rho$ is the bacterial density and $c$ is the concentration of signalling molecules.

\subsubsection{Critical transition between dilute and dense state  (2015/2021)}
\label{ssec:chemo_growth}

We now allow for bacterial proliferation regulated by a carrying capacity. Gelimson and Golestanian~\cite{gelimson_prl15} studied this system   in the regime of a phase transition between a dilute but finite density and dense, collapsed state. The situation is, {\it a priori}, similar to Model A dynamics, {\it i.e.}, the hydrodynamics of a nonconserved scalar variable. 
As we shall see, the long-ranged chemotactic field, as an additional hydrodynamic variable, however, changes the UC and leads to distinct scaling behaviour. 

Taking the quasi-static limit for the chemical field $c$, and neglecting chemical degradation, yields the constraint
\begin{equation}
-\nabla^2 c = \alpha \rho \ ,
\end{equation}
which allows the chemical field to be eliminated in favour of effective long-ranged interactions between density fluctuations~\cite{gelimson_prl15}. Alternatively, one may retain $c$ as an auxiliary field, as done in later studies~\cite{mahdisoltani_prr21,zinati_epl22}. We choose to follow these later studies here. Expanding the density field around the carrying capacity then leads to the following long-ranged scalar field theory:
\begin{equation}
\partial_t \delta \rho
= D_\rho \nabla^2 \delta \rho
 - \theta \delta \rho
- \nu_1 \nabla \cdot \left[
\delta \rho \nabla c
\right]
-\nu_2 \nabla^2(\nabla c)^2
- \frac{\lambda}{2}\,\delta \rho^2
+ f \, .
\end{equation}

The critical point is obtained by simultaneously tuning $\theta$ and $\lambda$ to zero.   Since the density of the dilute phase
remains finite, the noise can be taken as purely additive. 

The nonlinearity proportional to $\nu_2$ was overlooked in the original study~\cite{gelimson_prl15}. However, later work showed that it is both symmetry-allowed and RG relevant, and is generically generated under the RG flow~\cite{zinati_epl22}. This term can also be interpreted as arising from a polarity-induced chemotactic mechanism studied independently by Mahdisoltani \emph{et al.}~\cite{mahdisoltani_prr21}. Once this additional nonlinearity is included, the fixed point identified in Ref.~\cite{gelimson_prl15} ceases to exist~\cite{zinati_epl22}. The correct UC was instead identified by Mahdisoltani, Ben Alì Zinati, Duclut, Gambassi and Golestanian~\cite{mahdisoltani_prr21}. Using dynamical RG, they showed that the critical point has upper critical dimension $d_c=6$ and is controlled by the coupled nonlinearities
\begin{equation}
\nabla \cdot \left[
\delta \rho \nabla c
\right]
\sep
\nabla^2(\nabla c)^2 \ .
\end{equation}

The one-loop RG flow exhibits a hyperbolic line of stable fixed points. At these fixed points, an emergent Galilean symmetry, together with the nonrenormalization of the noise amplitude, determines the critical exponents exactly. These fixed points define the UC governing the critical transition between regulated and unregulated chemotactic growth.

\subsubsection{Generic scale invariance in conserved chemotactic systems (2021)}
\label{ssec:phase_chemo_growth}

We now return to the same model discussed in {\bf  Section~\ref{ssec:chemo_growth}}, but assume that the number of chemotactic bacteria is locally conserved. This immediately eliminates the nonconservative terms $\theta$ and $\lambda$, while rendering the noise conserved. The resulting model was also analysed by Mahdisoltani, Ben Alì Zinati, Duclut, Gambassi and Golestanian~\cite{mahdisoltani_prr21}, but now describes the generic scale invariant phase of the conserved system rather than a critical point. Despite the different conservation law, the underlying analysis remains  structurally similar: the one-loop RG calculation identifies a hyperbolic line of stable fixed points, along which an emergent Galilean symmetry, together with the nonrenormalization of the conserved noise amplitude, determines the scaling exponents exactly.  Due to the different noise statistics, however, both the scaling exponents and the upper critical dimension ($d_c=4$) are modified, giving rise to a distinct UC.

\subsubsection{Absorbing-state transition between living and extinct states (2023)}
\label{ssec:chemo_percolation}

van der Kolk, Rasshofer, Swiderski, Haldar, Basu and Frey~\cite{vanderkolk_prl23} considered the same class of chemotactic models near the zero-density, or extinction, state. In this regime, the chemical field $c$ obeys the same reaction--diffusion equation as before, while the density field $\rho$ denotes the actual particle density and satisfies
\begin{equation}
\partial_t \rho
= D_\rho \nabla^2 \rho
- \theta \rho
-\nu_1 \nabla \cdot (\rho \nabla c)
+ \nu_3 \rho \nabla^2 c
- \frac{ \lambda}{2}\, \rho^2
+ f \, .
\end{equation}
Here, $\theta$ is the control parameter for the absorbing-state transition, which occurs at $\theta=0$.

Since we are close to the extinction regime,
the noise is necessarily multiplicative,
\begin{equation}
\langle f(\mathbf r,t)\,f(\mathbf r',t') \rangle
\propto
\rho \,\delta^{d}(\mathbf r-\mathbf r')\,\delta(t-t') \ ,
\end{equation}
so that fluctuations vanish in the absorbing state $\rho=0$. In the absence of chemotaxis, this multiplicative-noise structure, together with the nonlinear saturation term $\rho^2$, gives the directed-percolation (DP) universality class, equivalently Reggeon field theory~\cite{cardy_jpa80}.

Using a perturbative RG expansion about the upper critical dimension $d_c=4$, the authors showed that, in the critical regime and in the limit of infinitely fast chemical diffusion,
$w \equiv \frac{D_c}{D_c+D_\rho} = 1$,
the relevant nonlinearities are
\begin{equation}
\nabla \cdot (\rho \nabla c)
\sep
\rho \nabla^2 c
\sep
\rho^2 \, .
\end{equation}
Their coupled RG flow yields two distinct chemotaxis-modified absorbing-state UCs: a \emph{chemoattractive} (CA) UC for attractive chemotaxis and a \emph{chemorepellent} (CR) UC for repulsive chemotaxis. These two classes differ not only in their fixed-point structure but also in their transport behaviour: attraction slows density spreading and gives subdiffusive critical dynamics, $z>2$, whereas repulsion enhances spreading and gives superdiffusive dynamics, $z<2$.

\subsection{Driven $O(N)$ model with inertia (2024)}
\label{ssec:driven_ON}

A final class of systems in which a new critical UC has recently been identified consists of driven $O(N)$ models with inertial dynamics. Specifically, Daviet, Zelle, Rosch and Diehl~\cite{daviet_prl24} studied a classical $N$-component spin model with intrinsically nonequilibrium, dissipative, and inertial dynamics. Although originally motivated by driven open quantum systems, the authors emphasised that the model also admits purely classical realisations, most notably oscillatory media near the onset of self-sustained oscillations --- a \emph{Hopf instability} in the language of dynamical systems. Unlike the Toner--Tu model, the order parameter is not coupled to physical space through advection or self-transport, and hence convective nonlinearities are absent.

Following {\bf Recipe~1}, the generic universal EOM for the spin field $\vec{\phi}$ takes the form
\begin{equation}
\Big[
\partial_t^2
+ (2\gamma - \nu\nabla^2)\partial_t
+ (r - \mu\nabla^2)
+ u\,|\vec{\phi}|^2\,\partial_t
+ \lambda\,|\vec{\phi}|^2
+ u_0\,\partial_t|\vec{\phi}|^2
\Big]\vec{\phi}
= \bff \ ,
\end{equation}
where $\gamma$ controls the linear damping, $r$ tunes the oscillatory instability, and $\bff$ denotes Gaussian nonconserving noise. In addition to the usual cubic saturation term proportional to $\lambda$, symmetry permits genuinely nonequilibrium nonlinearities of the form $
|\vec{\phi}|^2\,\partial_t \vec{\phi}$ and $
\partial_t|\vec{\phi}|^2 \vec{\phi}$.

In the hydrodynamic limit, the inertial term $\partial_t^2\vec{\phi}$ would normally be subleading compared to the dissipative term proportional to $\gamma \partial_t \vec{\phi}$. The key observation of the authors is that the damping coefficient $\gamma$ can itself be tuned to zero, thereby rendering the inertial dynamics dominant and driving the system towards a finite-frequency critical instability.

Focusing on the symmetry-broken regime ($r>0$) precisely at the onset of oscillations ($\gamma=0$), the authors carried out a perturbative RG analysis. They showed that below the upper critical dimension $d_c=4$, the transition into the rotating oscillatory phase is governed by a new UC controlled by the coupled nonlinearities
\beq
|\vec{\phi}|^2 \vec{\phi}
\sep
|\vec{\phi}|^2\,\partial_t \vec{\phi}
\sep
\partial_t|\vec{\phi}|^2 \vec{\phi}\ .
\eeq
The associated critical exponents were computed to two-loop order.

\subsection{Multicritical phenomena}

Thus far, we have identified nine new nonequilibrium phases and nine distinct critical phenomena uncovered through RG analyses. Beyond these, recent work has also revealed a rich variety of genuinely nonequilibrium \emph{multicritical} phenomena. Rather than reviewing each example in detail, we summarise the currently known multicritical points and their associated universality classes in Table~\ref{tab:multi}.

\begin{table}
 \renewcommand{\arraystretch}{2}
\begin{tabular}{c|c|c|c|c|c}
Ref. & System & Relevant terms & $d_c$ & num.~new FPs & Method 
\\
\hline
\hline
Legrand \& Lee\cite{legrand_newton26}&
\makecell{Mal.~PAF with an easy axis\\(`Transverse' Lifshitz point)} &$v_x \pp_x v_x$   & 7 & 1 & 1-loop
 \\
\hline
Wong \& Lee\cite{wong_a25}&
PAF with an easy axis &$\partial_x g_x^2 \ , \ \delta\rho^2 g_x$   & 4 & 2 & 1-loop
 \\
\hline
Jentsch \& Lee\cite{jentsch_prl24}&
\makecell{Constrained PAF\\ in the ordered phase}  &\makecell{$(\bg_\perp \cdot \vnab_\perp) \bg_\perp \ , \ \bg_\perp  (\vnab_\perp\cdot \bg_\perp) \ ,$\\$
\vnab_\perp |\bg_\perp|^2 \  , \ \bg_\perp |\bg_\perp|^2$}  & $\frac{11}{3}$ & 1 &  NP-FRG
 \\
\hline
Young et al. \cite{young_prx20,young_pre26}&
\makecell{Driven \& coupled\\ $O(N_1) \times O(N_2)$ model} & \makecell{$|\vec{\phi}_1|^2\vec{\phi}_2 \ , \
|\vec{\phi}_2|^2\vec{\phi}_1 \ , $\\$ {   |\vec{\phi}_1|^2\vec{\phi}_1 \ , \ |\vec{\phi}_2|^2\vec{\phi}_2}$ } & 4 & \makecell{$N_1$ \& $N_2$ dependent,\\ tabulated in paper} & 2-loop
 \\
\hline
Jentsch \& Lee\cite{jentsch_prr23}&
PAF &$\delta \rho^2 \bg \ , \
\delta \rho^2 \vnab \delta \rho$  & 6 & 3 & NP-FRG
 \\
\hline
\end{tabular}
\caption{
{\it New universality classes associated with nonequilibrium multicritical phenomena uncovered since 2015, ordered in reverse chronological order.} The columns list: (i) the reference(s) in which the multicritical behaviour was identified; (ii) the physical system or hydrodynamic model; (iii) the relevant nonlinear term(s) controlling the RG fixed point(s); (iv) the upper critical dimension $d_c$
; (v) the number of distinct new fixed points identified; and (vi) the analytical method used in the corresponding study.
}
\label{tab:multi}
\end{table}

\subsection{A `periodic table' of new universality classes}
\label{sec:periodic}

To further elucidate the diversity of universality classes (UCs) encountered above, we now attempt to organise them in a manner loosely analogous to the periodic table of chemical elements. In chemistry, elements can be linearly ordered by a single integer --- the number of protons in the nucleus. For UCs, however, no such one-dimensional ordering exists: the defining characteristics are inherently multi-dimensional, involving the hydrodynamic degrees of freedom, symmetries, conservation laws, and constraints (SCC).

To construct a compact overview of the newly discovered UCs, we therefore classify them according to two primary criteria: (i) the nature of the hydrodynamic variables involved, and (ii) whether the UC governs a phase or a critical point (following {\bf Recipe~3}). Multicritical points are omitted for clarity. Additional colour coding indicates the broader physical settings --- namely, polar active fluids (with or without quenched disorder), active matter, interfacial dynamics, chemotaxis, and driven spin systems --- to which the various UCs belong.

As discussed in {\bf Section~\ref{sec:phases_crit}}, this organisation is itself not unique: the same UC may naturally appear in different categories depending on the underlying microscopic physics and the route by which the hydrodynamic description is constructed.

\begin{figure}
  \includegraphics[width=\textwidth]{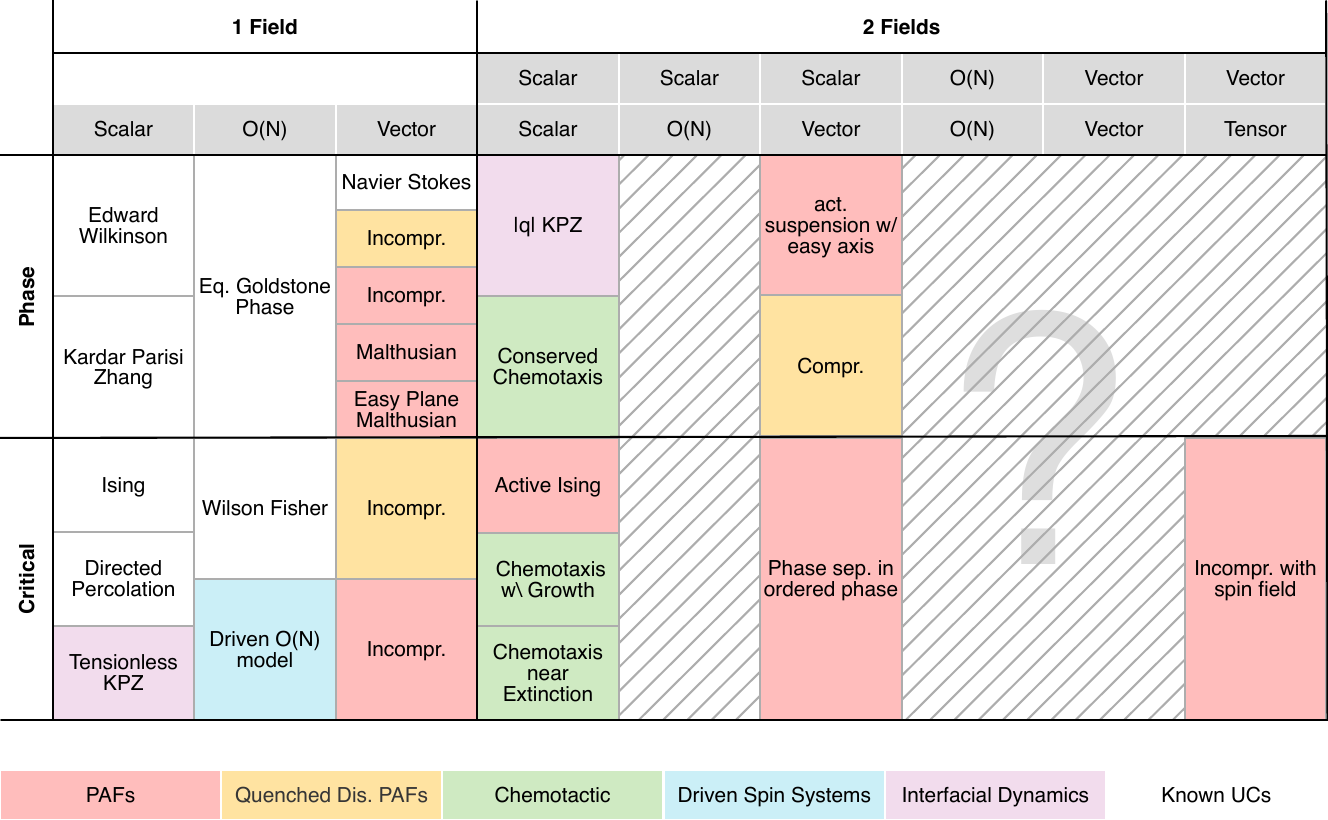}
  \caption{ 
{\it A ``periodic table'' of universality classes.} 
As discussed in {\bf Section~\ref{sec:periodic}}, the space of UCs is intrinsically high-dimensional. For clarity, we therefore classify the known UCs according to (i) the number and type of hydrodynamic variables involved, and (ii) whether they govern a phase or a critical point. The question mark highlights regions of this space where additional UCs may yet await discovery. 
  }
  
  \label{fig:periodictab}
\end{figure}

%% file: references.bib
@article{nesbitt_njp21,
	title = {Uncovering novel phase transitions in dense dry polar active fluids using a lattice {Boltzmann} method},
	volume = {23},
	url = {http://arxiv.org/abs/1902.00530},
	doi = {10.1088/1367-2630/abd8c0},
	number = {4},
	journal = {New Journal of Physics},
	publisher = {arXiv},
	author = {Nesbitt, David and Pruessner, Gunnar and Lee, Chiu Fan},
	month = apr,
	year = {2021},
	pages = {043047--043047},
}

@article{chen_prl22b,
	title = {Incompressible {Polar} {Active} {Fluids} with {Quenched} {Random} {Field} {Disorder} in {Dimensions} d {\textgreater} 2},
	volume = {129},
	issn = {0031-9007, 1079-7114},
	url = {https://link.aps.org/doi/10.1103/PhysRevLett.129.198001},
	doi = {10.1103/PhysRevLett.129.198001},
	number = {19},
	urldate = {2022-11-01},
	journal = {Physical Review Letters},
	author = {Chen, Leiming and Lee, Chiu Fan and Maitra, Ananyo and Toner, John},
	month = oct,
	year = {2022},
	pages = {198001},
}

@article{wilson_physrep74,
	title = {The renormalization group and the $\epsilon$ expansion},
	volume = {12},
	copyright = {https://www.elsevier.com/tdm/userlicense/1.0/},
	issn = {03701573},
	url = {https://linkinghub.elsevier.com/retrieve/pii/0370157374900234},
	doi = {10.1016/0370-1573(74)90023-4},
	number = {2},
	urldate = {2026-04-19},
	journal = {Physics Reports},
	author = {Wilson, K and Kogut, J},
	month = aug,
	year = {1974},
	pages = {75--199},
}

@article{young_pre26,
	title = {Nonequilibrium universality of the nonreciprocally coupled {O} ( n 1 ) × {O} ( n 2 ) model},
	volume = {113},
	issn = {2470-0045, 2470-0053},
	url = {https://link.aps.org/doi/10.1103/n4l9-p6xn},
	doi = {10.1103/n4l9-p6xn},
	language = {en},
	number = {3},
	urldate = {2026-06-06},
	journal = {Physical Review E},
	author = {Young, Jeremy T. and Gorshkov, Alexey V. and Maghrebi, Mohammad},
	month = mar,
	year = {2026},
	pages = {034132},
}

@article{kadanoff_annphys00,
	title = {Hydrodynamic {Equations} and {Correlation} {Functions}},
	volume = {281},
	url = {https://www.sciencedirect.com/science/article/pii/S0003491600960238?via%3Dihub},
	doi = {10.1006/aphy.2000.6023},
	number = {1-2},
	journal = {Annals of Physics},
	publisher = {Academic Press},
	author = {Kadanoff, Leo P. and Martin, Paul C.},
	month = apr,
	year = {2000},
	pages = {800--852},
}

@article{dupuis_pr21,
	title = {The nonperturbative functional renormalization group and its applications},
	volume = {910},
	url = {http://arxiv.org/abs/2006.04853},
	doi = {10.1016/j.physrep.2021.01.001},
	journal = {Physics Reports},
	author = {Dupuis, N. and Canet, L. and Eichhorn, A. and Metzner, W. and Pawlowski, J.M. and Tissier, M. and Wschebor, N.},
	month = may,
	year = {2021},
	pages = {1--114},
}

@article{alert_annrev20,
	title = {Physical {Models} of {Collective} {Cell} {Migration}},
	volume = {11},
	issn = {1947-5454, 1947-5462},
	url = {https://www.annualreviews.org/doi/10.1146/annurev-conmatphys-031218-013516},
	doi = {10.1146/annurev-conmatphys-031218-013516},
	number = {1},
	urldate = {2026-06-04},
	journal = {Annual Review of Condensed Matter Physics},
	author = {Alert, Ricard and Trepat, Xavier},
	month = mar,
	year = {2020},
	pages = {77--101},
}

@article{keber_science14,
	title = {Topology and dynamics of active nematic vesicles},
	volume = {345},
	issn = {0036-8075, 1095-9203},
	url = {https://www.science.org/doi/10.1126/science.1254784},
	doi = {10.1126/science.1254784},
	number = {6201},
	urldate = {2026-06-02},
	journal = {Science},
	author = {Keber, Felix C. and Loiseau, Etienne and Sanchez, Tim and DeCamp, Stephen J. and Giomi, Luca and Bowick, Mark J. and Marchetti, M. Cristina and Dogic, Zvonimir and Bausch, Andreas R.},
	month = sep,
	year = {2014},
	pages = {1135--1139},
}

@article{geyer_prx19,
	title = {Freezing a {Flock}: {Motility}-{Induced} {Phase} {Separation} in {Polar} {Active} {Liquids}},
	volume = {9},
	url = {https://link.aps.org/doi/10.1103/PhysRevX.9.031043},
	doi = {10.1103/PhysRevX.9.031043},
	number = {3},
	journal = {Physical Review X},
	author = {Geyer, Delphine and Martin, David and Tailleur, Julien and Bartolo, Denis},
	month = sep,
	year = {2019},
	pages = {031043--031043},
}

@article{decamp_natmat15,
	title = {Orientational order of motile defects in active nematics},
	volume = {14},
	url = {www.nature.com/naturematerials},
	doi = {10.1038/nmat4387},
	number = {11},
	journal = {Nature Materials},
	publisher = {Nature Publishing Group},
	author = {DeCamp, Stephen J. and Redner, Gabriel S. and Baskaran, Aparna and Hagan, Michael F. and Dogic, Zvonimir},
	month = nov,
	year = {2015},
	pages = {1110--1115},
}

@article{sanchez_nature12,
	title = {Spontaneous motion in hierarchically assembled active matter},
	volume = {491},
	copyright = {http://www.springer.com/tdm},
	issn = {0028-0836, 1476-4687},
	url = {https://www.nature.com/articles/nature11591},
	doi = {10.1038/nature11591},
	language = {en},
	number = {7424},
	urldate = {2026-06-04},
	journal = {Nature},
	author = {Sanchez, Tim and Chen, Daniel T. N. and DeCamp, Stephen J. and Heymann, Michael and Dogic, Zvonimir},
	month = nov,
	year = {2012},
	pages = {431--434},
}

@article{attanasi_prl14,
	title = {Finite-{Size} {Scaling} as a {Way} to {Probe} {Near}-{Criticality} in {Natural} {Swarms}},
	volume = {113},
	url = {http://link.aps.org/doi/10.1103/PhysRevLett.113.238102},
	number = {23},
	journal = {Physical Review Letters},
	publisher = {American Physical Society},
	author = {Attanasi, Alessandro and Cavagna, Andrea and Del Castello, Lorenzo and Giardina, Irene and Melillo, Stefania and Parisi, Leonardo and Pohl, Oliver and Rossaro, Bruno and Shen, Edward and Silvestri, Edmondo and Viale, Massimiliano},
	year = {2014},
	pages = {238102--238102},
}

@article{cavagna_natphys17,
	title = {Dynamic scaling in natural swarms},
	volume = {13},
	issn = {1745-2473, 1745-2481},
	url = {https://www.nature.com/articles/nphys4153},
	doi = {10.1038/nphys4153},
	number = {9},
	urldate = {2026-05-19},
	journal = {Nature Physics},
	author = {Cavagna, Andrea and Conti, Daniele and Creato, Chiara and Del Castello, Lorenzo and Giardina, Irene and Grigera, Tomas S. and Melillo, Stefania and Parisi, Leonardo and Viale, Massimiliano},
	month = sep,
	year = {2017},
	pages = {914--918},
}

@article{cavagna_pnas10,
	title = {Scale-free correlations in starling flocks},
	volume = {107},
	issn = {0027-8424, 1091-6490},
	url = {https://pnas.org/doi/full/10.1073/pnas.1005766107},
	doi = {10.1073/pnas.1005766107},
	number = {26},
	urldate = {2026-06-02},
	journal = {Proceedings of the National Academy of Sciences},
	author = {Cavagna, Andrea and Cimarelli, Alessio and Giardina, Irene and Parisi, Giorgio and Santagati, Raffaele and Stefanini, Fabio and Viale, Massimiliano},
	month = jun,
	year = {2010},
	pages = {11865--11870},
}

@article{weber_rpp19,
	title = {Physics of active emulsions},
	volume = {82},
	url = {https://iopscience.iop.org/article/10.1088/1361-6633/ab052b},
	doi = {10.1088/1361-6633/ab052b},
	number = {6},
	journal = {Reports on Progress in Physics},
	publisher = {Institute of Physics Publishing},
	author = {Weber, Christoph A. and Zwicker, David and Jülicher, Frank and Lee, Chiu Fan},
	month = jun,
	year = {2019},
	pages = {064601--064601},
}

@article{aditisimha_prl02,
	title = {Hydrodynamic {Fluctuations} and {Instabilities} in {Ordered} {Suspensions} of {Self}-{Propelled} {Particles}},
	volume = {89},
	url = {http://dx.doi.org/10.1103/PhysRevLett.89.058101},
	doi = {10.1103/PhysRevLett.89.058101},
	number = {5},
	journal = {Physical Review Letters},
	publisher = {American Physical Society},
	author = {Aditi Simha, R. and Ramaswamy, Sriram},
	month = jul,
	year = {2002},
	pages = {058101--058101},
}

@article{besse_prl23,
  title = {Interface Roughening in Nonequilibrium Phase-Separated Systems},
  author = {Besse, M. and Fausti, G. and Cates, M. E. and Delamotte, B. and Nardini, C.},
  journal = {Phys. Rev. Lett.},
  volume = {130},
  issue = {18},
  pages = {187102},
  numpages = {6},
  year = {2023},
  month = {May},
  publisher = {American Physical Society},
  doi = {10.1103/PhysRevLett.130.187102},
  url = {https://link.aps.org/doi/10.1103/PhysRevLett.130.187102}
}

@Article{cavagna_natphys23,
author={Cavagna, Andrea
and Di Carlo, Luca
and Giardina, Irene
and Grigera, Tom{\'a}s S.
and Melillo, Stefania
and Parisi, Leonardo
and Pisegna, Giulia
and Scandolo, Mattia},
title={Natural swarms in 3.99 dimensions},
journal={Nature Physics},
year={2023},
month={Jul},
day={01},
volume={19},
number={7},
pages={1043-1049},
issn={1745-2481},
doi={10.1038/s41567-023-02028-0},
url={https://doi.org/10.1038/s41567-023-02028-0}
}

@Article{chen_natcomm16,
author={Chen, Leiming
and Lee, Chiu Fan
and Toner, John},
title={Mapping two-dimensional polar active fluids to two-dimensional soap and one-dimensional sandblasting},
journal={Nature Communications},
year={2016},
month={Jul},
day={25},
volume={7},
number={1},
pages={12215},
issn={2041-1723},
doi={10.1038/ncomms12215},
url={https://doi.org/10.1038/ncomms12215}
}

@article{chen_njp15,
doi = {10.1088/1367-2630/17/4/042002},
url = {https://dx.doi.org/10.1088/1367-2630/17/4/042002},
year = {2015},
month = {apr},
publisher = {IOP Publishing},
volume = {17},
number = {4},
pages = {042002},
author = {Chen, Leiming and Toner, John and Lee, Chiu Fan},
title = {Critical phenomenon of the order–disorder transition in incompressible active fluids},
journal = {New Journal of Physics}
}

@article{forster_pra77,
  title = {Large-distance and long-time properties of a randomly stirred fluid},
  author = {Forster, Dieter and Nelson, David R. and Stephen, Michael J.},
  journal = {Phys. Rev. A},
  volume = {16},
  issue = {2},
  pages = {732--749},
  numpages = {0},
  year = {1977},
  month = {Aug},
  publisher = {American Physical Society},
  doi = {10.1103/PhysRevA.16.732},
  url = {https://link.aps.org/doi/10.1103/PhysRevA.16.732}
}

@article{toner_prl18,
  title = {Swarming in the Dirt: Ordered Flocks with Quenched Disorder},
  author = {Toner, John and Guttenberg, Nicholas and Tu, Yuhai},
  journal = {Phys. Rev. Lett.},
  volume = {121},
  issue = {24},
  pages = {248002},
  numpages = {6},
  year = {2018},
  month = {Dec},
  publisher = {American Physical Society},
  doi = {10.1103/PhysRevLett.121.248002},
  url = {https://link.aps.org/doi/10.1103/PhysRevLett.121.248002}
}

@article{toner_pre18,
  title = {Hydrodynamic theory of flocking in the presence of quenched disorder},
  author = {Toner, John and Guttenberg, Nicholas and Tu, Yuhai},
  journal = {Phys. Rev. E},
  volume = {98},
  issue = {6},
  pages = {062604},
  numpages = {19},
  year = {2018},
  month = {Dec},
  publisher = {American Physical Society},
  doi = {10.1103/PhysRevE.98.062604},
  url = {https://link.aps.org/doi/10.1103/PhysRevE.98.062604}
}

@article{miller_pre24,
  title = {Phase separation in ordered polar active fluids: A completely different universality class from that of equilibrium fluids},
  author = {Miller, Maxx and Toner, John},
  journal = {Phys. Rev. E},
  volume = {110},
  issue = {5},
  pages = {054607},
  numpages = {21},
  year = {2024},
  month = {Nov},
  publisher = {American Physical Society},
  doi = {10.1103/PhysRevE.110.054607},
  url = {https://link.aps.org/doi/10.1103/PhysRevE.110.054607}
}

@article{canet_jstatmech25,
doi = {10.1088/1742-5468/ae1e75},
url = {https://doi.org/10.1088/1742-5468/ae1e75},
year = {2025},
month = {dec},
publisher = {IOP Publishing},
volume = {2025},
number = {12},
pages = {124003},
author = {Canet, Léonie},
title = {The non-perturbative sides of the Kardar–Parisi–Zhang equation},
journal = {Journal of Statistical Mechanics: Theory and Experiment}
}

@article{solon_prl13,
  title = {Revisiting the Flocking Transition Using Active Spins},
  author = {Solon, A. P. and Tailleur, J.},
  journal = {Phys. Rev. Lett.},
  volume = {111},
  issue = {7},
  pages = {078101},
  numpages = {5},
  year = {2013},
  month = {Aug},
  publisher = {American Physical Society},
  doi = {10.1103/PhysRevLett.111.078101},
  url = {https://link.aps.org/doi/10.1103/PhysRevLett.111.078101}
}

@article{solon_pre15,
  title = {Flocking with discrete symmetry: The two-dimensional active Ising model},
  author = {Solon, A. P. and Tailleur, J.},
  journal = {Phys. Rev. E},
  volume = {92},
  issue = {4},
  pages = {042119},
  numpages = {18},
  year = {2015},
  month = {Oct},
  publisher = {American Physical Society},
  doi = {10.1103/PhysRevE.92.042119},
  url = {https://link.aps.org/doi/10.1103/PhysRevE.92.042119}
}

@misc{wong_a25,
      title={New universality classes govern the critical and multicritical behavior of an active Ising model}, 
      author={Matthew Wong and Chiu Fan Lee},
      year={2025},
      eprint={2507.06068},
      archivePrefix={arXiv},
      primaryClass={cond-mat.stat-mech},
      url={https://arxiv.org/abs/2507.06068}, 
}

@article{toner_prl12,
  title = {Birth, Death, and Flight: A Theory of Malthusian Flocks},
  author = {Toner, John},
  journal = {Phys. Rev. Lett.},
  volume = {108},
  issue = {8},
  pages = {088102},
  numpages = {5},
  year = {2012},
  month = {Feb},
  publisher = {American Physical Society},
  doi = {10.1103/PhysRevLett.108.088102},
  url = {https://link.aps.org/doi/10.1103/PhysRevLett.108.088102}
}

@article{chen_prl20,
  title = {Moving, Reproducing, and Dying Beyond Flatland: Malthusian Flocks in Dimensions $d>2$},
  author = {Chen, Leiming and Lee, Chiu Fan and Toner, John},
  journal = {Phys. Rev. Lett.},
  volume = {125},
  issue = {9},
  pages = {098003},
  numpages = {6},
  year = {2020},
  month = {Aug},
  publisher = {American Physical Society},
  doi = {10.1103/PhysRevLett.125.098003},
  url = {https://link.aps.org/doi/10.1103/PhysRevLett.125.098003}
}

@article{chen_pre20,
  title = {Universality class for a nonequilibrium state of matter: A $d=4-\epsilon$ expansion study of Malthusian flocks},
  author = {Chen, Leiming and Lee, Chiu Fan and Toner, John},
  journal = {Phys. Rev. E},
  volume = {102},
  issue = {2},
  pages = {022610},
  numpages = {40},
  year = {2020},
  month = {Aug},
  publisher = {American Physical Society},
  doi = {10.1103/PhysRevE.102.022610},
  url = {https://link.aps.org/doi/10.1103/PhysRevE.102.022610}
}

@article{toner_pre24,
  title = {Birth, death, and horizontal flight: Malthusian flocks with an easy plane in three dimensions},
  author = {Toner, John},
  journal = {Phys. Rev. E},
  volume = {110},
  issue = {6},
  pages = {064604},
  numpages = {6},
  year = {2024},
  month = {Dec},
  publisher = {American Physical Society},
  doi = {10.1103/PhysRevE.110.064604},
  url = {https://link.aps.org/doi/10.1103/PhysRevE.110.064604}
}

@article{chen_njp18,
doi = {10.1088/1367-2630/aaec31},
url = {https://doi.org/10.1088/1367-2630/aaec31},
year = {2018},
month = {nov},
publisher = {IOP Publishing},
volume = {20},
number = {11},
pages = {113035},
author = {Chen, Leiming and Lee, Chiu Fan and Toner, John},
title = {Incompressible polar active fluids in the moving phase in dimensions $d>2$},
journal = {New Journal of Physics}
}

@article{toner_prl95,
  title = {Long-Range Order in a Two-Dimensional Dynamical $\mathrm{XY}$ Model: How Birds Fly Together},
  author = {Toner, John and Tu, Yuhai},
  journal = {Phys. Rev. Lett.},
  volume = {75},
  issue = {23},
  pages = {4326--4329},
  numpages = {0},
  year = {1995},
  month = {Dec},
  publisher = {American Physical Society},
  doi = {10.1103/PhysRevLett.75.4326},
  url = {https://link.aps.org/doi/10.1103/PhysRevLett.75.4326}
}

@article{chen_pre24,
  title = {Dynamics of packed swarms: Time-displaced correlators of two-dimensional incompressible flocks},
  author = {Chen, Leiming and Lee, Chiu Fan and Maitra, Ananyo and Toner, John},
  journal = {Phys. Rev. E},
  volume = {109},
  issue = {1},
  pages = {L012601},
  numpages = {6},
  year = {2024},
  month = {Jan},
  publisher = {American Physical Society},
  doi = {10.1103/PhysRevE.109.L012601},
  url = {https://link.aps.org/doi/10.1103/PhysRevE.109.L012601}
}

@article{chen_pre22,
  title = {Hydrodynamic theory of two-dimensional incompressible polar active fluids with quenched and annealed disorder},
  author = {Chen, Leiming and Lee, Chiu Fan and Maitra, Ananyo and Toner, John},
  journal = {Phys. Rev. E},
  volume = {106},
  issue = {4},
  pages = {044608},
  numpages = {29},
  year = {2022},
  month = {Oct},
  publisher = {American Physical Society},
  doi = {10.1103/PhysRevE.106.044608},
  url = {https://link.aps.org/doi/10.1103/PhysRevE.106.044608}
}

@article{chen_prl22,
  title = {Packed Swarms on Dirt: Two-Dimensional Incompressible Flocks with Quenched and Annealed Disorder},
  author = {Chen, Leiming and Lee, Chiu Fan and Maitra, Ananyo and Toner, John},
  journal = {Phys. Rev. Lett.},
  volume = {129},
  issue = {18},
  pages = {188004},
  numpages = {7},
  year = {2022},
  month = {Oct},
  publisher = {American Physical Society},
  doi = {10.1103/PhysRevLett.129.188004},
  url = {https://link.aps.org/doi/10.1103/PhysRevLett.129.188004}
}

@article{zinati_pre22,
  title = {Dense polar active fluids in a disordered environment},
  author = {Zinati, Riccardo Ben Al\`{\i} and Besse, Marc and Tarjus, Gilles and Tissier, Matthieu},
  journal = {Phys. Rev. E},
  volume = {105},
  issue = {6},
  pages = {064605},
  numpages = {10},
  year = {2022},
  month = {Jun},
  publisher = {American Physical Society},
  doi = {10.1103/PhysRevE.105.064605},
  url = {https://link.aps.org/doi/10.1103/PhysRevE.105.064605}
}

@article{dadhichi_prl26,
  title = {Superdiffusion and Antidiffusion in an Aligned Active Suspension},
  author = {Dadhichi, Lokrshi Prawar and Sahoo, Suvendra K. and Kumar, K. Vijay and Ramaswamy, Sriram},
  journal = {Phys. Rev. Lett.},
  volume = {136},
  issue = {12},
  pages = {128302},
  numpages = {9},
  year = {2026},
  month = {Mar},
  publisher = {American Physical Society},
  doi = {10.1103/d61l-1tyh},
  url = {https://link.aps.org/doi/10.1103/d61l-1tyh}
}

@article{tjhung_prx18,
  title = {Cluster Phases and Bubbly Phase Separation in Active Fluids: Reversal of the Ostwald Process},
  author = {Tjhung, Elsen and Nardini, Cesare and Cates, Michael E.},
  journal = {Phys. Rev. X},
  volume = {8},
  issue = {3},
  pages = {031080},
  numpages = {19},
  year = {2018},
  month = {Sep},
  publisher = {American Physical Society},
  doi = {10.1103/PhysRevX.8.031080},
  url = {https://link.aps.org/doi/10.1103/PhysRevX.8.031080}
}

@Article{wittkowski_natcomm14,
author={Wittkowski, Raphael
and Tiribocchi, Adriano
and Stenhammar, Joakim
and Allen, Rosalind J.
and Marenduzzo, Davide
and Cates, Michael E.},
title={Scalar $\phi$4 field theory for active-particle phase separation},
journal={Nature Communications},
year={2014},
month={Jul},
day={10},
volume={5},
number={1},
pages={4351},
issn={2041-1723},
doi={10.1038/ncomms5351},
url={https://doi.org/10.1038/ncomms5351}
}

@article{cates_annurev15,
   author = "Cates, Michael E. and Tailleur, Julien",
   title = "Motility-Induced Phase Separation", 
   journal= "Annual Review of Condensed Matter Physics",
   year = "2015",
   volume = "6",
   number = "Volume 6, 2015",
   pages = "219-244",
   doi = "https://doi.org/10.1146/annurev-conmatphys-031214-014710",
   url = "https://www.annualreviews.org/content/journals/10.1146/annurev-conmatphys-031214-014710",
   publisher = "Annual Reviews",
   issn = "1947-5462",
   type = "Journal Article",
  }

@article{fily_prl12,
  title = {Athermal Phase Separation of Self-Propelled Particles with No Alignment},
  author = {Fily, Yaouen and Marchetti, M. Cristina},
  journal = {Phys. Rev. Lett.},
  volume = {108},
  issue = {23},
  pages = {235702},
  numpages = {5},
  year = {2012},
  month = {Jun},
  publisher = {American Physical Society},
  doi = {10.1103/PhysRevLett.108.235702},
  url = {https://link.aps.org/doi/10.1103/PhysRevLett.108.235702}
}

@Article{lee_soft17,
author ="Lee, Chiu Fan",
title  ="Interface stability{,} interface fluctuations{,} and the Gibbs–Thomson relationship in motility-induced phase separations",
journal  ="Soft Matter",
year  ="2017",
volume  ="13",
issue  ="2",
pages  ="376-385",
publisher  ="The Royal Society of Chemistry",
doi  ="10.1039/C6SM01978A",
url  ="http://dx.doi.org/10.1039/C6SM01978A"}

@article{gelimson_prl15,
  title = {Collective Dynamics of Dividing Chemotactic Cells},
  author = {Gelimson, Anatolij and Golestanian, Ramin},
  journal = {Phys. Rev. Lett.},
  volume = {114},
  issue = {2},
  pages = {028101},
  numpages = {5},
  year = {2015},
  month = {Jan},
  publisher = {American Physical Society},
  doi = {10.1103/PhysRevLett.114.028101},
  url = {https://link.aps.org/doi/10.1103/PhysRevLett.114.028101}
}

@article{mahdisoltani_prr21,
  title = {Nonequilibrium polarity-induced chemotaxis: Emergent Galilean symmetry and exact scaling exponents},
  author = {Mahdisoltani, Saeed and Zinati, Riccardo Ben Al\`{\i} and Duclut, Charlie and Gambassi, Andrea and Golestanian, Ramin},
  journal = {Phys. Rev. Res.},
  volume = {3},
  issue = {1},
  pages = {013100},
  numpages = {22},
  year = {2021},
  month = {Jan},
  publisher = {American Physical Society},
  doi = {10.1103/PhysRevResearch.3.013100},
  url = {https://link.aps.org/doi/10.1103/PhysRevResearch.3.013100}
}

@article{zinati_epl22,
doi = {10.1209/0295-5075/ac48c9},
url = {https://doi.org/10.1209/0295-5075/ac48c9},
year = {2022},
month = {mar},
publisher = {EDP Sciences, IOP Publishing and Società Italiana di Fisica},
volume = {136},
number = {5},
pages = {50003},
author = {Ben Alì Zinati, Riccardo and Duclut, Charlie and Mahdisoltani, Saeed and Gambassi, Andrea and Golestanian, Ramin},
title = {Stochastic dynamics of chemotactic colonies with logistic growth},
journal = {Europhysics Letters}
}

@article{vanderkolk_prl23,
  title = {Anomalous Collective Dynamics of Autochemotactic Populations},
  author = {van der Kolk, Jasper and Ra\ss{}hofer, Florian and Swiderski, Richard and Haldar, Astik and Basu, Abhik and Frey, Erwin},
  journal = {Phys. Rev. Lett.},
  volume = {131},
  issue = {8},
  pages = {088201},
  numpages = {7},
  year = {2023},
  month = {Aug},
  publisher = {American Physical Society},
  doi = {10.1103/PhysRevLett.131.088201},
  url = {https://link.aps.org/doi/10.1103/PhysRevLett.131.088201}
}

@article{cardy_jpa80,
doi = {10.1088/0305-4470/13/12/002},
url = {https://doi.org/10.1088/0305-4470/13/12/002},
year = {1980},
month = {dec},
publisher = {},
volume = {13},
number = {12},
pages = {L423},
author = {J L Cardy and R L Sugar},
title = {Directed percolation and Reggeon field theory},
journal = {Journal of Physics A: Mathematical and General}
}

@article{daviet_prl24,
  title = {Nonequilibrium Criticality at the Onset of Time-Crystalline Order},
  author = {Daviet, Romain and Zelle, Carl Philipp and Rosch, Achim and Diehl, Sebastian},
  journal = {Phys. Rev. Lett.},
  volume = {132},
  issue = {16},
  pages = {167102},
  numpages = {7},
  year = {2024},
  month = {Apr},
  publisher = {American Physical Society},
  doi = {10.1103/PhysRevLett.132.167102},
  url = {https://link.aps.org/doi/10.1103/PhysRevLett.132.167102}
}

@article{legrand_newton26,
	title = {Universal behavior at the {Lifshitz} points of an active {Malthusian} {Ising} model},
	issn = {29506360},
	url = {https://linkinghub.elsevier.com/retrieve/pii/S2950636026001659},
	doi = {10.1016/j.newton.2026.100563},
	urldate = {2026-06-15},
	journal = {Newton},
	author = {Legrand, Gabriel and Lee, Chiu Fan},
	month = jun,
	year = {2026},
	pages = {100563},
}

@article{jentsch_prr23,
  title = {Critical phenomena in compressible polar active fluids: Dynamical and functional renormalization group studies},
  author = {Jentsch, Patrick and Lee, Chiu Fan},
  journal = {Phys. Rev. Res.},
  volume = {5},
  issue = {2},
  pages = {023061},
  numpages = {24},
  year = {2023},
  month = {Apr},
  publisher = {American Physical Society},
  doi = {10.1103/PhysRevResearch.5.023061},
  url = {https://link.aps.org/doi/10.1103/PhysRevResearch.5.023061}
}

@article{young_prx20,
  title = {Nonequilibrium Fixed Points of Coupled Ising Models},
  author = {Young, Jeremy T. and Gorshkov, Alexey V. and Foss-Feig, Michael and Maghrebi, Mohammad F.},
  journal = {Phys. Rev. X},
  volume = {10},
  issue = {1},
  pages = {011039},
  numpages = {32},
  year = {2020},
  month = {Feb},
  publisher = {American Physical Society},
  doi = {10.1103/PhysRevX.10.011039},
  url = {https://link.aps.org/doi/10.1103/PhysRevX.10.011039}
}

@article{jentsch_prl24,
  title = {New Universality Class Describes Vicsek's Flocking Phase in Physical Dimensions},
  author = {Jentsch, Patrick and Lee, Chiu Fan},
  journal = {Phys. Rev. Lett.},
  volume = {133},
  issue = {12},
  pages = {128301},
  numpages = {7},
  year = {2024},
  month = {Sep},
  publisher = {American Physical Society},
  doi = {10.1103/PhysRevLett.133.128301},
  url = {https://link.aps.org/doi/10.1103/PhysRevLett.133.128301}
}

@misc{hayakawa_a25,
      title={Emergence of chiral multi-armed spirals in an open system of migrating cells under continuous cell supply}, 
      author={Masayuki Hayakawa and Biplab Bhattacherjee and Hidekazu Kuwayama and Tatsuo Shibata},
      year={2025},
      eprint={2511.16074},
      archivePrefix={arXiv},
      primaryClass={cond-mat.soft},
      url={https://arxiv.org/abs/2511.16074}, 
}

@article{chen_pre26,
  title = {Only the ambidextrous can flock: Two-dimensional chiral Malthusian flocks, time cholesterics, and the Kardar-Parisi-Zhang equation},
  author = {Chen, Leiming and Lee, Chiu Fan and Toner, John},
  journal = {Phys. Rev. E},
  volume = {113},
  issue = {4},
  pages = {045419},
  numpages = {30},
  year = {2026},
  month = {Apr},
  publisher = {American Physical Society},
  doi = {10.1103/sj5y-8dsd},
  url = {https://link.aps.org/doi/10.1103/sj5y-8dsd}
}

@article{bricard_nature13,
author={Bricard, Antoine
and Caussin, Jean-Baptiste
and Desreumaux, Nicolas
and Dauchot, Olivier
and Bartolo, Denis},
title={Emergence of macroscopic directed motion in populations of motile colloids},
journal={Nature},
year={2013},
month={Nov},
day={01},
volume={503},
number={7474},
pages={95-98},
issn={1476-4687},
doi={10.1038/nature12673},
url={https://doi.org/10.1038/nature12673}
}

@article{toner_pre98,
  title = {Flocks, herds, and schools: A quantitative theory of flocking},
  author = {Toner, John and Tu, Yuhai},
  journal = {Phys. Rev. E},
  volume = {58},
  issue = {4},
  pages = {4828--4858},
  numpages = {0},
  year = {1998},
  month = {Oct},
  publisher = {American Physical Society},
  doi = {10.1103/PhysRevE.58.4828},
  url = {https://link.aps.org/doi/10.1103/PhysRevE.58.4828}
}

@article{toner_pre12,
  title = {Reanalysis of the hydrodynamic theory of fluid, polar-ordered flocks},
  author = {Toner, John},
  journal = {Phys. Rev. E},
  volume = {86},
  issue = {3},
  pages = {031918},
  numpages = {9},
  year = {2012},
  month = {Sep},
  publisher = {American Physical Society},
  doi = {10.1103/PhysRevE.86.031918},
  url = {https://link.aps.org/doi/10.1103/PhysRevE.86.031918}
}

@article{chate_prl24,
  title = {Dynamic Scaling of Two-Dimensional Polar Flocks},
  author = {Chat\'e, Hugues and Solon, Alexandre},
  journal = {Phys. Rev. Lett.},
  volume = {132},
  issue = {26},
  pages = {268302},
  numpages = {6},
  year = {2024},
  month = {Jun},
  publisher = {American Physical Society},
  doi = {10.1103/PhysRevLett.132.268302},
  url = {https://link.aps.org/doi/10.1103/PhysRevLett.132.268302}
}

@misc{chen_a25,
      title={The inconvenient truth about flocks}, 
      author={Leiming Chen and Patrick Jentsch and Chiu Fan Lee and Ananyo Maitra and Sriram Ramaswamy and John Toner},
      year={2025},
      eprint={2503.17064},
      archivePrefix={arXiv},
      primaryClass={cond-mat.soft},
      url={https://arxiv.org/abs/2503.17064}, 
}

@article{hohenberg_rmp77,
  title = {Theory of dynamic critical phenomena},
  author = {Hohenberg, P. C. and Halperin, B. I.},
  journal = {Rev. Mod. Phys.},
  volume = {49},
  issue = {3},
  pages = {435--479},
  numpages = {0},
  year = {1977},
  month = {Jul},
  publisher = {American Physical Society},
  doi = {10.1103/RevModPhys.49.435},
  url = {https://link.aps.org/doi/10.1103/RevModPhys.49.435}
}

@article{kardar_prl86,
  title = {Dynamic Scaling of Growing Interfaces},
  author = {Kardar, Mehran and Parisi, Giorgio and Zhang, Yi-Cheng},
  journal = {Phys. Rev. Lett.},
  volume = {56},
  issue = {9},
  pages = {889--892},
  numpages = {0},
  year = {1986},
  month = {Mar},
  publisher = {American Physical Society},
  doi = {10.1103/PhysRevLett.56.889},
  url = {https://link.aps.org/doi/10.1103/PhysRevLett.56.889}
}

@article{vicsek_prl95,
  title = {Novel Type of Phase Transition in a System of Self-Driven Particles},
  author = {Vicsek, Tam\'as and Czir\'ok, Andr\'as and Ben-Jacob, Eshel and Cohen, Inon and Shochet, Ofer},
  journal = {Phys. Rev. Lett.},
  volume = {75},
  issue = {6},
  pages = {1226--1229},
  numpages = {0},
  year = {1995},
  month = {Aug},
  publisher = {American Physical Society},
  doi = {10.1103/PhysRevLett.75.1226},
  url = {https://link.aps.org/doi/10.1103/PhysRevLett.75.1226}
}

@BOOK{stone_b09,
  title     = "Mathematics for physics",
  author    = "Stone, Michael and Goldbart, Paul",
  publisher = "Cambridge University Press",
  month     =  jul,
  year      =  2009,
  address   = "Cambridge, England",
  language  = "en"
}

@article{edwards_rspa82,
    author = {Edwards, Samuel Frederick and Wilkinson, D. R.},
    title = {The surface statistics of a granular aggregate},
    journal = {Proceedings of the Royal Society of London. A. Mathematical and Physical Sciences},
    volume = {381},
    number = {1780},
    pages = {17-31},
    year = {1982},
    month = {05},
    issn = {0080-4630},
    doi = {10.1098/rspa.1982.0056},
    url = {https://doi.org/10.1098/rspa.1982.0056},
    eprint = {https://royalsocietypublishing.org/rspa/article-pdf/381/1780/17/64766/rspa.1982.0056.pdf},
}

@article{fontaine_prl23,
  title = {Unpredicted Scaling of the One-Dimensional Kardar-Parisi-Zhang Equation},
  author = {Fontaine, C\^ome and Vercesi, Francesco and Brachet, Marc and Canet, L\'eonie},
  journal = {Phys. Rev. Lett.},
  volume = {131},
  issue = {24},
  pages = {247101},
  numpages = {6},
  year = {2023},
  month = {Dec},
  publisher = {American Physical Society},
  doi = {10.1103/PhysRevLett.131.247101},
  url = {https://link.aps.org/doi/10.1103/PhysRevLett.131.247101}
}

@article{cartes_prsa22,
    author = {Cartes, C. and Tirapegui, E. and Pandit, R. and Brachet, M.},
    title = {The Galerkin-truncated Burgers equation: crossover from inviscid-thermalized to Kardar–Parisi–Zhang scaling},
    journal = {Philosophical Transactions of the Royal Society A: Mathematical, Physical and Engineering Sciences},
    volume = {380},
    number = {2219},
    pages = {20210090},
    year = {2022},
    month = {01},
    issn = {1364-503X},
    doi = {10.1098/rsta.2021.0090},
    url = {https://doi.org/10.1098/rsta.2021.0090},
    eprint = {https://royalsocietypublishing.org/rsta/article-pdf/doi/10.1098/rsta.2021.0090/1322947/rsta.2021.0090.pdf},
}

@article{rodriguez_fernandez_pre22,
  title = {Anomalous ballistic scaling in the tensionless or inviscid Kardar-Parisi-Zhang equation},
  author = {Rodr\'{\i}guez-Fern\'andez, Enrique and Santalla, Silvia N. and Castro, Mario and Cuerno, Rodolfo},
  journal = {Phys. Rev. E},
  volume = {106},
  issue = {2},
  pages = {024802},
  numpages = {9},
  year = {2022},
  month = {Aug},
  publisher = {American Physical Society},
  doi = {10.1103/PhysRevE.106.024802},
  url = {https://link.aps.org/doi/10.1103/PhysRevE.106.024802}
}

@article{gosteva_pre24,
  title = {Inviscid fixed point of the multidimensional Burgers--Kardar-Parisi-Zhang equation},
  author = {Gosteva, Liubov and Tarpin, Malo and Wschebor, Nicol\'as and Canet, L\'eonie},
  journal = {Phys. Rev. E},
  volume = {110},
  issue = {5},
  pages = {054118},
  numpages = {17},
  year = {2024},
  month = {Nov},
  publisher = {American Physical Society},
  doi = {10.1103/PhysRevE.110.054118},
  url = {https://link.aps.org/doi/10.1103/PhysRevE.110.054118}
}

@article{keller_jtb70,
title = {Initiation of slime mold aggregation viewed as an instability},
journal = {Journal of Theoretical Biology},
volume = {26},
number = {3},
pages = {399-415},
year = {1970},
issn = {0022-5193},
doi = {https://doi.org/10.1016/0022-5193(70)90092-5},
url = {https://www.sciencedirect.com/science/article/pii/0022519370900925},
author = {Evelyn F. Keller and Lee A. Segel}
}
